\documentclass{aa}
\usepackage{graphicx}
\usepackage{natbib}
\usepackage{txfonts}
\usepackage{subfig}
\usepackage{float}
\usepackage[backref,breaklinks,colorlinks,citecolor=blue]{hyperref}
\usepackage[all]{hypcap}

\bibpunct{(}{)}{;}{a}{}{,} 

\begin{document}

\title{CSI 2264: Probing the inner disks of AA Tau-like systems in \newline NGC 2264\thanks{Based on
data from the \textit{Spitzer} and \textit{CoRoT} missions, as well as the Canada France Hawaii
Telescope (CFHT) MegaCam CCD, the European Southern Observatory (ESO) Very Large Telescope, and the
U.S. Naval Observatory. The \textit{CoRoT} space mission was developed and operated by the French
space agency CNES, with participation of ESA's RSSD and Science Programmes, Austria, Belgium, Brazil,
Germany, and Spain. MegaCam is a joint project of CFHT and CEA/DAPNIA, at the Canada-France-Hawaii
Telescope (CFHT), operated by the National Research Council (NRC) of Canada, the Institut National
des Sciences de l'Univers of the Centre National de la Recherche Scientifique (CNRS) of France, and
the University of Hawaii. 
} \thanks{Figures \ref{fig:app1} - \ref{fig:app4} are only available in electronic form.} }

\author{P. T. McGinnis\inst{1} \and S. H. P. Alencar\inst{1} \and M. M. Guimar\~aes\inst{2} \and 
 A. P. Sousa\inst{1} \and J. Stauffer\inst{3} \and J. Bouvier\inst{4,5} \and L. Rebull\inst{3} \and N. N. J. Fonseca\inst{1,4} 
\and L. Venuti\inst{4} \and L. Hillenbrand\inst{6} \and A. M. Cody \inst{3} \and P. S. Teixeira\inst{7}  
\and S. Aigrain\inst{8} \and F. Favata\inst{9} \and G. F\"ur\'esz\inst{10} \and F. J. Vrba\inst{11} 
\and E. Flaccomio\inst{12} \and N. J. Turner\inst{13} \and J. F. Gameiro\inst{14} \and C. Dougados\inst{4}
\and W. Herbst\inst{15} \and M. Morales-Calder\'on\inst{16} \and G. Micela\inst{12}}  

 \institute{Departamento de F\'{\i}sica -  ICEx - UFMG, Av. Ant\^onio Carlos, 6627, 30270-901 
 Belo Horizonte, MG, Brazil\\
\email{\href{mailto:pauline@fisica.ufmg.br}{pauline@fisica.ufmg.br}}
 \and Departamento de F\'{\i}sica e Matem\'atica - UFSJ - Rodovia MG 443, KM 7, 36420-000, Ouro Branco, MG, Brazil
 \and \textit{Spitzer} Science Center, California Institute of Technology, Pasadena, CA 91125, USA
 \and Univ. Grenoble Alpes, IPAG, F-38000 Grenoble, France
 \and CNRS, IPAG, F-38000 Grenoble, France
 \and Astronomy Department, California Institute of Technology, Pasadena, CA 91125, USA
 \and University of Vienna, Department of Astrophysics, T\"urkenschanzstr. 17, 1180 Vienna, Austria
 \and Department of Astrophysics, Denys Wilkinson Building, University of Oxford, Oxford OX1 3RH, UK
 \and European Space Agency, 8-10 rue Mario Nikis, F-75738 Paris, Cedex 15, France
 \and MIT Kavli Institute for Astrophysics and Space Research, 77 Mass Ave 37-582f, Cambridge, MA 02139
 \and U.S. Naval Observatory, Flagstaff Station, 10391 West Naval Observatory Road, Flagstaff, AZ 86001, USA
 \and INAF - Observatorio Astronomico di Palermo, Piazza del Parlamento 1, 90134, Palermo, Italy
 \and Jet Propulsion Laboratory, California Institute of Technology, Pasadena, CA 91109, USA
 \and Instituto de Astrof\'{\i}sica e Ci\^encias Espaciais and Faculdade de Ci\^encias, 
  Universidade do Porto, Rua das Estrelas, PT4150-762 Porto, Portugal
 \and Astronomy Department, Wesleyan University, Middletown, CT 06459, USA
 \and Centro de Astrobiolog\'{\i}a, Departamento de Astrof\'{\i}sica, INTA-CSIC, PO BOX 78, E-28691, ESAC 
  Campus, Villanueva de la Ca\~nada, Madrid, Spain
 }

\date{Received 8 December 2014; Accepted 6 February 2015}

\abstract
{The classical T Tauri star (CTTS) AA Tau has presented photometric variability that was attributed to an inner 
disk warp, caused by the interaction between the inner disk and an inclined magnetosphere. Previous studies 
of the young cluster NGC 2264 have shown that similar photometric behavior is common among CTTS.}
{The goal of this work is to investigate the main causes of the observed photometric variability of CTTS 
in NGC 2264 that present AA Tau-like light curves, and verify if an inner disk warp could be responsible 
for their observed variability.}
{In order to understand the mechanism causing these stars' photometric behavior, we investigate veiling
variability in their spectra and $u-r$ color variations and estimate parameters of the inner disk warp 
using an occultation model proposed for AA Tau. We also compare infrared \textit{Spitzer} IRAC and optical 
CoRoT light curves to analyze the dust responsible for the occultations.} 
{AA Tau-like variability proved to be transient on a timescale of a few years.
We ascribe this variability to stable accretion regimes and aperiodic variability to unstable accretion 
regimes and show that a transition, and even coexistence, between the two is common. We find evidence of hot 
spots associated with occultations, indicating that the occulting structures could be located at the base of 
accretion columns. We find average values of warp maximum height of 0.23 times its radial location, consistent 
with AA Tau, with variations of on average 11\% between rotation cycles. We also show that extinction laws in 
the inner disk indicate the presence of grains larger than interstellar grains.}
{The inner disk warp scenario is consistent with observations for all but one star with AA Tau-like variability 
in our sample. AA Tau-like systems are fairly common, comprising 14\% of CTTS observed in NGC 2264, though 
this number increases to 35\% among systems of mass $0.7 M_{\odot} \lesssim M \lesssim 2.0 M_{\odot}$. Assuming 
random inclinations, we estimate 
that nearly all systems in this mass range likely possess an inner disk warp. We attribute this 
to a possible change in magnetic field configurations among stars of lower mass.}


\titlerunning{Probing the inner disks of AA Tau-like systems in NGC 2264}

\authorrunning{P. T. McGinnis et al.}

\maketitle

\section{Introduction}\label{sec:intro} 

Magnetospheric accretion models account for most of the observed characteristics of classical T Tauri 
stars (CTTS), such as their strong ultraviolet (UV) and optical excess, as well as broad emission lines 
and forbidden emission lines \citep{shu94, hartmann94, muzerolle01, kurosawa06, lima10}. These stars 
have also been observed to have strong magnetic fields \citep{johns-krull01} and X-ray emission 
\citep[e.g.,][]{preibisch05}. Magnetohydrodynamics (MHD) models suggest that the stellar magnetosphere 
truncates the inner circumstellar disk at a distance of a few stellar radii, where the disk's matter 
stress equals the magnetic stress, forcing the material within this truncation radius to be accreted 
onto the star, following the magnetic field lines \citep{bessolaz08}. This forms what are called 
accretion columns, which are responsible for the broad emission lines observed, particularly those of 
the hydrogen Balmer series. When the free-falling material reaches the stellar surface, accretion shocks 
result in hot spots near the magnetic poles of the stars, causing the observed UV and optical excess. 
Part of the material from the disk can also be ejected in a disk wind, resulting in forbidden emission 
lines. These stars also present a strong infrared (IR) excess with regard to the photospheric flux because 
of the absorption and re-emission at longer wavelengths of stellar radiation by the circumstellar disk.

\begin{figure}
\centering
\includegraphics[width=9cm]{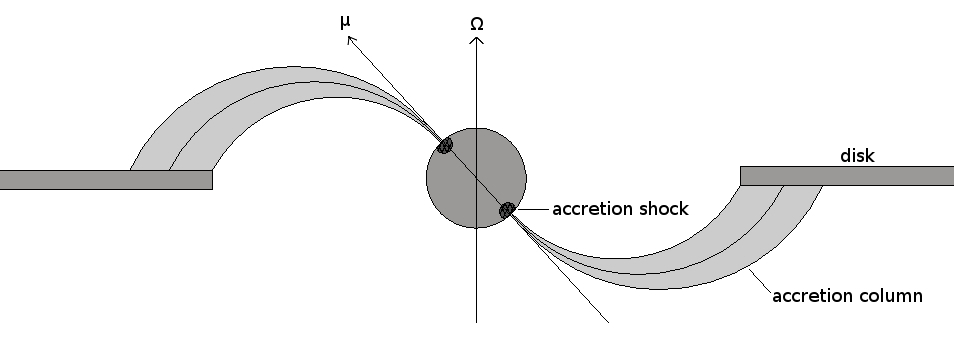}
\caption{When there is a misalignment between the axis of the (predominately dipolar) magnetic field 
($\mu$) and the stellar rotation axis ($\Omega$), magnetospheric accretion may occur via the stable 
accretion mechanism, where two main accretion columns form at the regions where the stellar magnetic 
pole is closest to the disk. Two main hot spots due to accretion shocks are located near the magnetic 
poles.
}\label{fig:maginclined}
\end{figure}

Magnetohydrodynamics models \citep[e.g.,][]{kulkarni09,romanova09,kurosawa13,romanova13} have 
proposed that CTTS can accrete matter from their circumstellar disks through a stable or an unstable regime. 
In the stable regime, a misalignment between the star's magnetic field and the star's rotation axis causes 
the material present in the region near the disk's truncation radius to fall onto the star via two main 
accretion columns, following the magnetic field lines connecting the disk to the star (Fig. 
\ref{fig:maginclined}). This interaction may also lift dust above the disk midplane, creating an optically 
thick warp in the inner disk region near the truncation radius, located at the base of these accretion columns. 
At the other end of each accretion column, an accretion shock on the stellar surface results in a hot spot, 
located at a high latitude. Observations of the magnetic fields of CTTS have shown that this misalignment 
between the star's magnetic field and the star's rotation axis is common \citep{gregory12}.

Magnetohydrodynamics models also predict that the stellar magnetic field interacts with the disk at and near the 
co-rotation radius. Differential rotation between the star and the disk, and throughout the accretion 
columns, results in a distortion of the magnetic field lines, causing them to inflate and eventually 
disconnect and reconnect, restoring the original field configuration \citep{zanni13}. 

Accretion through an unstable regime occurs when Rayleigh-Taylor instabilities between an accretion disk 
and a stellar magnetosphere cause matter to accrete onto the star via streams or tongues in random 
locations around the star, causing stochastic photometric variability \citep[see Fig. 1 in][]{kurosawa13}.

If a star undergoing stable accretion is seen at a medium or low inclination, its light curve should 
show periodic variability due to the modulation of its main hot spot. If it is seen at a high 
inclination (yet not so high that the flared outer disk completely occults the star), the inner disk 
warp will occult the stellar photosphere periodically, causing flux dips in the star's light curve.
If a star undergoing unstable accretion is seen at a medium or low inclination, its light curve will 
show stochastic variability due to the many hot spots on the stellar surface \citep{stauffer14a}. These 
hot spots are due to the various accretion shocks around the star. If these stars are seen at high 
inclinations, we can expect to observe stochastic occultations of the photosphere by dust lifted 
above the disk plane near the base of the accretion tongues whenever one passes in front of our line 
of sight, leading to light curves dominated by aperiodic extinction events. These aperiodic extinction 
events could also be due to occultation by dust that is lifted above the disk plane as a result of 
magnetorotational instabilities \citep{turner10}.

During a study of the photometric and spectroscopic variability of the classical T Tauri star AA Tauri, 
conducted between 1995 and 2007 \citep{bouvier99,bouvier03,bouvier07}, a simple geometrical model was 
presented to explain the behavior of AA Tau's light curve, characterized by a relatively constant level 
of brightness interrupted by quasi-periodic dips of around 1.4 magnitudes  
\citep[see Fig. 9 in][]{bouvier99}.
This variability was attributed to periodic occultations by an inner disk warp spatially associated 
with two major accretion columns, as in the stable accretion regime. This result provided great support 
for MHD models. 

It is interesting to note that after over 20 years of observations often showing the same quasi-periodic 
behavior, AA Tau's light curve suddenly changed quite drastically when its average brightness level 
decreased by over 2 magnitudes in 2011 \citep{bouvier13}, at which point it lost its periodicity. This 
indicates that, though the inner disk warp structure may be stable over the course of many years, it can 
disappear in a timescale of less than one year. 

There are many possible scenarios in which a warp can appear within the circumstellar disk of a T Tauri 
star besides the one described here \citep[see, e.g.,][]{terquem00, flaherty10}. 
In this paper, we treat the case of classical T Tauri stars, which are actively subject to ongoing magnetospheric 
accretion, with inner disk warps located at or near the disk's co-rotation radius (see discussion in Sect. 
\ref{sec:model}). It is reasonable to assume that a warp at this location is due to the interaction between an inclined 
magnetosphere and the inner disk, and therefore associated with the accretion columns, as was considered for 
AA Tau in \citet{bouvier07}. For the stars in this paper, we will consider this to be the main mechanism 
responsible for the inner disk warp.

\subsection{The occultation model proposed for AA Tau}\label{sec:introoccmodel}

\citet{bouvier99} considered that the dynamic interaction between the inclined magnetic field of the 
star and its inner disk region would cause the material in this part of the disk to be lifted, 
preferentially where the stellar magnetic dipole is closest to the disk, resulting in two arch-shaped 
walls located opposite to each other.  
This warp, located at or near the co-rotation radius ($R_{co}$) of the accretion disk, eclipses the stellar 
photosphere when it passes through the observer's line of sight. It has an azimuthal extension of $\phi_w$ 
and a maximum height of $h_{max}$ from the disk's mid plane, which occults the stellar photosphere 
at phase 0.5, provoking a minimum in the light curve. 

The model uses simple geometric principles to reproduce a synthetic light curve where an optically thick 
clump of given maximum height $h_{max}$ and azimuthal extension $\phi_w$, present at the co-rotation 
radius $R_{co}$, occults the stellar photosphere of a system seen at inclination $i$. The height of the 
clump is assumed to vary according to 

\begin{equation}
h(\phi)=h_{max}\left|\cos\left(\frac{\pi(\phi-\phi_0)}{\phi_w}\right)\right|
\end{equation}

\noindent
for values of $-\phi_w/2 \leq \phi \leq \phi_w/2$, where $\phi_0$ is the azimuth of maximum disk height. 
The clump height is equal to zero for all values of $|\phi| > \phi_w/2$.
In the case of AA Tau, the warp found to best reproduce its light curve had a maximum height of 
$h_{max}=0.3 R_{co}$, and an azimuthal extension of $360 ^{\circ}$.

\subsection{AA Tau-like light curves in NGC 2264}\label{sec:introaataungc}

In a study of CTTS in the young cluster NGC 2264 using light curves from the CoRoT satellite 
\citep[Convection, Rotation and planetary Transits,][]{baglin06}, various CTTS with light curves 
similar to that of AA Tau were found \citep{alencar10}. NGC 2264 is a young stellar cluster of 
approximately 3 Myr situated in the Orion spiral arm, around 760 pc from the Sun \citep[for 
a review of the cluster, see][]{dahm08}. Of the stars in the region observed by the CoRoT satellite 
in 2008, 301 were confirmed as members of the cluster, and of these 83 were classified as CTTS 
by \citet{alencar10}. 

\begin{figure}[t]
\centering
\includegraphics[width=9cm]{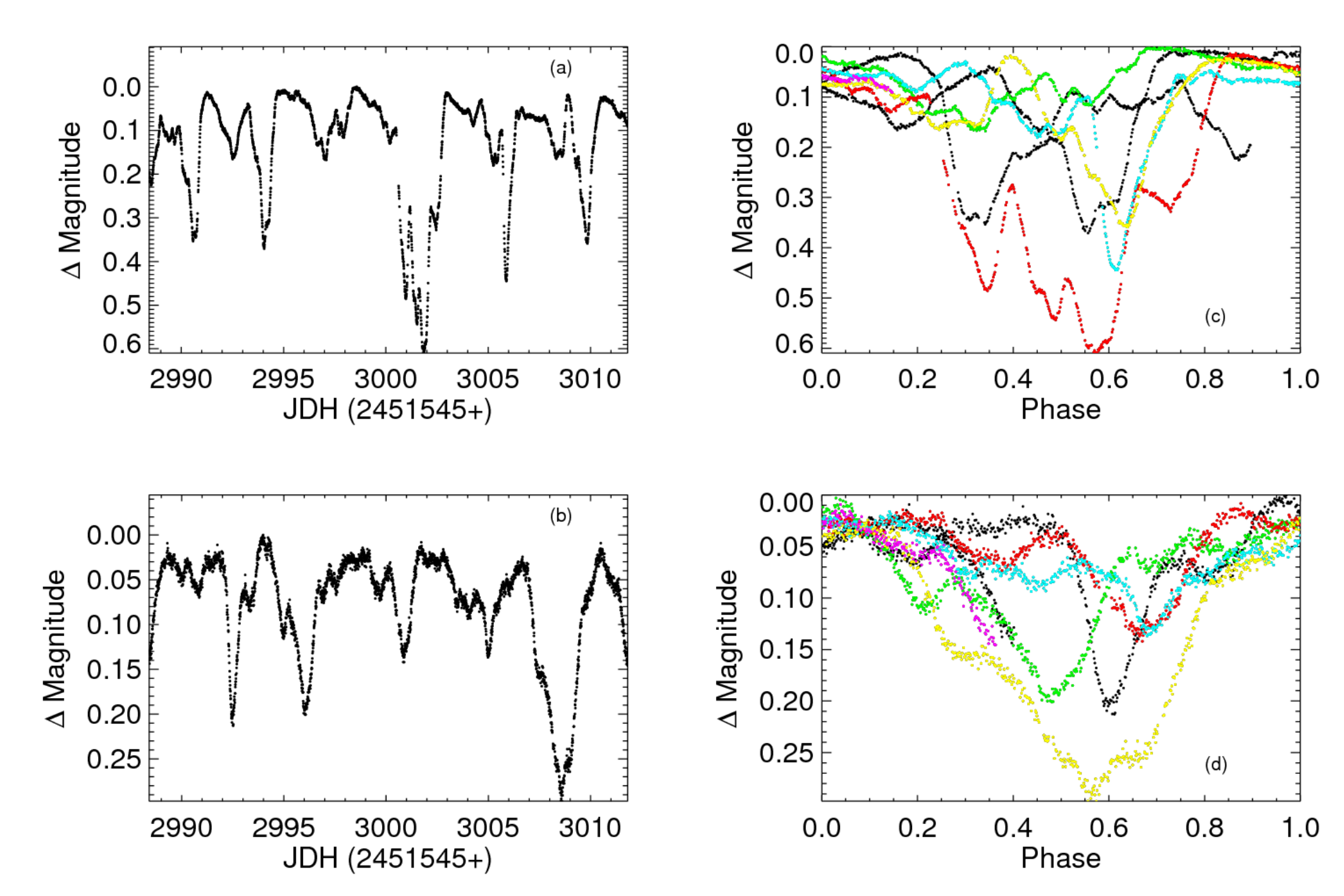}
\caption{Two examples of AA Tau-like light curves found for CTTS in NGC 2264 by \citet{alencar10}. 
(c) and (d) show the light curves (a) and (b), respectively, folded in phase. Different colors 
correspond to different rotation cycles. 
}\label{fig:aatauex}
\end{figure}

The 83 CTTS were separated into three groups, according to the morphology
of their light curves. The first group consisted of 28 periodic, nearly sinusoidal light
curves, whose variabilities were attributed to long-lived cold spots; the second group consisted 
of 23 light curves with the same characteristics as those of AA Tau, periodic but with
varying amplitude and shape of dips; and the third group consisted of 32 non-periodic 
light curves. Two examples of the AA Tau-like light curves, shown in full and folded in phase, can 
be seen in Fig. \ref{fig:aatauex}.

In the same study that identified these light curves, \textit{Spitzer} IRAC single epoch photometry 
was used to determine whether there is dust in the inner disk region of these stars, by identifying 
near-infrared excess emission. Of the 83 CTTS, 68 were also observed by \textit{Spitzer}. All of the 
stars with AA Tau-like light curves that were observed by \textit{Spitzer} and CoRoT were shown 
to present dust in the inner disk region. This is consistent with the assumption that the photometric 
behavior of these stars is due to obscuration by optically thick material in their inner disks.

In order to better understand the physical processes that govern CTTS variability, 
a new observational campaign of NGC 2264 was organized in December 2011. The Coordinated Synoptic 
Investigation of NGC 2264 \citep[CSI 2264,][]{cody14}\footnote{The CoRoT and Spitzer light curves 
for all probable NGC 2264 members, as well as our broad band photometry for these stars, are available 
at \url{http://irsa.ipac.caltech.edu/data/SPITZER/CSI2264}.} is composed of simultaneous observations on 
fifteen telescopes (eleven ground-based and four space-based telescopes), including the CoRoT 
satellite, the \textit{Spitzer} Infrared Array Camera \citep[IRAC,][]{fazio04}, the 
Canada-France-Hawaii Telescope (CFHT) MegaCam \citep{boulade03}, and the FLAMES multi-object 
spectrograph on the Very Large Telescope \citep[VLT,][]{pasquini02}. This campaign provided us 
with simultaneous photometric and spectroscopic information in a wide range of wavelengths. New 
CTTS were identified according to their UV excess \citep{venuti14}, and various new stars with 
AA Tau-like light curves were found, in addition to stars with newly identified types of photometric 
behavior \citep{stauffer14a, stauffer14b}.

Because of the large number of previous studies of NGC 2264, it was possible to establish reliable 
criteria for cluster membership, as discussed in \citet{alencar10}. As probable cluster members, 
we used photometric H$\alpha$ and variability from \citet{lamm04} following their criteria; X-ray 
detection from \citet{ramirez04} and \citet{flaccomio06} and location on the cluster sequence in 
the $(I,R-I)$ diagram, when available; spectroscopic H$\alpha$ equivalent width greater than 
a spectral type dependent threshold \citep{white03}; and H$\alpha$ emission line width at 10\% 
intensity greater than 270 km/s, as proposed by \citet{white03} to identify accreting T Tauri stars.

In this study we present a multi-wavelength analysis of the CTTS members of NGC 2264 that appear 
to have extinction dominated photometric behavior. In Sect. \ref{sec:obs} we present the observations 
used in this study, while in Sect. \ref{sec:analysis} we show the analysis of our data and our results.  
In Sect. \ref{sec:discuss} we present a discussion on the proposed cause of these stars' photometric 
variabilities, and in Sect. \ref{sec:conc} we present our conclusions.

\section{Observations}\label{sec:obs}

\subsection{CoRoT data}\label{sec:corotobs}

The first CoRoT short run occurred in March 2008, when the young cluster NGC 2264 was 
observed for 23 days uninterruptedly. Nearly four years later, during the CSI 2264 campaign, 
the cluster was observed once again by CoRoT from 2011 December 1, to 2012 January 3, 
during the satellite's fifth short run. CoRoT's field of view is larger than the full extent 
of the cluster, which fits entirely into one of the CCDs designed for studying exoplanets. 
Stars of magnitude down to R $\sim$
18 were observed, with a cadence of 512 seconds. In some cases the data were taken with
exposure times of 32 seconds to provide high cadence light curves. All of the light
curves from the first epoch were re-binned to 512 seconds.

CoRoT observes in a white light bandpass of approximately 3700\AA \space to
10000\AA. It can, in principle, provide three-color photometry by passing the                    
stellar light through a low-resolution spectral dispersing prism, but for the purpose of this 
paper we used only the integrated white light flux.

After undergoing standard CoRoT pipeline procedures, detailed in \cite{samadi06}, the data 
were subsequently processed to remove unwanted effects, such as those caused by Earth 
eclipses, and to remove measurements affected by the South Atlantic Anomaly, using a 
sigma-clipping filter. Care was taken not to remove flaring events. In some cases, jumps
due to detector temperature jumps were present in the photometry, and when possible they
were removed manually \citep[for more details on the correction of CoRoT light curve 
systematics, see][]{cody14}.

\subsection{\textit{Spitzer} IRAC data}\label{sec:iracobs}

The \textit{Spitzer} Space Telescope mapped a region of $\sim 0.8^{\circ} \times 0.8^{\circ}$ centered at 
R.A. 06:40:45.0, declination +09:40:40, from 2011 December 3, to 2012 January 1. Targets were 
observed approximately twelve times a day in the IRAC $3.6 \mu \mathrm{m}$ and $4.5 \mu \mathrm{m}$ 
channels in its Warm Mission mode. Most of the objects have data from both bands, but around 
40\% of the objects fall near the edges of the mapping region, and only have data from one band.
A staring mode was also used around the beginning of the run for a region near the center of the 
cluster. A cadence of approximately 15 seconds was obtained during four blocks of 20, 26, 16, and 
19 hours, on 2011 December 3, December 5-6, December 7-8, and December 8-9. 

The IRAC data reduction procedures and production of IRAC light curves are explained in detail 
in \citet{cody14} and \citet{rebull14}.  

\subsection{CFHT MegaCam data}\label{sec:cfhtobs}

NGC 2264 was observed using the Canada France Hawaii Telescope's wide-field imager MegaCam 
\citep{boulade03} between 2012 February 14 and 2012 February 28. The MegaCam's field of view spans a full 
$1\times1$ square degree, covering the cluster with only one pointing. Data were taken in the 
\emph{u} and \emph{r} bands on 30 epochs. Reduction of these data is explained in \citet{venuti14}.

\subsection{USNO data}\label{sec:usnoobs}

On the order of 900 epochs of I-band photometric observations of NGC 2264 were obtained at the US Naval 
Observatory 40" telescope from December 2011 to March 2012. A field of view of 23'x23', 
centered approximately on the cluster center, was observed. The data underwent standard reduction 
procedures using bias and dome flat images that were obtained at the beginning of each night. 
Aperture photometry was performed for all stars in each image, and a set of non-variable stars 
was identified in order to establish zero-points for each CCD frame and construct light curves of 
the targets using differential photometry.

\subsection{Spectroscopy}\label{sec:specobs}

For some stars in our sample, 20-22 epochs of VLT FLAMES spectra were obtained between 2011 
December 4, and 2012 February 29. Of these, we have four to six epochs of simultaneous FLAMES   
spectroscopy and CoRoT photometry. 

FLAMES can obtain medium- and high-resolution spectra for multiple objects over a field of 
view 25 arcmin in diameter. Since this is much smaller than the extent of NGC 2264, even with two 
pointings, only the central part of the cluster was observed (one pointing centered at R.A. 06:41:04.8, 
declination +09:45:00, the other centered at R.A. 06:40:58.8, declination +09:21:54). 
We used the HR15N grating of the GIRAFFE/MEDUSA 
mode, which covers H$\alpha$ and Li 6707\AA \space with a resolution of 17,000. A few of the brighter 
stars were observed with the UVES (Ultraviolet and Visual Echelle Spectrograph) red mode, which 
covers the region from 4800\AA \space to 6800\AA \space with a resolution of 38,700. Only one of the stars 
treated in this paper was observed with UVES.

Where FLAMES spectra were not available, we used high-resolution Hectochelle spectra from 2004-2005 
provided by Gabor F\H ur\'esz \citep{furesz06}. They used the 190\AA \space wide order centered at H$\alpha$, 
with a resolution of R$\sim 34000$. These spectra were used for eight stars.

\section{Data analysis and results}\label{sec:analysis} 

\subsection{Initial selection of candidates for extinction dominated light curves}\label{sec:selec}

Throughout this paper we are concerned with stars that present extinction events due to 
circumstellar material eclipsing the stellar photosphere. This requires that the stars in our sample 
be classical T Tauri stars, with at least some optically thick dust still present in the disk.  
Therefore, after selecting the probable members of NGC 2264 following criteria described in 
\citet{alencar10}, a subset of stars with CoRoT observations in either epoch was classified as 
CTTS based on the following criteria: H$\alpha$ width at $10\%$ intensity greater than 270 km/s, 
H$\alpha$ equivalent width higher than a spectral type dependent criterion proposed by \citet{white03}, 
or U-V excess greater than a spectral type dependent threshold described in \citet{alencar10}. 
A new group of CTTS candidates proposed by \citet{venuti14} based on UV excess measured with the CFHT 
Megacam during the CSI 2264 campaign was also taken into consideration. We also consider the 
two stars Mon-56 and Mon-14132 as candidate CTTS since they have IR excess and photometric 
variability characteristic of the CTTS studied in this paper, though we do not have enough spectral 
information to classify them as CTTS and they present no considerable UV excess. 

Of the $\sim$ 300 confirmed members observed in 2008, we classified 95 as CTTS or CTTS candidates, 
12 more than had been considered by \citet{alencar10}. Of the $\sim$ 500 confirmed members observed 
in 2011, 148 were classified as CTTS or candidates. Of all of these CTTS and CTTS candidates, 84 
were observed in both epochs. Therefore our initial sample of possible CTTS is of 159 cluster members, 
observed with CoRoT in either 2008, 2011, or both. Sousa et al. (in prep.) performed an in-depth study of all 
of these CTTS, focusing on the VLT spectroscopy.

\begin{figure*}[ht]
\centering
\includegraphics[width=15cm]{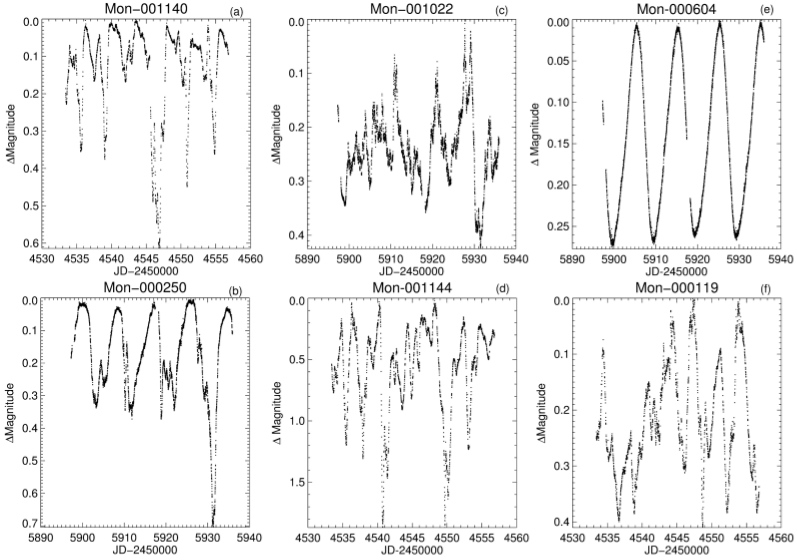}
\caption{Examples of different light curve morphologies found in CTTS in NGC 2264. (a) and (b) show 
periodic light curves attributed to circumstellar extinction, (c) shows an accretion burst 
dominated light curve, (d) shows a light curve dominated by aperiodic extinction, (e) shows a 
spot-like light curve, and (f) shows a light curve whose morphology is very complex, probably as 
a result of a combination of processes.
}\label{fig:morph}
\end{figure*}

The CoRoT light curves of these 159 stars were separated into different groups based on their visual 
morphologies. This was done based solely on the CoRoT light curves, with no color information. 
One of these groups consists of AA Tau-like light curves, which present a relatively stable maximum
brightness level interrupted by quasi-periodic flux dips that are neither spot-like nor eclipse-like 
in nature. The term quasi-periodic means that the flux dips present a stable period but vary in shape and/or
amplitude between rotation cycles. They vary by on average 25\% in width and 30\% in 
amplitude from one rotation cycle to the next. Their periods range from 3 to 10 days and are consistent 
with Keplerian rotation near the co-rotation radius (see Fig. \ref{fig:per} and discussion in Sect. 
\ref{sec:model}). They are usually accompanied by other narrower, shallower flux dips that are not necessarily 
periodic and can have significant structures within the main flux dips.

\citet{bouvier07} showed that the photometric behavior of the star AA Tau varied significantly between 
observations separated by a few years, showing at times the quasi-periodic flux dips attributed to the 
inner disk warp, at times more than one flux dip per stellar rotation period, and at times with almost no 
variability \citep[see Fig. 15 in ][]{bouvier07}. The light curves we have described as AA Tau-like 
are based on the observations of AA Tau from 1995. There are, however, some light curves that are similar
to the 1995 AA Tau light curve, but that either do not have a well-defined maximum brightness, 
or show relatively weak periodicity (for example, see star Mon-1054 in Fig. \ref{fig:mon1054}). 
\citet{cody14} do not classify these as AA Tau-like light curves, but they are important in the context 
of this paper because they may represent systems that are undergoing events similar to those that AA 
Tau underwent during its other phases. Therefore, we consider stars that present these
light curves in one CoRoT observation or the other to be candidate AA Tau systems along with the stars
that have classical AA Tau-like light curves. We analyze their photometric variability in the same way
as we do the AA Tau-like light curves, in order to establish whether the stable accretion mechanism
viewed at a high inclination could also be responsible for their variability.

Another group that is of interest to this study is composed of light curves with similar characteristics 
to the AA Tau-like light curves, but that show no obvious periodicity. These light curves appear to be 
dominated by aperiodic extinction events. These two groups can easily be distinguished from the periodic, 
nearly sinusoidal, variability attributed to long-lived cold spots, or the nearly symmetric, narrow flux 
bursts described in \citet{stauffer14a} and attributed to stochastic accretion events. A few examples of 
different types of variability are shown in Fig. \ref{fig:morph}. For a detailed discussion of these 
and other light curve morphologies, including a statistical analysis of symmetry in the flux and 
periodicity, see \citet{cody14}.

\begin{figure*}[hbt]
\centering
\includegraphics[width=6.5cm]{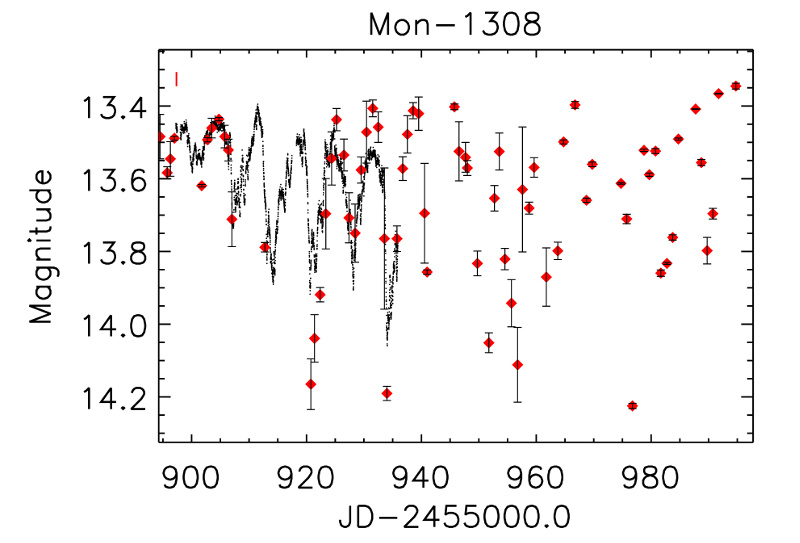}
\includegraphics[width=6.5cm]{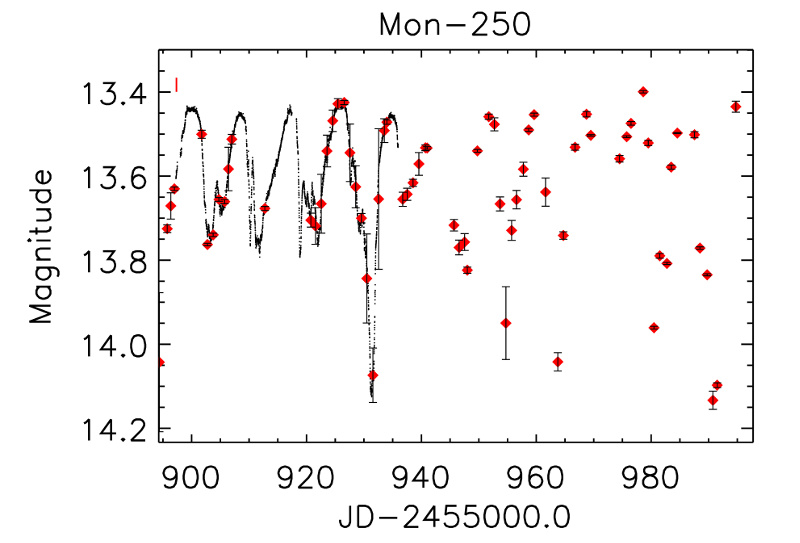}
\caption{I-band and CoRoT photometry, shifted in magnitude to coincide, for the AA Tau-like 
stars Mon-1308 and Mon-250. The high cadence space-based observations allow the 
AA Tau-like flux dips to be easily identified, while with even high quality ground-based 
photometry it is difficult to obtain enough information for the variability to be quantified.
}\label{fig:ci}
\end{figure*}

Fig. \ref{fig:ci} shows the CoRoT light curve and USNO I-band light curve, shifted to coincide in 
magnitude, of two AA Tau-like stars. It is evident from this figure, though the I-band photometry coincides 
closely with the CoRoT photometry, that the lower cadence makes it difficult for AA Tau-like behavior 
to be identified using only ground-based data. This shows the importance of high cadence space-based 
photometry to better understand the phenomena that occur among these young stellar objects. 

Table \ref{table:uxori2} shows information on the stars in our sample that are considered to have 
AA Tau-like or aperiodic extinction dominated light curves in either epoch. The CSIMon ID is an internal 
naming scheme devised for the CSI 2264 campaign \citep[see][]{cody14}, and comprises all cluster 
members, candidates and field stars in the NGC 2264 field of view (for brevity, throughout the text, 
we eliminate `CSI' and the leading zeros and replace, for instance, CSIMon-000250 with Mon-250). 
The CoRoT ID is the identification number assigned by the CoRoT satellite to all objects it has 
observed. In some cases where the same star was observed in both epochs, two different CoRoT IDs 
were assigned to the same star. For these stars we present the ID from the 2008 observation. The 
2MASS identification is also presented. 

The sixth and seventh columns of Table \ref{table:uxori2} show how the light curve was classified in 
one epoch or the other. Those labeled with AAT present AA Tau-like light curves, those labeled with 
AATc are those that were not classified as AA Tau-like by \citet{cody14}, but that we consider to be 
candidate AA Tau systems, and those labeled as Ap show aperiodic light curves. For six stars, 
a possible AA Tau-like variability was found in both epochs, meaning the mechanism responsible 
for this variability can be stable on a timescale of a few years. Four stars that present AA Tau-like 
variability were only observed in one epoch. There are thirteen stars that 
were observed in both epochs and classified as AA Tau-like in one and aperiodic in the other. Of these, 
five displayed AA Tau-like behavior in 2008, yet in 2011 presented aperiodic variability, 
while the other eight showed the opposite trend. This points to a possible change from a stable 
accretion regime to one of unstable accretion (or vice-versa) in a matter of less than four 
years. This is discussed in Sect. \ref{sec:stab}. Another five stars were classified as having an aperiodic 
extinction dominated light curve in both epochs. Five more stars with aperiodic extinction dominated 
light curves were only observed in 2011.  

It is interesting to note that, though many stars suffered a transition between AA Tau-like and aperiodic 
extinction dominated light curves, no star whose light curve was classified as AA Tau-like in one CoRoT 
observing run was classified as spot-like in 
the other. This could be due to inclinations, since high inclinations would favor extinction events 
and possibly hide any variability due to spots on the stellar surface. 

\begin{sidewaystable*}[p]
\begin{center}
\caption{Stars with AA Tau-like and variable extinction dominated light curves. 
}\label{table:uxori2} 
\small{
\begin{tabular}{l c c c c c c c c c c c c}
\hline
\hline
 CSIMon ID & CoRoT ID & 2MASS & R. A. & DEC & Sp type & LC type & LC type & P (days) & P (days) & $v\sin i$ & $\alpha_{IRAC}$ & [F(disk)/F(star)]$_{4.5 \mu \mathrm{m}}$ \\ 
           &          &       &       &     &               & (2008)  & (2011)  & (2008)   & (2011)   &    (km/s) &  &  \\ 
\hline
CSIMon-000056\tablefootmark{a} & 223994760 & 06415315+0950474 & 6:41:53.16 & +9:50:47.4 & K5 & AAT & AAT & 5.71 & 5.86 & 11.1$\pm$2.2 & -1.21 & 1.0 \\ 
CSIMon-000126  & 616895876 & 06405783+0941201 & 6:40:57.84 &  +9:41:20.1 & M0 &  --  & Ap   &   --  & 0.35\tablefootmark{b} & 14.0$\pm$2.0 & -1.04 & 4.8 \\ 
CSIMon-000250* & 223980688 & 06410050+0945031 & 6:41:00.50 &  +9:45:03.1 & K3 & AATc & AAT  &  8.32 & 8.93 &  9.8$\pm$2.2 & -2.10 & 2.1 \\ 
CSIMon-000296* & 500007209 & 06405059+0954573 & 6:40:50.59 &  +9:54:57.3 & K2 & Ap   & AAT  &  2.51\tablefootmark{b} & 3.91 & 17.3$\pm$2.6 & -2.14 & 1.6 \\ 
CSIMon-000297* & 223976747 & 06404516+0928444 & 6:40:45.16 &  +9:28:44.4 & K2 & AAT  & Ap   &  3.16 &  --  & 30.8$\pm$1.8 & -2.24 & 1.6 \\ 
CSIMon-000314  & 500007930 & 06404459+0932261 & 6:40:44.59 &  +9:32:26.2 & M3 & Ap   & Ap   &   --  & 3.38\tablefootmark{b} & 24.0$\pm$2.0 & -2.61 & 0.7 \\ 
CSIMon-000325\tablefootmark{a} & 605538641 & 06405934+0955201 & 6:40:59.34 & +9:55:20.2 & G5V\tablefootmark{c} &-- & Ap & -- & -- & --      & -1.38 & -- \\  
CSIMon-000358  & 400007959 & 06410673+0947275 & 6:41:06.73 &  +9:47:27.6 & M3 & Ap   & AAT  &   --  & 5.86 &  --          & -1.55 & 1.3 \\ 
CSIMon-000379* & 223981811 & 06410497+0950460 & 6:41:04.97 &  +9:50:46.1 & K2 & Ap   & AATc &  3.73\tablefootmark{b} & 3.68 & 25.0$\pm$1.2 & -1.91 & 2.2 \\ 
CSIMon-000433  & 616919770 & 06410111+0934522 & 6:41:01.11 &  +9:34:52.2 & M1 &  --  & Ap   &   --  &  --  &  --          & -2.05 & -- \\  
CSIMon-000441* & 223980048 & 06405809+0936533 & 6:40:58.10 &  +9:36:53.3 & M2 & AAT  & Ap   &  4.06 &  --  &  --          & -1.41 & 1.0 \\ 
CSIMon-000456  & 616872585 & 06405154+0943242 & 6:40:51.54 &  +9:43:24.2 & K4 &  --  & AAT  &   --  & 5.03 & 19.3$\pm$1.5 & -1.62 & 2.5 \\ 
CSIMon-000498* & 500007120 & 06404750+0949289 & 6:40:47.50 &  +9:49:28.9 & K3 & AATc & AAT  &  4.23 & 4.28 & 24.1$\pm$2.4 & -1.41 & 2.9 \\ 
CSIMon-000619  & 603402479 & 06411475+0934134 & 6:41:14.75 &  +9:34:13.4 & K8.5 & Ap   & Ap   &   --  &  --  &  --          & -1.11 & 11.5 \\ 
CSIMon-000654* & 500007610 & 06405949+0929517 & 6:40:59.50 &  +9:29:51.7 & M3 & AATc & Ap   &  4.66 &  --  & 18.0$\pm$3.0 & -0.92 & -- \\ 
CSIMon-000660* & 223980693 & 06410051+0929159 & 6:41:00.51 &  +9:29:15.9 & K4 & AAT  & AAT  &  5.25 & 5.25 & 21.4$\pm$2.5 & -1.72 & 3.4 \\ 
CSIMon-000667* & 223987997 & 06412878+0938388 & 6:41:28.78 &  +9:38:38.8 & K3 & Ap   & Ap   &   --  &  --  & 24.0$\pm$2.0 & -0.54 & 1.5 \\ 
CSIMon-000717  & 616943877 & 06411511+0926443 & 6:41:15.11 &  +9:26:44.3 & M0.5&  --  & Ap   &   --  &  --  &  --          & -1.43 & 1.4 \\ 
CSIMon-000774$^{\dagger}$ & 223980264 & 06405884+0930573 & 6:40:58.84 &  +9:30:57.3 & K2.5 & AAT & Ap & 3.46 & -- & 30.7$\pm$1.3 & -1.25 & 2.1 \\ 
CSIMon-000811* & 500007460 & 06404321+0947072 & 6:40:43.21 &  +9:47:07.2 & K6 & Ap   & AAT  &   --  & 7.88 & 13.5$\pm$1.7 & -1.08 & 3.0 \\ 
CSIMon-000824* & 223981023 & 06410183+0938411 & 6:41:01.84 &  +9:38:41.1 & K4 & AAT  &  --  &  7.05 &  --  & 14.8$\pm$1.4 & -2.02 & -- \\ 
CSIMon-000928* & 223987178 & 06412562+0934429 & 6:41:25.62 &  +9:34:43.0 & M0 & AAT  & Ap   &  9.92 &  --  & 20.1$\pm$2.8 & -1.11 & 1.4 \\ 
CSIMon-001037  & 223973200 & 06403086+0934405 & 6:40:30.86 &  +9:34:40.5 & K1 & Ap   & Ap   &   --  & 8.93\tablefootmark{b} & 32.0$\pm$3.0 & -1.29 & 11.3 \\
CSIMon-001038  & 602095739 & 06402262+0949462 & 6:40:22.62 &  +9:49:46.3 & M0 &  --  & Ap   &   --  &  --  &  --          & -1.73 & 3.0 \\ 
CSIMon-001054* & 400007538 & 06403652+0950456 & 6:40:36.52 &  +9:50:45.6 & M2 & Ap   & AATc &   --  & 4.08 & 20.4$\pm$2.0 & -1.10 & 2.0 \\ 
CSIMon-001131  & 223957455 & 06393441+0954512 & 6:39:34.41 &  +9:54:51.2 & M2 & Ap   & AAT  &  5.15\tablefootmark{b} & 5.18 &  8.8$\pm$1.7 & -1.38 & 1.4 \\ 
CSIMon-001140  & 223959618 & 06394147+0946196 & 6:39:41.47 &  +9:46:19.7 & K4 & AAT  & AAT  &  3.87 & 3.90 & 19.8$\pm$2.3 & -2.37 & 0.8 \\ 
CSIMon-001144* & 223971231 & 06402309+0927423 & 6:40:23.09 &  +9:27:42.3 & K5 & Ap   & Ap   &   --  &  --  & 24.0$\pm$3.0 & -1.73 & 3.75 \\ 
CSIMon-001167* & 400007528 & 06403787+0934540 & 6:40:37.87 &  +9:34:54.0 & M3 & Ap   & AATc &  8.30\tablefootmark{b} & 8.78 & 11.3$\pm$2.1 & -1.87 & 1.2 \\ 
CSIMon-001296  & 223948127 & 06390374+0940234 & 6:39:03.75 &  +9:40:23.4 & K7 & Ap   & AAT  &   --  & 9.75 & 11.0$\pm$1.6 & -1.47 & -- \\ 
CSIMon-001308  & 223964667 & 06395924+0927245 & 6:39:59.24 &  +9:27:24.5 & M0 & AAT  & AAT  &  6.45 & 6.68 &  9.9$\pm$2.0 & -1.87 & 0.7 \\ 
CSIMon-006986  & 223950070 & 06390996+1005127 & 6:39:09.96 & +10:05:12.7 & K7:M1 & AATc &  --  &  6.16 &  --  &  --          & -1.22 & -- \\ 
CSIMon-014132\tablefootmark{a} & 602070634 & 06390355+0916159 & 6:39:03.55 & +9:16:16.0 & M3.5 & -- & AAT & -- & 4.48 & --   & -1.50 & -- \\ 
\hline
\end{tabular}
\tablefoot{
The first three columns show the 
identification of each star using our internal CSI 2264 naming scheme, the identification assigned to each 
star by the CoRoT satellite, and the respective 2MASS identification. In the cases where a star was 
observed twice by CoRoT and assigned a different identification in each epoch, we use the one given in 
the first epoch. 
An asterisk following the CSIMon ID indicates that we have FLAMES/GIRAFFE spectra, and a dagger 
indicates that we have FLAMES/UVES spectra, both simultaneous with the second CoRoT epoch. 
The fourth and fifth columns show each star's coordinates, and the sixth column shows their 
spectral types, taken from \citet{venuti14}, except where specified. The 
seventh and eighth columns show the classification given to the light curve (LC) in each epoch (2008 
March observations and 2011 December observations). The ninth and tenth columns show the 
periods that were measured for each CoRoT light curve in the first (2008) and second (2011) epochs. 
The eleventh column shows the values of projected rotational velocity $v\sin i$ measured using FLAMES 
spectra when available, or Hectochelle spectra otherwise. The twelfth column shows the $\alpha_{IRAC}$ 
index calculated for each star, which corresponds to the slope of the spectral energy distribution 
between $3.6 \mu\mathrm{m}$ and $8 \mu\mathrm{m}$. The thirteenth column shows the ratio of disk flux to 
stellar flux at $4.5 \mu\mathrm{m}$, estimated from each star's SED. \\
\tablefoottext{a}{There is not enough spectral information to classify these stars as CTTS, though they have 
IR excess and show variability characteristic of CTTS.}
\tablefoottext{b}{Low probability periods found during a periodogram analysis of the light curves.}
\tablefoottext{c}{Spectral type from \citet{young78}.}
}
}
\end{center}
\end{sidewaystable*}

Another group of stars with what appear to be short duration extinction events present in their 
CoRoT light curves is studied by \citet{stauffer14b}. These stars present narrow flux dips that are 
generally shallower than in AA Tau-like light curves, and nearly Gaussian shaped, while AA Tau-like 
light curves usually show very irregularly shaped dips, often with significant structure in the 
minima. The narrow dip stars are periodic over timescales of tens of days, with periods 
that are consistent with Keplerian rotation periods of the inner disk region. \citet{stauffer14b} 
argue that the inner disk warp is not adequate to explain these flux dips because of their short 
durations and regular shapes, and propose alternate scenarios to account for the variabilities. 
Two of the stars in our sample of AA Tau-like light curves are considered in their paper as well. They 
are Mon-56 and Mon-1131, and they present a rather complex photometric behavior. Therefore they 
are studied therein through different points of view, in order to investigate other possible causes 
for their variabilities. 
There are also some stars with AA Tau-like light curves that at times present these narrow 
flux dips along with the broader ones. These features of their light curves are also discussed 
in \citet{stauffer14b}.

\subsection{Veiling and variability}\label{sec:veil}

\subsubsection{Correlation between photometric variability and veiling}\label{sec:veil1}

Veiling occurs as a result of a strong UV and optical excess produced in the accretion shocks on the stellar 
surface of CTTS. It is present in spectra when the hot spot caused by these accretion shocks is 
visible to the observer. This UV excess results in an added blue continuum to the photosphere's 
spectra, resulting in somewhat shallower absorption lines than those from a pure photosphere. 
For the CTTS in NGC 2264 for which we obtained FLAMES spectra, we measured the veiling and its 
variability, using the equivalent widths of the Li I 6707.8\AA \space and Fe I 6633.4\AA \space lines. These 
lines were chosen since they are the absorption lines with the best signal-to-noise ratio in our 
spectra. As veiling increases, the equivalent widths we measure will decrease.

\begin{figure*}[t]
\centering
\hspace{0.1cm}
\includegraphics[height=4.0cm]{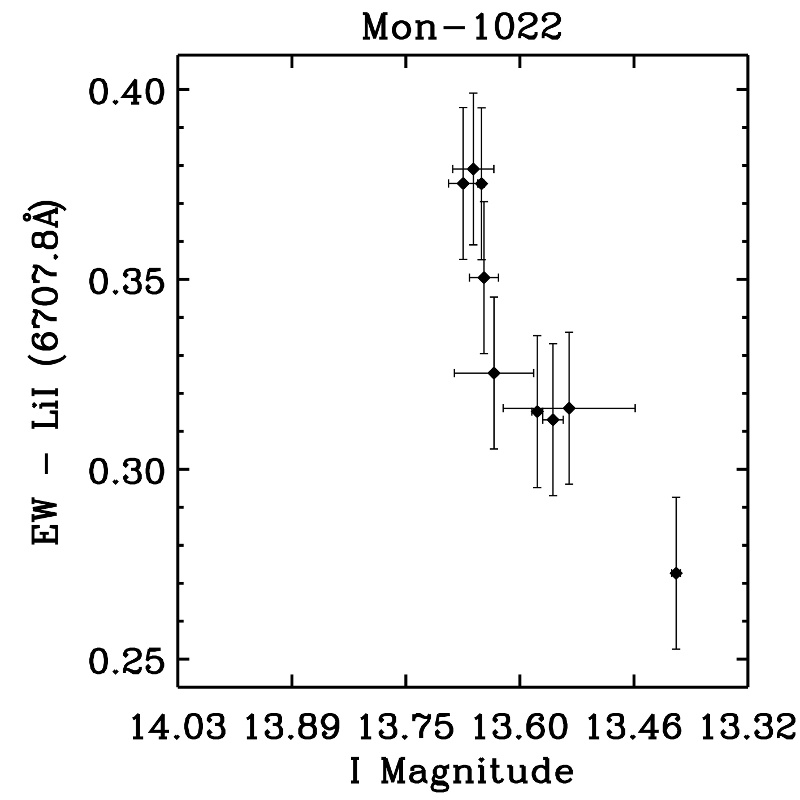}
\hspace{1.8cm}
\includegraphics[height=4.0cm]{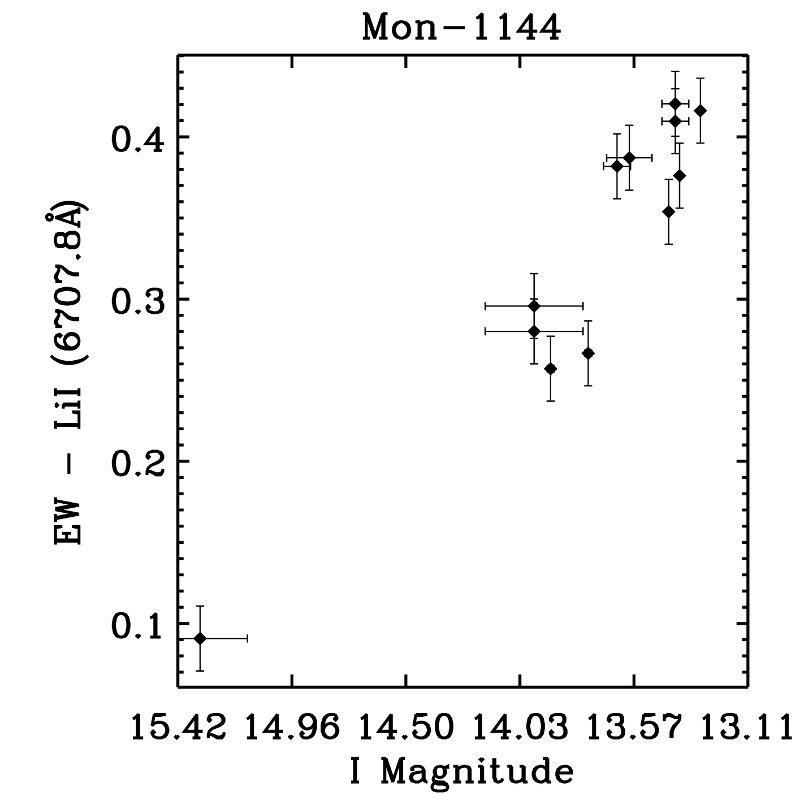}
\hspace{1.8cm}
\includegraphics[height=4.0cm]{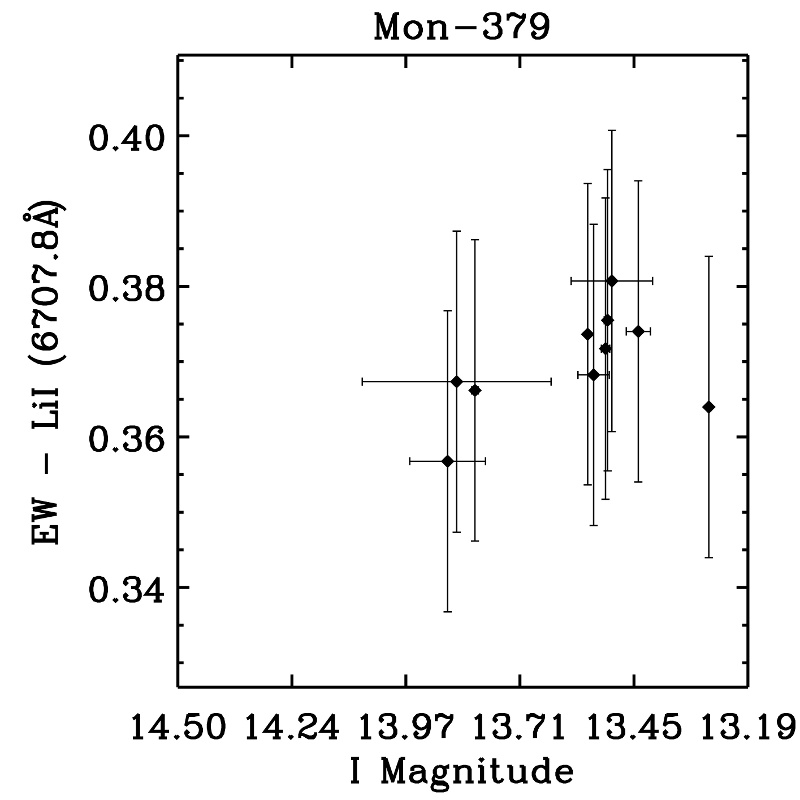}
\hspace{2.0cm}
\includegraphics[height=4.0cm]{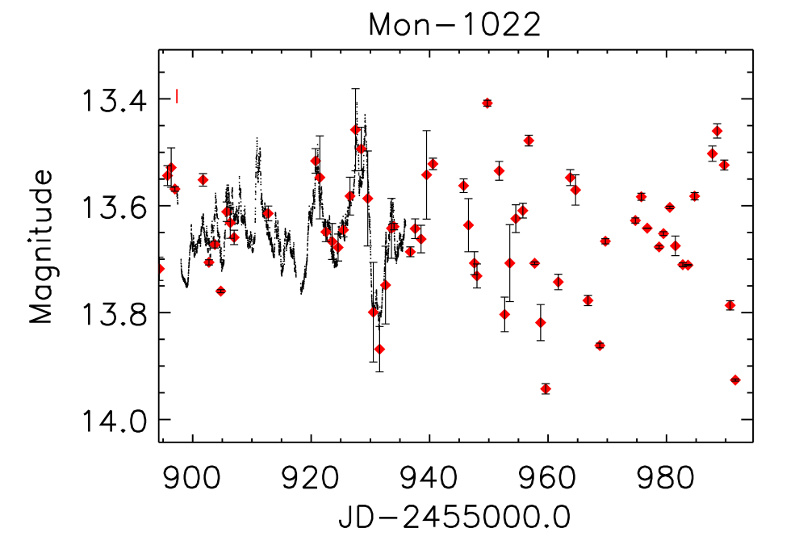}
\includegraphics[height=4.0cm]{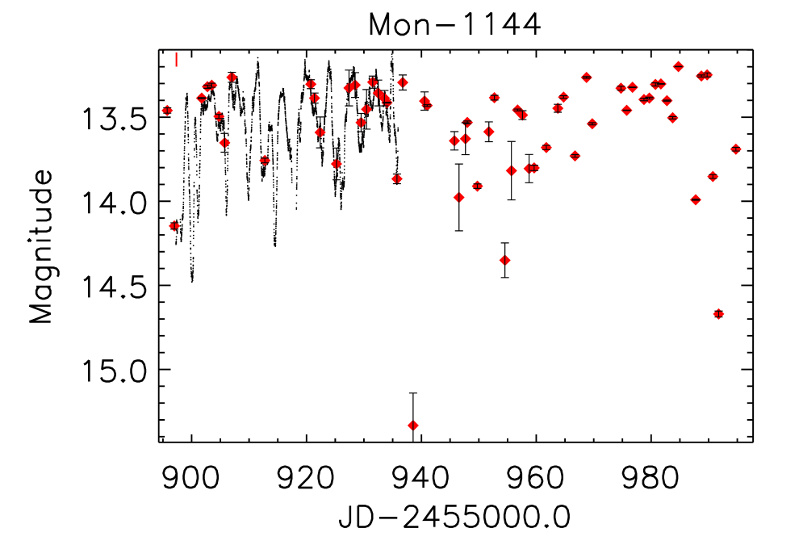}
\includegraphics[height=4.0cm]{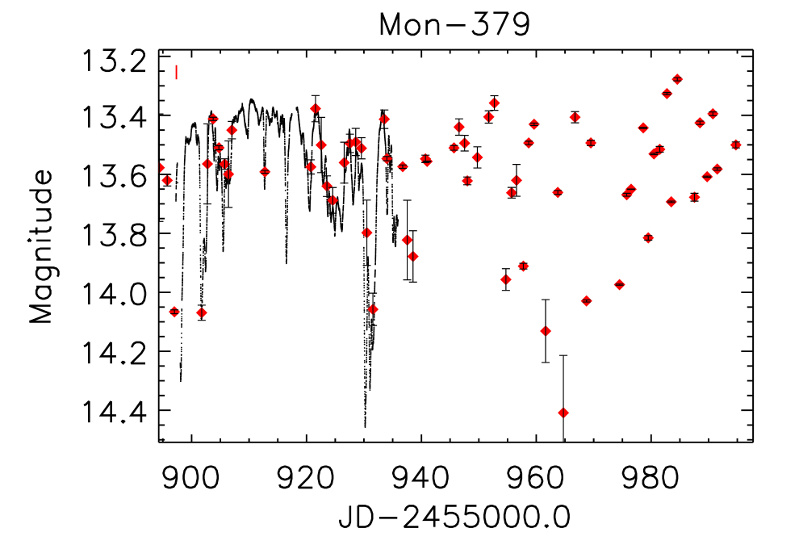}
\caption{Top: plots of LiIEW vs. I magnitude for three different stars. Bottom: I-band and CoRoT light 
curves, shifted in magnitude to coincide, of the same three stars. Left panels: the star has an accretion-burst 
dominated light curve, and veiling variability that supports the variable accretion scenario. Middle panels: the 
star has a light curve that presents aperiodic extinction events, and veiling variability that supports a 
scenario in which the occulting structures are associated with accretion shocks. Right panels: the star has 
a light curve with AA Tau-like characteristics, while its near lack of veiling variability can neither 
confirm nor refute the hypothesis that it is an AA Tau-like system. 
}\label{fig:veiling}
\end{figure*}

Many of these CTTS show a certain amount of variability in their measured veiling. We verified whether 
there was a correlation between this variability and the photometric variability. In principle, a 
correlation or anti-correlation should exist in some cases, depending on the physical processes occurring 
on the stellar photosphere. If, for instance, the variability is due to obscuration of the photosphere by 
an inner disk warp that is associated with accretion columns, we can expect two possible scenarios. 
When the accretion shocks on the stellar surface are not visible, or the veiling is too small to be 
accurately measured, there will be no considerable variability in the observed line equivalent widths. 
If, on the other hand, accretion shocks appear in our line of sight and produce measurable veiling, they 
should appear along with the inner disk warp, since both are associated with the accretion columns. In 
this case, when the inner disk warp occults the star causing brightness minima, the veiling should 
increase thanks to the appearance of the hot spots caused by accretion shocks. Therefore, for an AA Tau-like 
light curve, the equivalent width of Li I (LiIEW) or Fe I (FeIEW) should either remain relatively constant 
or increase with increasing brightness. In the latter case, this could be evidence that the star's 
photometric variability is due to obscuration by an inner disk warp that is spatially associated 
with the accretion columns. 

For photometric variability that is due mainly to a configuration of hot spots on the stellar surface, 
the veiling should increase with increasing brightness, and therefore FeIEW and LiIEW should decrease 
with increasing brightness. This is the case for the accretion-driven flux bursts discussed by 
\citet{stauffer14a}. The reoccurring narrow increases in brightness are likely due to an increase in 
accretion luminosity and are therefore associated with an increase in veiling. 

If the star's photometric variability is due simply to the rotational modulation of a stable, cold spot 
on the photosphere, the veiling would not be expected to vary, since cold spots do not cause veiling. 

Measures of veiling using the FeI 6633.4\AA \space line are much less accurate than LiI 6707.8\AA \space, 
since it is shallower in these stars and therefore suffers more from noise in the spectra. For this reason  
veiling was only measured using FeI for a small subset of these stars with FLAMES spectroscopy. In 
these cases the observed tendencies agree with those determined using the LiI line.

We made plots of LiIEW vs. USNO I-band magnitudes for the stars in our sample where both data were available. 
The I-band light curves were used rather than the CoRoT light curves because they had better time coverage, 
despite the lower cadence. There are at most six FLAMES spectra of each star obtained simultaneously with 
the CoRoT observations, whereas the USNO observations spanned the entire FLAMES campaign. In order to construct 
the plots, I-band measurements taken on the same night were averaged and matched with the equivalent width 
found using the FLAMES spectrum taken on that night. The USNO and VLT are separated by little more than 
$40^{\circ}$ in longitude, so the observations were likely taken within a few hours of each other on most 
nights.

\begin{figure}[th]
\centering
\includegraphics[width=4.0cm]{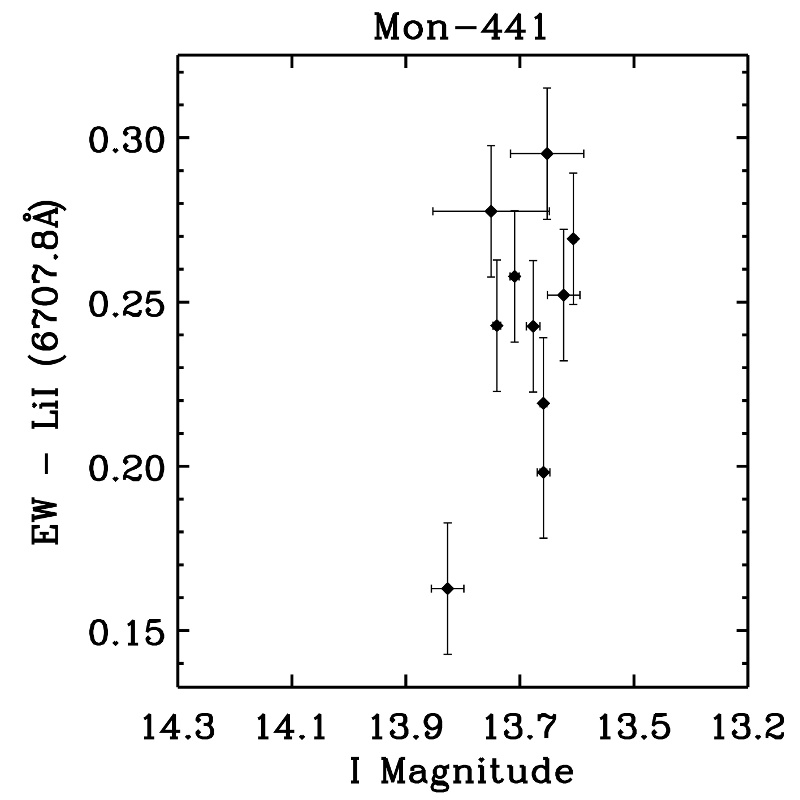}
\includegraphics[width=4.0cm]{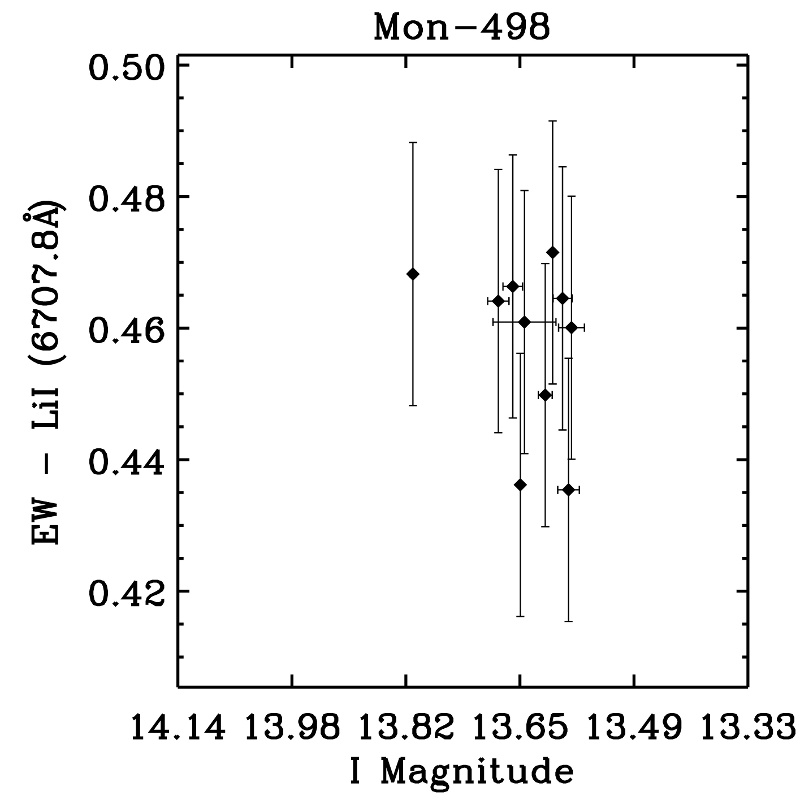}
\includegraphics[width=4.0cm]{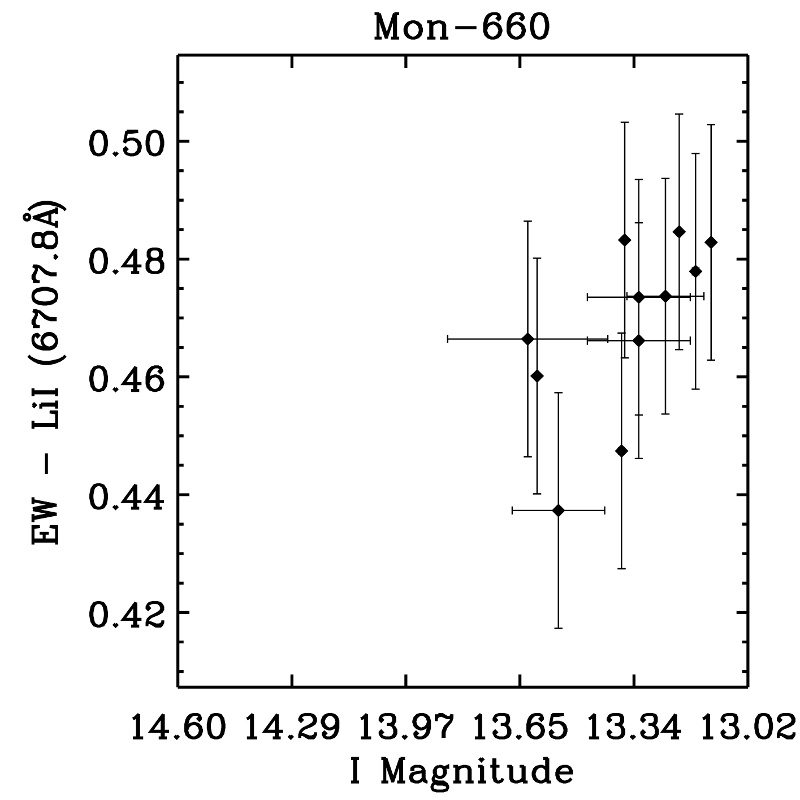}
\includegraphics[width=4.0cm]{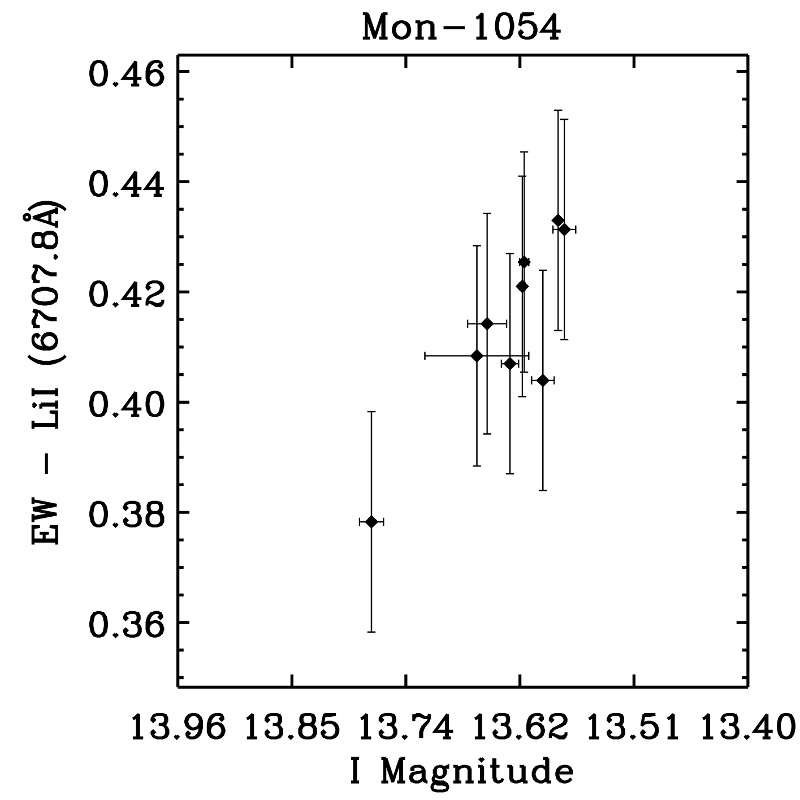}
\caption{Plots of LiIEW vs. I magnitude for some of the stars in our sample. The star Mon-1054 is an 
example of a candidate AA Tau system that shows increasing LiIEW with decreasing magnitude. The star 
Mon-441 shows similar behavior, though its photometric behavior is aperiodic. The star Mon-660 is an 
AA Tau-like system that appears to show increasing LiIEW with decreasing magnitude, though the 
uncertainties are quite large, and the star Mon-498 is an example of a system with very irregular 
veiling that shows no obvious pattern.
}\label{fig:veiling2}
\end{figure}

Fig. \ref{fig:veiling} shows plots of LiIEW vs. I-band magnitudes (top panels) for three different 
examples, along with the corresponding CoRoT and I-band light curves (bottom panels). The left panels of 
Fig. \ref{fig:veiling} show a star whose veiling increases with increasing brightness. We can see from 
its light curve that it is dominated by flux bursts attributed to variable accretion events. The middle 
panels of Fig. \ref{fig:veiling} show a star whose veiling increases with decreasing brightness, a 
tendency we attribute to occultation of the photosphere by an inner disk warp associated with the accretion 
mechanism. Its light curve morphology shows variability that could be explained by this scenario. The right 
panels of Fig. \ref{fig:veiling} show a star whose veiling does not vary considerably, which can mean the 
photometric variability is due to a configuration of cold spots on the surface, or to obscuration by an 
inner disk warp that either fully occults the accretion shocks associated with it, leaving the veiling 
unaltered during eclipses, or the accretion is relatively low, and only a small hot spot is present, not 
capable of generating a large UV excess or veiling. The light curve agrees with the inner disk warp scenario, 
since cold spots would result in a much more regular photometric variability, but the veiling can 
neither confirm nor disprove this hypothesis. This star also presents little UV excess, consistent 
with either of these scenarios. 

The star Mon-811 was outside the USNO field, and therefore we have no I-band information for it. For 
this one star, we made a plot of LiIEW vs. CoRoT flux counts instead. This plot was not nearly as well 
sampled, since very few FLAMES spectra were taken during the CoRoT observations, but it shows a tendency 
similar to the middle panels of Fig. \ref{fig:veiling}.

Fig. \ref{fig:veiling2} shows plots of LiI equivalent width vs. I magnitude for some of the AA 
Tau-like or aperiodic extinction stars in our sample (plots of the missing stars can be found in 
the online material, Fig. \ref{fig:app1}). In many cases the veiling variability is very irregular, 
and we cannot see a clear pattern (e.g., Mon-498). For the star Mon-379 (shown in the right panels of 
Fig. \ref{fig:veiling}), the veiling is relatively constant. 
For six stars, we see an apparent decrease in veiling (increase in LiI equivalent width) as the star's 
brightness increases, which could indicate that the mechanism responsible for their photometric variability 
is obscuration by an inner disk structure associated with accretion columns. This is the case for 
Mon-250, Mon-441, Mon-660, Mon-811, Mon-1054, and Mon-1144. The last has the largest variability 
amplitude in veiling ($\sim 0.3 \text{\AA}$), meaning there should be significant hot spots on the stellar surface. 
It is the star with the highest UV excess among the stars in our sample where veiling was measured, which 
could be expected, since the hot spots associated with the accretion shocks on the surface of the star should 
produce a considerable UV excess when visible.

Two of the stars that show an increase in veiling during the minima in their light curves have aperiodic 
photometric variability (Mon-441 and Mon-1144). This provides support for our hypothesis that the 
occultations in these light curves are caused by dust lifted above the disk plane at the base 
of unstable accretion streams, since these should be associated with accretion shocks on the stellar 
surface in a manner analogous to the stable accretion funnels. 
If these stars are undergoing unstable accretion, they should show no obvious periodicity in their 
light curves, but if they are seen at high enough inclinations, they should share other similarities 
with AA Tau-like systems, such as the one we see here.

\subsubsection{The specific case of Mon-250}\label{sec:mon250} 

\begin{figure}[t]
\centering
\includegraphics[width=9.0cm]{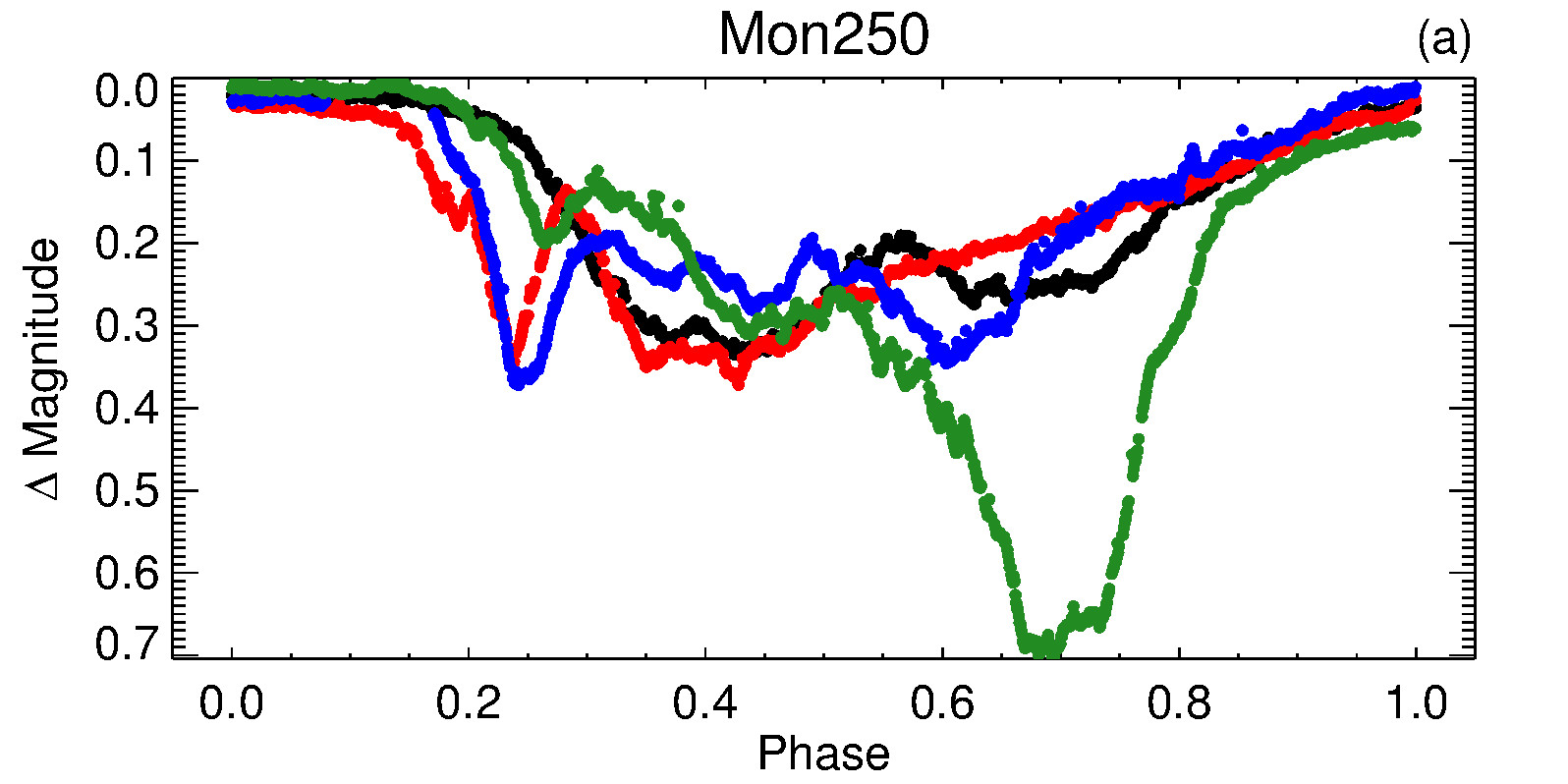}
\includegraphics[width=9.0cm]{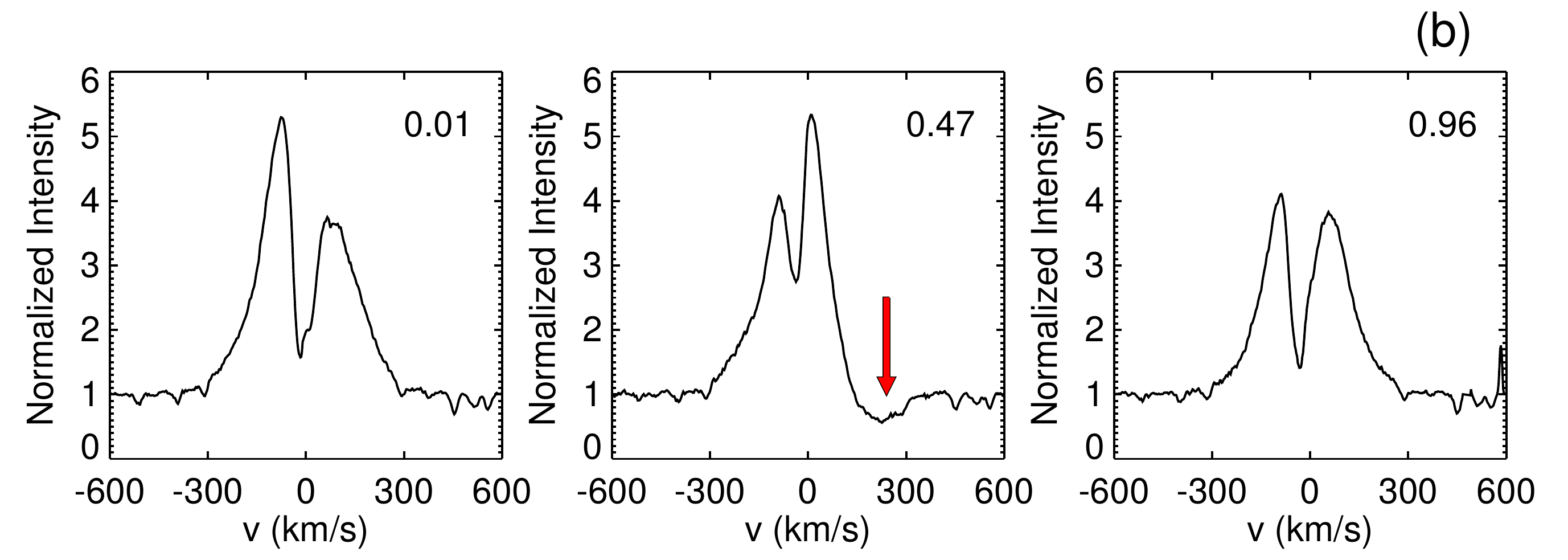}
\includegraphics[width=9.0cm]{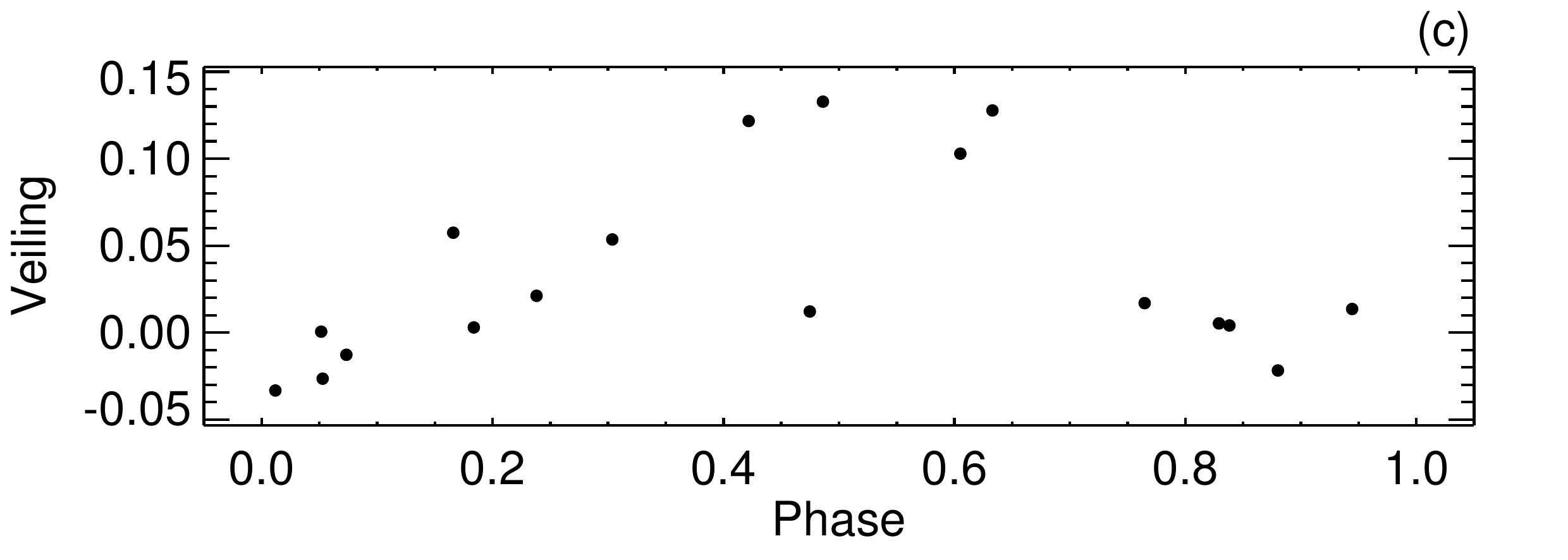}
\includegraphics[width=9.0cm]{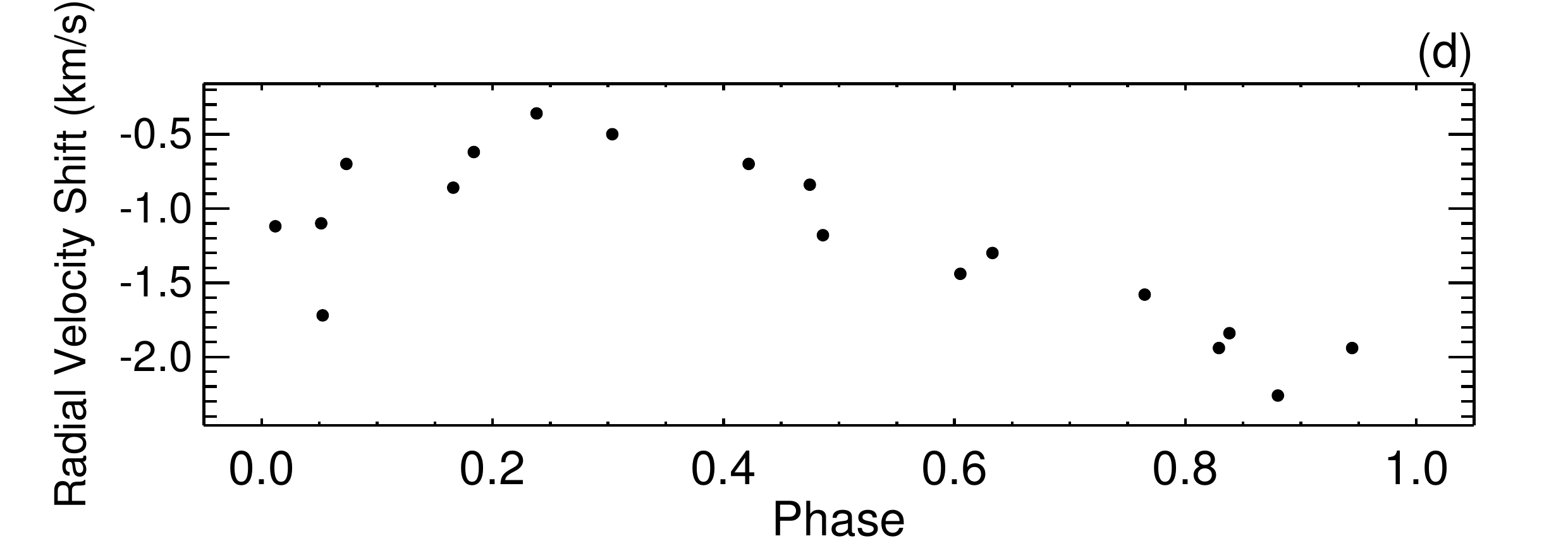}
\caption{Mon-250. a) 2011 Corot light curve folded in phase. b) H$\alpha$ line profile during 
different rotational phases, indicated in each panel. A redshifted absorption component seen only near phase 
0.5 is indicated by a red arrow. c) Measured veiling folded in phase. d) Shift in radial 
velocity measured from photospheric lines in the spectra, folded in phase. In all four panels, 
the same period of 8.6 days, determined from the radial velocity variation of photospheric lines, was used to 
calculate the phase. The same initial date was also used.
}\label{fig:veilingmon250}
\end{figure}

We analyzed the H$\alpha$ line profile of the star Mon-250 during different rotational phases (Fig. 
\ref{fig:veilingmon250}b), and note a redshifted absorption component (indicated by an arrow in the figure) 
that appears in several spectra during the light curve minima that is not present in other phases. This is 
clear evidence that the accretion funnel and hot spot appear in front of our line of sight during the 
photometric flux dips, since the infalling material responsible for the absorption is moving away from us. 

This star's veiling variability also points to the appearance of accretion shocks during occultations. For this 
particular star, we were able to measure veiling in all 20 FLAMES epochs by comparing line depths to 
those of a weak line T Tauri star of the same spectral type, to determine what fraction of the observed 
continuum should be due to the hot spot. Fig. \ref{fig:veilingmon250}c shows a plot of veiling 
folded in phase, using the rotation period determined from radial velocity shifts of photospheric 
lines in the spectra (Fig. \ref{fig:veilingmon250}d). It can be seen that there is almost always an 
increase in veiling as the light curve undergoes a minimum in brightness. This is evidence of the 
appearance of a hot spot on the stellar surface, one large enough to produce measurable veiling, during 
occultations. The rotation period found from the variability of radial velocities measured in the 
Mon-250 photospheric lines was of $8.6 \pm 0.5$ days, consistent with the period found from the CoRoT 
light curve observed at the same epoch of $8.93 \pm 0.50$ days. This shows that the occulting structure 
is located at or near the disk's co-rotation radius, as was the case for AA Tau. Unfortunately, this type 
of analysis was only possible for this star since it was the only one with a high enough variability 
amplitude in veiling and radial velocity of photospheric lines to draw significant conclusions.

We attribute these phenomena, which were also observed for AA Tau \citep{bouvier07}, to the 
appearance of a major accretion funnel and accretion shock associated with it, at the moment of occultation 
by the inner disk warp. This shows the association between the inner disk warp and the stable accretion 
columns, and provides support to our hypothesis that in these cases, the same interaction between the 
inclined stellar magnetic field and the inner disk that is causing the material to accrete onto the star 
via two major accretion columns is also responsible for lifting dust from the disk plane, creating a 
warp in the disk near the co-rotation radius.

\subsection{Stable and unstable accretion regimes}\label{sec:stab}

In this section, we discuss the main differences between the quasi-periodic AA Tau-like stars 
and the stars whose light curves are dominated by circumstellar extinction, but have no obvious 
periodicity. We propose that the former may be undergoing accretion via a stable regime, while the 
latter could be accreting via an unstable regime. \citet{kurosawa13} cite the parameters that are 
important in determining through which of these regimes a star will accrete. They are mass accretion 
rate ($\dot{M}_{acc}$), truncation radius, stellar rotation rate, the inclination angle $\beta$ 
between the stellar rotation axis and the axis of the magnetic field ($\Omega$ and $\mu$, respectively, 
in Fig. \ref{fig:maginclined}), and the $\alpha$-parameter, which describes viscosity in the disk.

\citet{kurosawa13} mention a possible transition between regimes, which can occur if one or a few of 
the factors that determine instability change with time within a certain system. For instance, if the 
truncation radius $R_{mag}$ becomes smaller than the co-rotation radius $R_{co}$, due for example to a 
change in the stellar magnetic field, accretion flows tend to become unstable. 
The mass accretion rate can also affect the ratio $R_{mag}/R_{co}$, since a higher mass accretion rate 
leads to a smaller truncation radius, therefore leading to a more unstable configuration. The viscosity 
parameter ($\alpha$) controls the global mass accretion rate. The higher $\alpha$ is, the higher the 
mass accretion rate is, leading to more instabilities.
According to \citet{kulkarni09}, the transition from a stable to an unstable regime occurs around 
critical values of mass accretion rates. They state typical critical values for stars with 
$R_{mag} \approx 2-3 R_*$ to be around $\dot{M}_{acc} = 1.7 \times 10^{-7}$, and for stars with 
$R_{mag} \approx 4-5 R_*$, around $\dot{M}_{acc} = 2.1 \times 10^{-8}$. 

A smaller angle $\beta$ also leads to less stable configurations, since at large $\beta$ the magnetic 
poles become closer to the inner disk rim, lowering the potential barrier in the vertical direction 
that the material must overcome to be lifted into the accretion funnels. According to 
\citet{kulkarni09}, values of $\beta > 25^{\circ}$ usually lead to stable configurations 
\citep[see also][]{romanova14}.

\citet{venuti14} showed that mass accretion rates in CTTS can vary considerably over time.
They calculated mass accretion rates for stars in NGC 2264 using data from 2010 December and 2012 January, 
the latter being nearly simultaneous with the CoRoT observations. They show that many of the stars 
in our sample had mass accretion rates vary by up to 50\% in little over one year. 
Unfortunately, there is no estimate of mass accretion rate simultaneous with the 2008 CoRoT observations, 
but it is possible that a variation such as this contributed to a transition between accretion regimes.
However, \citet{venuti14} also show that long term changes in mass accretion rates are not necessarily 
larger than short term changes. 

Magnetic field configurations may also be responsible for these transitions.
It is possible for the magnetic field of a CTTS to change significantly in a few years. 
This has been observed, for example, in the CTTS V2129 Oph \citep{donati11}. Its magnetic field was 
measured in two different epochs, nearly four years apart, and the magnetic field strength varied greatly 
from one epoch to the other. The dipolar component was found to be about 3 times stronger 
in the second epoch, and the octupolar component 1.5 times stronger than four years earlier. 
\citet{donati11} found that the ratio $R_{mag}/R_{co}$ nearly doubled in that time. 

\begin{figure}[t]
\centering
\includegraphics[width=4.0cm]{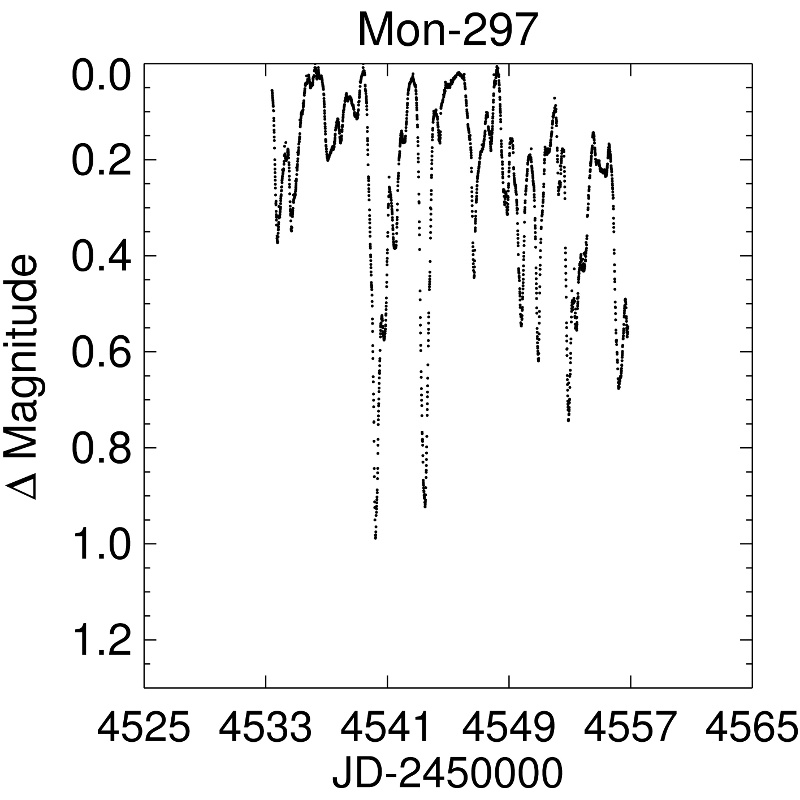}
\includegraphics[width=4.0cm]{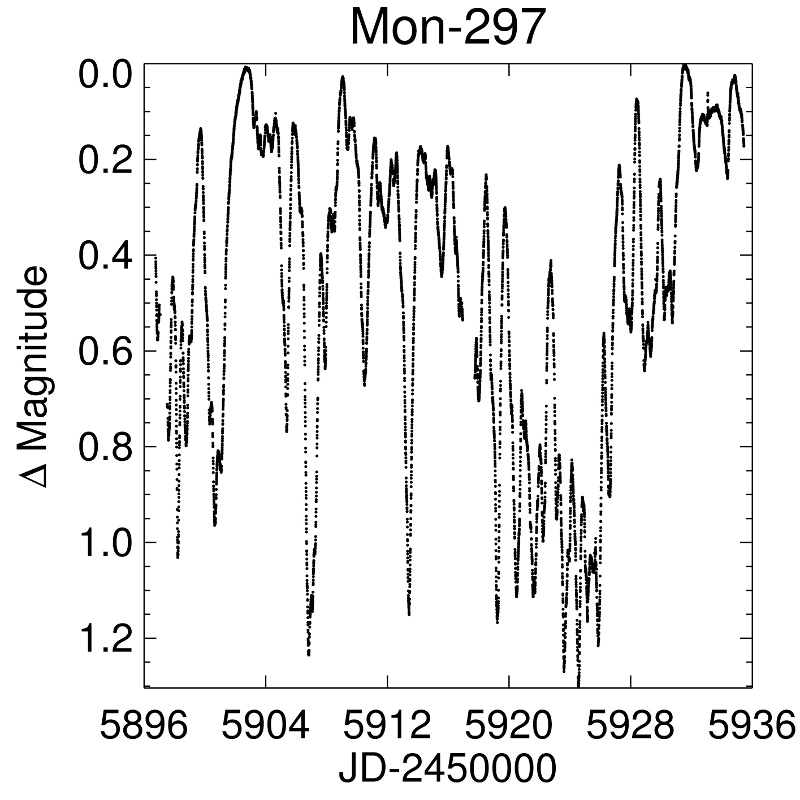}
\includegraphics[width=4.0cm]{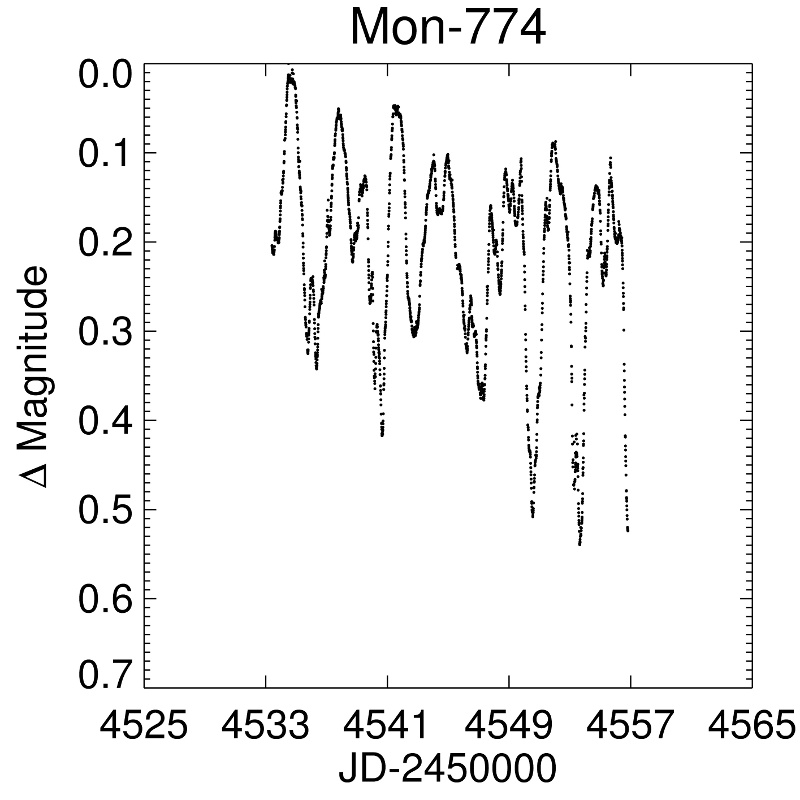}
\includegraphics[width=4.0cm]{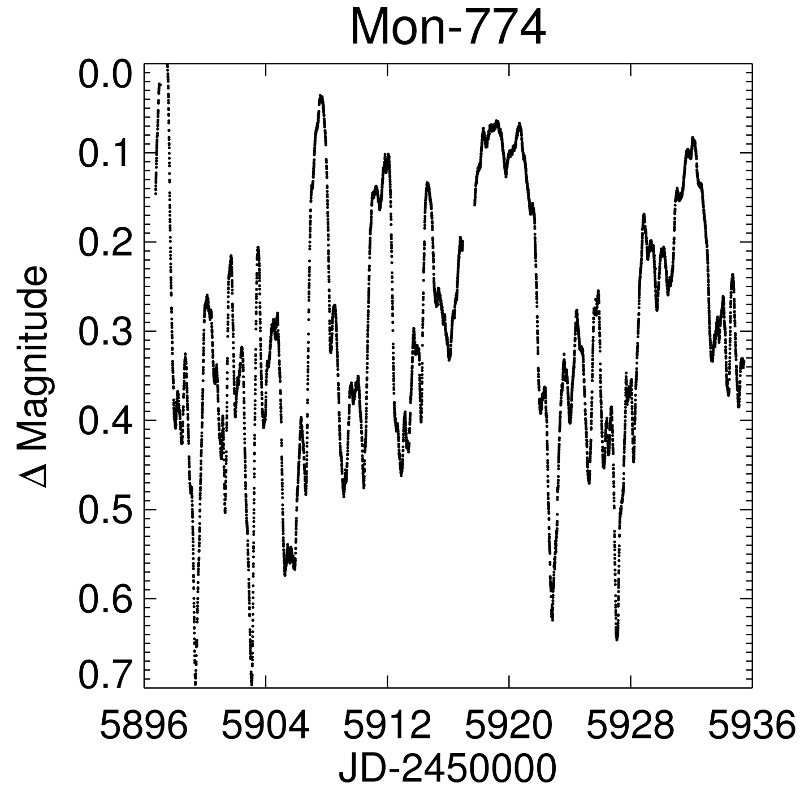}
\caption{Corot light curves of Mon-297 and Mon-774 during their AA Tau phases (left) and aperiodic 
phases (right).
}\label{fig:aat_unstab}
\end{figure}

In this study we have seen that 38\%$\substack{+14\%\\-11\%}$ (5/13) of the stars that were observed 
in 2008 as having photometric variability similar to AA Tau were observed again in 2011, showing 
aperiodic variability due to extinction from circumstellar material. Of the stars that were observed 
in 2008 as having aperiodic photometric variability due to circumstellar extinction, 
62\%$\substack{+11\%\\-15\%}$ (8/13) were observed again in 2011, then showing variability similar to 
AA Tau. If our proposed scenario is true, then this points to a transition from a stable to an 
unstable accretion regime, or vice-versa, in a matter of less than four years, for 57\%$\substack{+9\%\\-11\%}$ 
of our sample of possible AA Tau-like stars, i.e., (5+8)/23. An example of two of the stars that appear to 
have undergone this transition is shown in Fig. \ref{fig:aat_unstab}.

We also performed a periodogram analysis of the light curves of these 13 stars that changed between 
periodic and aperiodic photometric behavior. The periodograms of 
each epoch were analyzed separately and compared. For 11 stars, during the epoch of unstable accretion 
at least a weak signal was present at the period found for the AA Tau phase, believed to be the stellar 
rotation period. For eight of these, this peak was the only one 
present, or the strongest, despite being much weaker than the one present during the AA Tau phase.  
One exception is the star Mon-441, which shows a weak signal during its unstable phase, but this 
signal does not correspond to the period found in the AA Tau phase, or to a multiple of it. 

In most cases the variable depths of the flux dip amplitude in a star's light curve during its 
aperiodic phase are smaller than those measured in its AA Tau phase by 20\% to 60\%. However, for 4 of 
the 13 stars that underwent a transition between AA Tau-like and aperiodic light curve, the flux dip 
depths remained nearly the same in both epochs, and for 2 more stars the variability amplitude during 
their aperiodic phase was larger by 20\% to 30\%. 
This means that, at least in some cases, during the unstable accretion phase dust must be lifted above the disk 
plane at similar, and sometimes even higher, altitudes as during the stable accretion phase. 

\begin{figure}[thb]
\centering
\includegraphics[width=4.3cm]{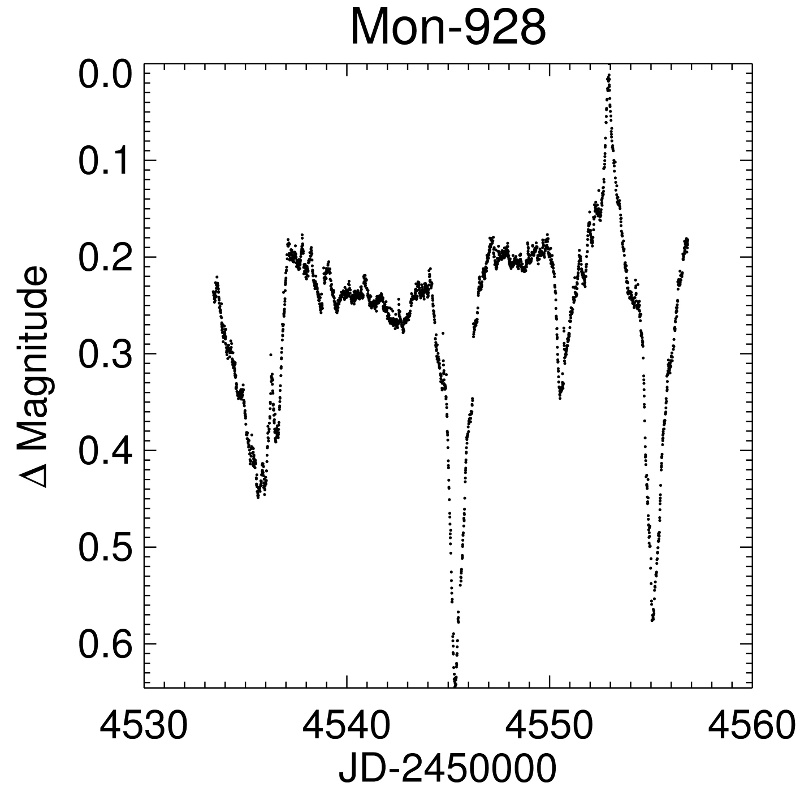}
\includegraphics[width=4.3cm]{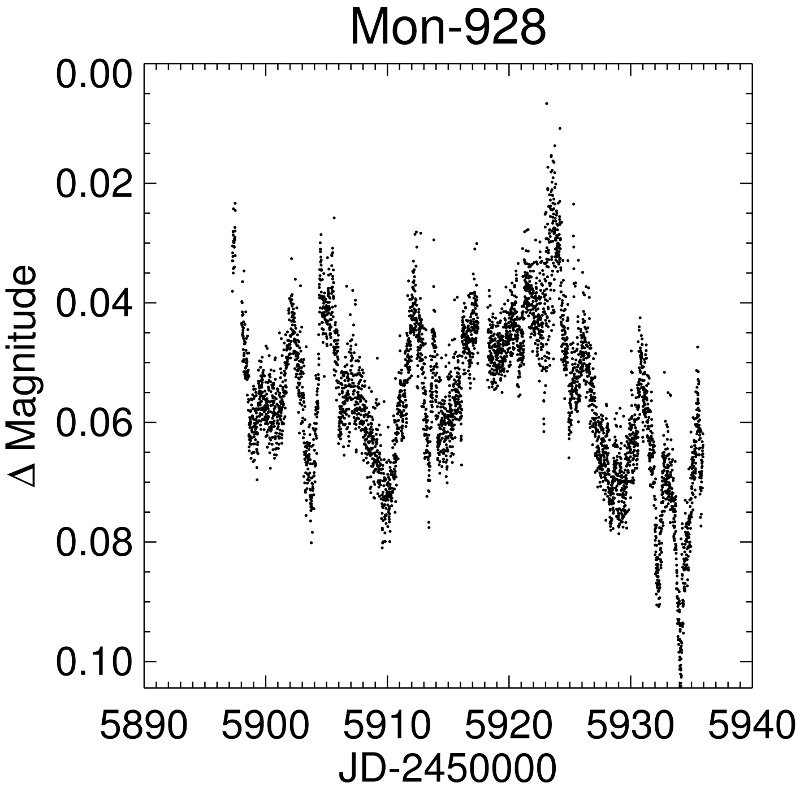}
\caption{CoRoT light curves of Mon-928 in 2008 (left) and 2011 (right). 
}\label{fig:928}
\end{figure}

As for the star Mon-928, the flux dips present in 2008 disappear entirely in 2011. The light curve 
we observe in the second epoch is not characteristic of circumstellar extinction. The period of 9.92 days 
found during the first epoch is very similar to periods previously measured for this star in the literature 
\citep{lamm05}, yet it is non-existent in the 2011 light curves (see Fig. \ref{fig:928}). One 
possibility is that in 2011 the inner disk warp did not reach high enough altitudes above the disk to occult 
the stellar photosphere. At the inclination of $71^{\circ}$ found by us for this system (see Sect. \ref{sec:model}), 
our simulations using the occultation model show that if the warp found to best reproduce the 
2008 CoRoT light curve decreases in height by at least 26\%, the occultations observed in 2008 would no 
longer occur. This amount of variation in warp height is well within the range we observe for these objects 
(see Sect. \ref{sec:model}). This star is also one of the two that appear to have undergone a small increase in 
mass accretion rate from 2010 December to 2012 January \citep{venuti14}; therefore, it is possible that a higher 
mass accretion rate led to a more unstable configuration during the 2011 observations, 
resulting in a less pronounced disk warp.

\citet{kurosawa13} also predict that for a more unstable regime, one should observe a higher frequency of flux 
peaks per stellar rotation period in the light curves of stars of medium or low inclinations thanks to the 
presence of a larger number of accretion shocks on the photosphere. It is fair to assume that, for systems 
seen at high inclinations, a higher frequency of flux dips would also be observed for more unstable scenarios 
because of the larger number of accretion streams present around the star, whose base could contain dust that 
can occult the photosphere. Of the stars in our sample that changed from one regime to the other, most show 
two to three flux dips per stellar rotation period during the unstable accretion phase, while some show up to 
seven flux dips per rotation period. This may be an indication of the degree of instability of each star-disk 
system. The two stars that show the highest frequencies of flux dips during their unstable accretion phases 
are the two that have the lowest masses in our sample of possible AA Tau-like stars. As is discussed in Sect. 
\ref{sec:other}, low mass stars appear to present more unstable accretion, possibly as a result of magnetic field 
configurations. These higher frequencies of flux dips can also be caused by dust lifted above the disk plane because of 
magnetorotational instabilities that do not necessarily lead to accretion streams \citep{turner10,hirose11}.

\begin{figure*}[t!]
\centering
\includegraphics[width=5.5cm]{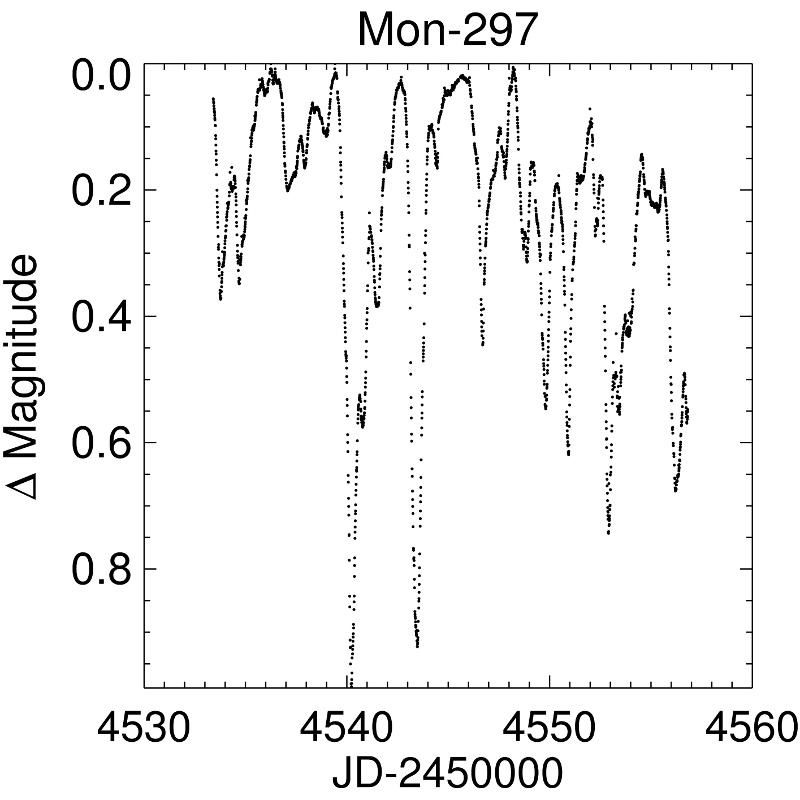}
\includegraphics[width=5.5cm]{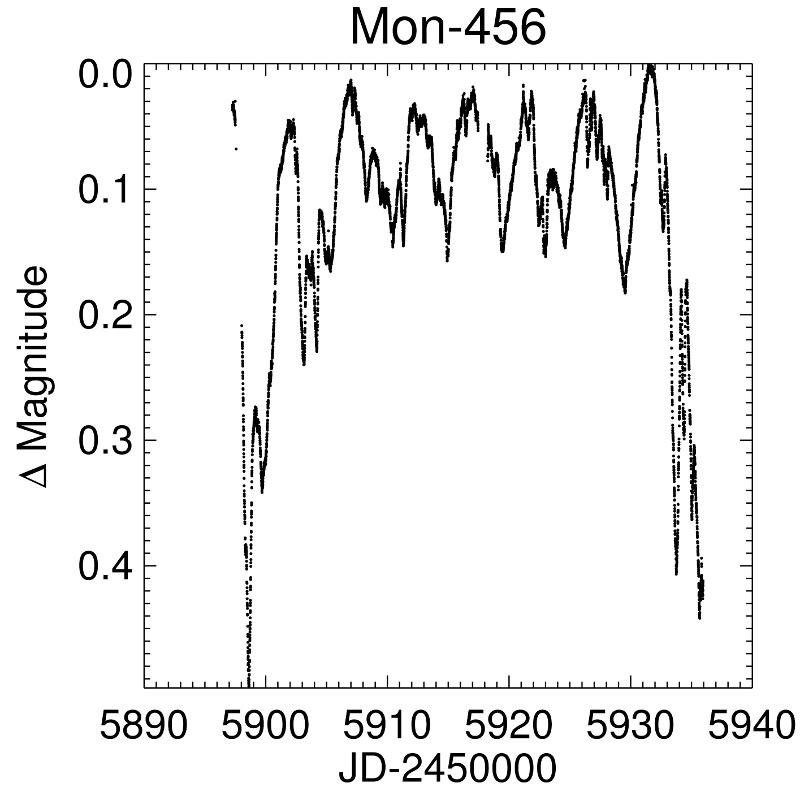}
\includegraphics[width=5.5cm]{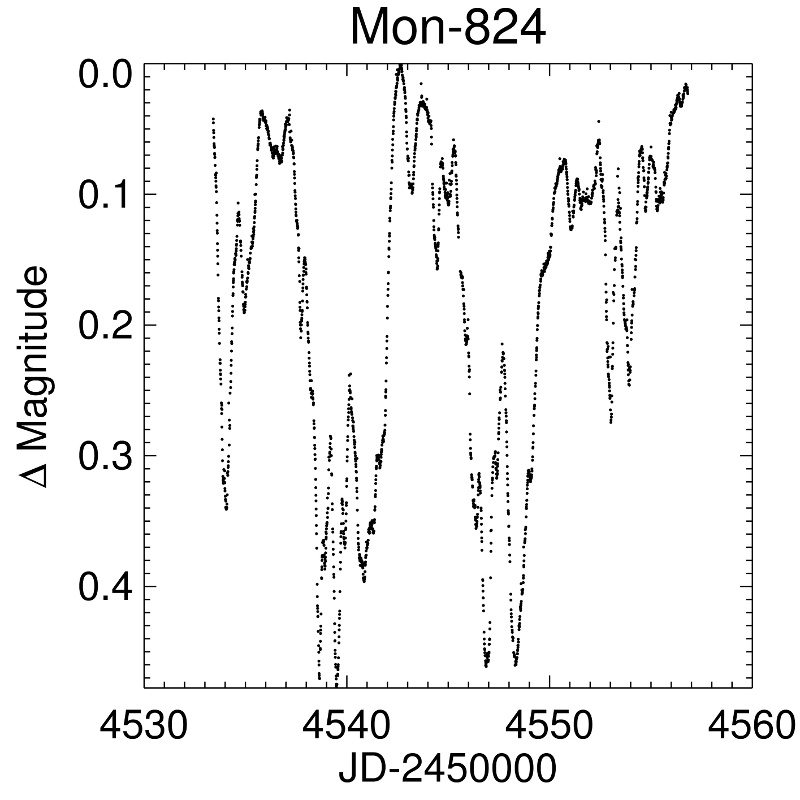}
\caption{CoRoT light curves of Mon-297, Mon-456, and Mon-824 during their AA Tau phase. 
These stars sometimes present two, or even three, flux dips per rotation period, and often 
secondary flux dips are present alongside the main flux dips, indicating a possible coexistence 
between unstable accretion streams and stable accretion funnels. 
}\label{fig:instabex}
\end{figure*}

Of the 23 stars that present AA Tau-like behavior, only six maintain this behavior in both 2008 and 2011. 
Many appear to show some signs of weak instability even during their AA Tau phase, sometimes presenting 
two or three flux dips in one rotation cycle. A few examples are Mon-297, Mon-456, and Mon-824 
(Fig. \ref{fig:instabex}). Most show many small secondary flux dips present in the light curve, 
indicating a possible coexistence of secondary accretion streams with the main accretion funnels.

\subsection{Comparing optical and infrared light curves}\label{sec:optir}

\begin{figure*}[bt]
\centering
\includegraphics[width=8cm]{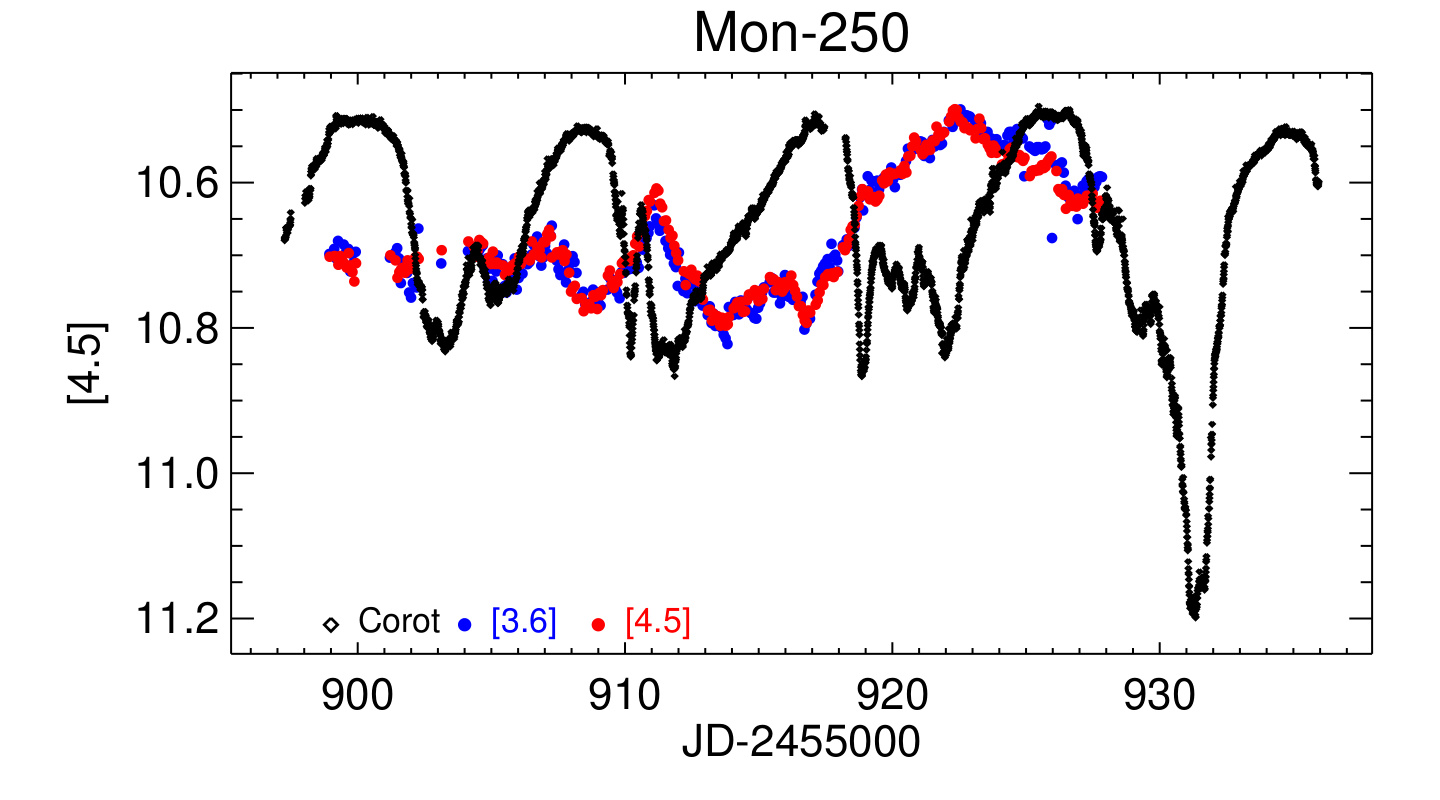}
\includegraphics[width=8cm]{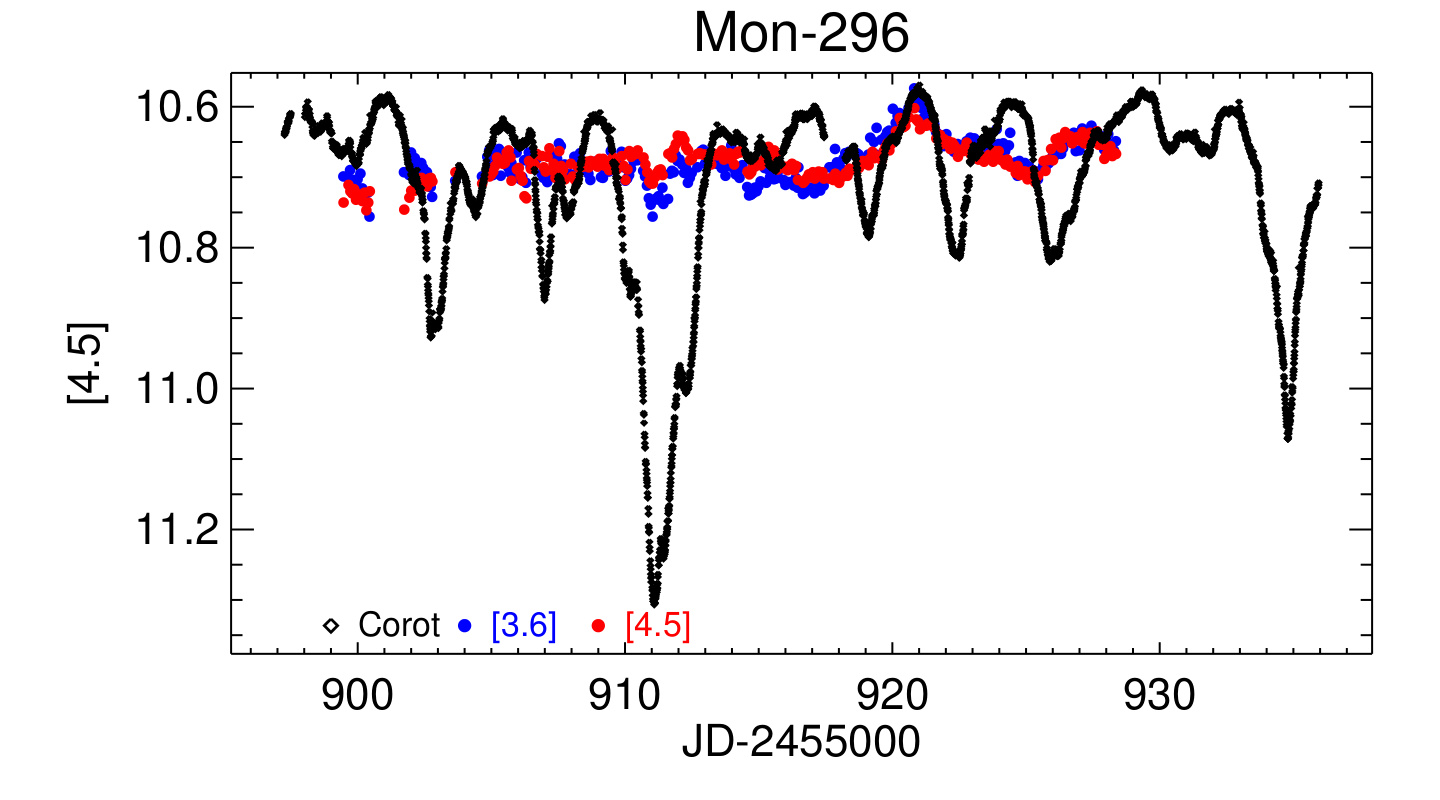}
\includegraphics[width=8cm]{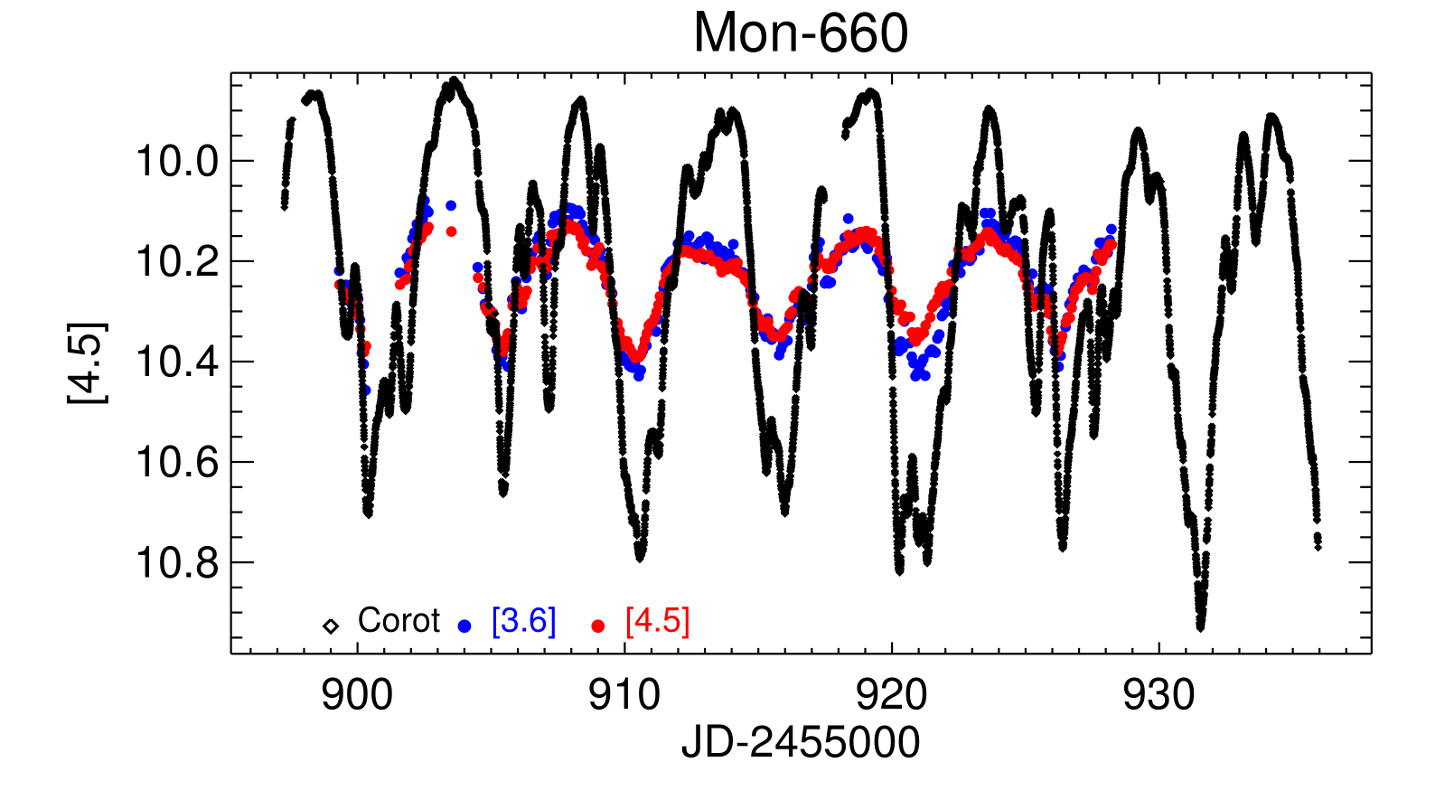}
\includegraphics[width=8cm]{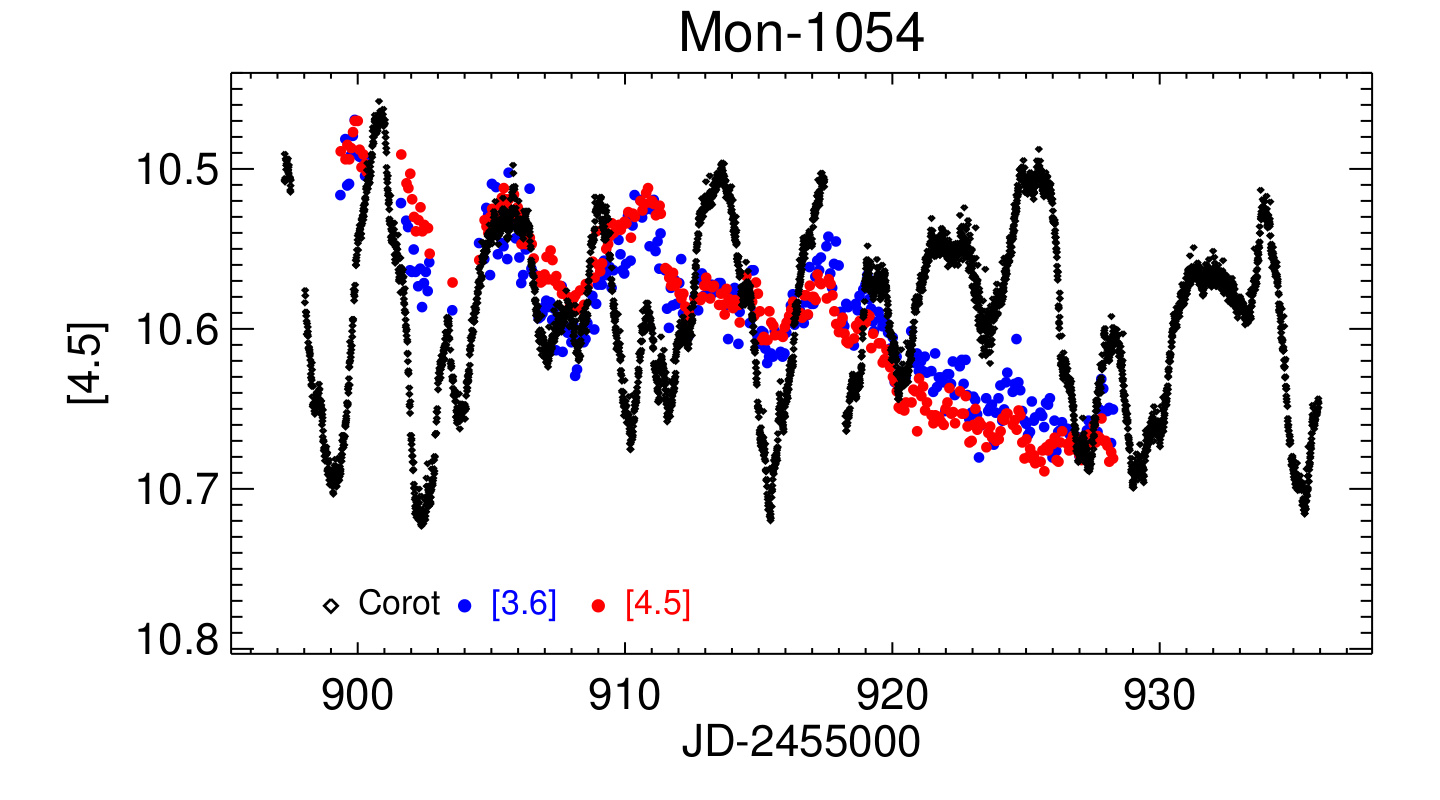}
\caption{\textit{Spitzer} IRAC $3.6 \mu m$ and $4.5 \mu \mathrm{m}$ light curves (blue and red filled 
circles, respectively) overplotted on CoRoT light curves (black circles). 
The CoRoT and IRAC $3.6 \mu m$ light curves were shifted in magnitude for easier comparison. 
}\label{fig:spitlcs}
\end{figure*}

Of the 33 stars in our sample of AA Tau-like or aperiodic extinction dominated CoRoT light curves, 
29 were observed almost simultaneously with \textit{Spitzer} in mapping mode. Fig. \ref{fig:spitlcs} 
shows a few examples of \textit{Spitzer} IRAC $3.6 \mu \mathrm{m}$ and $4.5 \mu \mathrm{m}$ light curves, 
plotted along with the respective CoRoT light curves. The CoRoT flux was transformed into magnitude 
using its greatest value as magnitude zero, then shifted to overlap the \textit{Spitzer} IRAC data. A 
figure showing the light curves of all other stars in our sample can be found in the online material, 
Fig. \ref{fig:app2}.

We see that 4 of the 29 stars show very similar behavior in both bands, and present AA Tau-like 
modulation in the infrared as well as the optical. These stars are Mon-660, Mon-811, 
Mon-1140, and Mon-1308. Another five show similar behavior, with minima and maxima generally 
coinciding in the infrared and optical, but these light curves do not overlap perfectly. These 
are Mon-297, Mon-314, Mon-433, Mon-441, and Mon-456.

There are nine stars that correlate in some parts of their infrared and optical light curves, while 
other parts behave differently, or even anti-correlate at times. Stars Mon-56, 
Mon-126, Mon-358, Mon-654, Mon-717, Mon-774, Mon-1054, Mon-1144, and Mon-1167 
show this behavior. The 10 stars Mon-296, Mon-325, Mon-379, Mon-498, Mon-619, 
Mon-667, Mon-928, Mon-1037, Mon-1038, and Mon-1131 have entirely different infrared 
and optical light curves, with very little or no features in common.

\begin{figure}[tb]
\centering
\includegraphics[width=8cm]{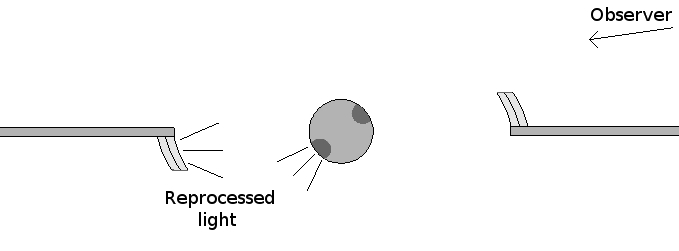}
\caption{When one side of the inner disk warp occults the stellar photosphere, the hot spot 
and photosphere on the opposite side illuminate the part of the warp on that side. The light 
is absorbed by the dust and is re-emitted at longer wavelengths.
}\label{fig:anticor}
\end{figure}

The star Mon-250 has \textit{Spitzer} IRAC light curves that almost anti-correlate with its CoRoT 
light curve. Most of the \textit{Spitzer} IRAC maxima coincide with CoRoT minima, meaning this 
variability cannot possibly be the consequence of a hot or cold spot on the stellar surface, since 
any variability on the surface should be present in all wavelengths. To explain why there is more 
IR emission during the optical eclipses, we propose the following scenario. The stable accretion 
regime assumes that there are two main accretion funnels opposite each other, one in each hemisphere. 
Each of these funnels is associated with a hot spot on the stellar surface, where the accretion shock 
is located. The base of each accretion funnel is associated with part of the inner disk warp. When 
one part of the warp occults the stellar photosphere in our line of sight, the hot spot opposite 
it, as well as the rest of the stellar photosphere on that side of the star, illuminates the 
corresponding part of the inner disk warp. The dust located there absorbs the light from the 
hot spot and photosphere and re-emits it at longer wavelengths, causing an increase in emission in 
the infrared (Fig. \ref{fig:anticor}). This emission in the infrared is predicted by the radiation 
transfer model of \citet{whitney13} when applied to an accretion hot spot associated with an inner 
disk warp. This behavior is also observed in a few other light curves, during one or two 
rotation cycles (see, e.g., Mon-1054 in Fig. \ref{fig:spitlcs}, between MJD 55909 and 
MJD 55913).

The $\alpha _{IRAC}$ index is a good indicator of the amount of dust present in the inner accretion 
disk. It corresponds to the slope of the spectral energy distribution (SED) between $3.6 \mu \mathrm{m}$ 
and $8 \mu \mathrm{m}$. Unfortunately, there were no \textit{Spitzer} IRAC observations in $8 \mu \mathrm{m}$ 
simultaneously with the 2011 CoRoT-\textit{Spitzer} campaign, since by this time \textit{Spitzer} was already 
operating in its warm mission. We do, however, have $\alpha _{IRAC}$ values from \citet{teixeira12} for 14 of 
these 29 stars with simultaneous CoRoT and \textit{Spitzer} light curves, 
and archival IRAC and Multiband Imaging Photometry for \textit{Spitzer} (MIPS) data \citep[for details see][]{cody14}, 
which enabled us to calculate values of $\alpha _{IRAC}$ for the stars that were not included in 
\citet{teixeira12}. These values are given in Table \ref{table:uxori2}. 
\citet{lada06} established a criterion to classify the inner disk structures of systems according to the 
$\alpha _{IRAC}$ index. Stars with $\alpha _{IRAC} < -2.56$ are classified as naked photospheres (systems 
devoid of dust within a few 0.1 AU), those with $-2.56 < \alpha _{IRAC} < -1.80$ have anemic disks (their 
inner disk region is optically thin), those with $-1.8 < \alpha _{IRAC} < -0.5$ have optically thick inner 
disks, stars with $-0.5 < \alpha _{IRAC} < 0.5$ are flat spectra sources, and stars with 
$\alpha _{IRAC} > 0.5$ are Class I protostellar objects. 

We note that the objects in our sample whose CoRoT and \textit{Spitzer} light curves correlate best 
have values of $\alpha _{IRAC}$ between -1.4 and -1.9, meaning they are generally in the 
transition region between optically thin and optically thick inner disks. If there is too much 
dust within a few AU of the star (higher values of $\alpha _{IRAC}$), the \textit{Spitzer} light curves 
would be dominated by emission from warm circumstellar dust, rather than eclipses of the stellar 
photosphere. On the other hand, if the portion of the disk occulting the star is not optically 
thick, then the flux dips at IRAC wavelengths could be much shallower than for the case where we assume
optically thick absorption at all bands.
This seems to be the case for Mon-296, whose \textit{Spitzer} light curve remains relatively steady 
except for a very shallow dip occurring at the same time as the optical light curve's lowest minimum, 
and a small burst near one of its maxima (top right panel in Fig. \ref{fig:spitlcs}). This star's 
$\alpha _{IRAC}$ shows that it has an anemic disk.
The star Mon-250, mentioned in the discussion above, also has an anemic disk. It has a value of 
$\alpha _{IRAC} = -2.10$, one of the lowest in our sample. 

Another useful quantity is the ratio of the disk flux to the stellar flux at 4.5 $\mu \mathrm{m}$, 
estimated from the slope of each star's SED as compared to an estimate of the photospheric flux based 
on the star’s spectral type and the Kurucz-Lejeune models \citep{lejeune97}, normalized to the observed 
$J$-band and assuming an appropriate amount of reddening customized to each object. This indicates 
what fraction of the 4.5 $\mu \mathrm{m}$ flux originates at the disk, and should relate directly to how 
well the light curves correlate. If this ratio is zero, then the optical and infrared light curves 
should coincide. If it is considerably larger, very little of the infrared flux comes from the star, 
and the flux dips present in the optical light curves should be absent from the infrared light curves. 
The values of $\mathrm{F}_{4.5 \mu \mathrm{m}}(\mathrm{disk})/\mathrm{F}_{4.5 \mu \mathrm{m}}(\mathrm{star})$ 
estimated for our stars are shown in Table \ref{table:uxori2}.

We see some of these predicted tendencies, but the relationship is not as direct as one would expect. 
The stars with the lowest fractions of disk flux at 4.5 $\mu \mathrm{m}$ tend to have well correlated 
light curves, and generally very similar variability amplitudes in the optical and infrared, while 
those with a very high value of disk flux per stellar flux have uncorrelated, or weakly correlated, 
optical and infrared light curves. For the intermediate values, however, we find all sorts of scenarios. 
Mon-660, for instance, has an estimated ratio of disk flux to stellar flux 
of $\mathrm{F}_{4.5 \mu \mathrm{m}}(\mathrm{disk})/\mathrm{F}_{4.5 \mu \mathrm{m}}(\mathrm{star})= 3.4$, 
and very strongly correlated light curves, while for Mon-928 
$\mathrm{F}_{4.5 \mu \mathrm{m}}(\mathrm{disk})/\mathrm{F}_{4.5 \mu \mathrm{m}}(\mathrm{star})= 1.4$,
and its optical and infrared light curves are completely different.
This could be due to uncertainties in the SED fits caused by poor or time variable data, or could 
simply be due to the time variable nature of the SEDs. There could have been some changes in the 
disk emission between the \textit{Spitzer} cryogenic mission, from whence the values used to plot the SEDs 
were taken, and the 2011 CoRoT and \textit{Spitzer} observing campaign.

For the stars where a correlation can be identified, we calculated which extinction law is necessary 
to transform the optical to the infrared light curve, assuming an optically thin regime. 
Extinction laws $A_{\lambda}/A_R$, where $\lambda = 3.6 \mu \mathrm{m}$ and $4.5 \mu \mathrm{m}$, 
were found to be between 0.28 and 0.77. According to \citet{cardelli89}, we can derive values of 
$A_{\lambda}/A_V$ for a given $R_V$ (where $R_V$ is defined as $A_V/E(B-V)$). A value of 
$R_V=3.1$, typically used for interstellar matter (ISM), translates to $A_{3.6}/A_V=0.05$ and 
$A_{4.5}/A_V=0.04$. Transforming to $A_R$, since the CoRoT filter is centered close to the 
center of the R Johnson filter, we would have $A_{3.6}/A_R=0.07$ and $A_{4.5}/A_R=0.05$. 
These extinction laws are much lower than those found for the stars in our 
sample, implying that the distribution of dust grains in the inner circumstellar disk region 
is quite different from ISM, possibly containing fewer small grains. In addition, the observed 
dip depth ratio should be corrected for the fraction of IR light emitted by the disk. This would 
only increase our values of $A_{\lambda}/A_R$. 

\subsection{Modeling the CoRoT light curves}\label{sec:model}

In this section we use the occultation model proposed for AA Tau in \citet{bouvier99} (described 
in Sect. \ref{sec:introoccmodel}) to attempt to reproduce the AA Tau-like light curves in our sample. If the optical 
photometric variability observed is due to periodic occultations by an inner disk warp, as we 
propose, then this geometrical model should be capable of reproducing the general trend of these 
light curves. We vary the model's main parameters in an attempt to obtain the configuration that 
most closely matches the observed variability. In this way we determine which values of warp 
height and azimuthal extension are needed to explain these stars' variability and how they must 
change from one rotation cycle to another in order to account for the changes in width and 
amplitude of each flux dip \citep[see also][for a detailed study of the light curve of V354 Mon - 
Mon-660]{fonseca14}. 

As was mentioned in Sect. \ref{sec:introoccmodel}, the parameters of the occultation model are the star-disk system's 
inclination $i$, the radius $R_w$ at which the warp is located, and the warp's maximum height 
$h_{max}$ and azimuthal extension $\phi_w$. 

We assume that the periods we measured from the CoRoT light curves are the Keplerian periods 
of the occulting material orbiting the star; therefore, if we know the star's mass, we can use 
Kepler's third law in the following form to calculate the radius at which the warp is located, 

\begin{equation}\label{eq:kepler}
R_w^3 = GM_* \left(\frac{P_{kep}}{2\pi}\right)^2 ,
\end{equation}

\noindent
where $M_*$ is the star's mass, $P_{kep}$ is the Keplerian rotation period, and $G$ is the 
gravitational constant.
To calculate the inclination we use the following relation between $v\sin i$, stellar radius 
$R_*$, and stellar rotation period $P_{rot}$:

\begin{equation}\label{eq:sini}
v\sin i = \frac{2 \pi R_*}{P_{rot}} \sin i .
\end{equation}

We used stellar masses and radii determined by \citet{venuti14}, who compared effective 
temperatures and bolometric luminosities placed on the Hertzsprung-Russell diagram to 
pre-main sequence model grids of \citet{siess00}.
The values are shown in Table \ref{table:param}, along with our values of Keplerian periods 
found from CoRoT light curves in each epoch, and the corresponding Keplerian rotation radius. 

Values of the stars' projected rotational velocities ($v\sin i$) were obtained comparing FLAMES 
spectra, when available, and Hectochelle spectra otherwise, to synthetic spectra using the 
program Spectroscopy Made Easy \citep[SME, ][]{valenti96} and Oleg Kochukhov's 
Binmag3\footnote{\url{http://www.astro.uu.se/~oleg/}}, as well as the spectrum synthesis code Synth3 
\citep{kochukhov07} and atomic line files extracted from the \emph{Vienna Atomic Line Database} 
(VALD)\footnote{\url{http://vald.astro.uu.se/}}
\citep{vald4,vald3,vald2,vald1}. We used photospheric lines to estimate the values of $v\sin i$ 
shown in Tables \ref{table:uxori2} and \ref{table:param}. 
For a few of our stars either no spectrum was available, or the signal-to-noise ratio of the 
photospheric lines was insufficient for this analysis and therefore no value is given. 

\begin{table*}[t]
\begin{center}
\caption{Stellar parameters initially calculated for the AA Tau-like candidates. 
}\label{table:param} 
\begin{tabular}{l c c c c c c c c c c}
\hline
\hline
CSIMon ID &   $L_{bol}$   &   $M_*$       &   $R_*$       &  $P_1$ &  $P_2$ & $v\sin i$ & $\sin i$ & Inclination  & $R_{w,1}$ & $R_{w,2}$ \\
          & ($L_{\odot}$) & ($M_{\odot}$) & ($R_{\odot}$) & (days) & (days) & (km/s)    &          & ($^{\circ}$) & ($R_*$)   & ($R_*$)   \\
\hline
 CSIMon-000056  & 0.77 &  1.16 &  1.52 &  5.71 &  5.86  & 11.1$\pm$2.2 &  0.85$\pm$0.38 & 58$\substack{+32\\-30}$ &  9.29 &  9.45 \\
 CSIMon-000250  & 1.23 &  1.35 &  1.63 &  8.32 &  8.93  &  9.8$\pm$2.2 &  1.02$\pm$0.49 & 62$\pm$28               & 11.71 & 12.28 \\
 CSIMon-000296  & 1.57 &  1.42 &  1.71 &   --  &  3.91  & 17.3$\pm$2.6 &  0.78$\pm$0.33 & 51$\substack{+39\\-24}$ &  5.11 &  6.86 \\
 CSIMon-000297  & 1.57 &  1.42 &  1.71 &  3.16 &   --   & 30.8$\pm$1.8 &  1.13$\pm$0.40 & 69$\pm$21               &  5.95 &   --  \\
 CSIMon-000358$\dagger$  & 0.22 & 0.29 & 1.39 & -- & 5.86 &  --        &      --        & --                      &   --  &  6.51 \\
 CSIMon-000379  & 2.12 &  1.60 &  1.99 &   --  &  3.68  & 25.0$\pm$1.2 &  0.91$\pm$0.30 & 66$\substack{+24\\-28}$ &  5.95 &  5.89 \\
 CSIMon-000441$\dagger$  & 0.50 & 0.36 & 1.92 & 4.06 & -- &  --        &      --        & --                      &  3.97 &   --  \\
 CSIMon-000456  & 1.58 &  1.41 &  2.00 &   --  &  5.03  & 19.3$\pm$1.5 &  0.96$\pm$0.21 & 74$\substack{+16\\-33}$ &   --  &  6.92 \\
 CSIMon-000498  & 3.85 &  1.90 &  2.88 &  4.23 &  4.28  & 24.1$\pm$2.4 &  0.71$\pm$0.26 & 45$\substack{+31\\-19}$ &  4.73 &  4.77 \\
 CSIMon-000654  & 0.47 &  0.30 &  2.03 &  4.66 &   --   & 18.0$\pm$3.0 &  0.82$\pm$0.37 & 55$\substack{+35\\-28}$ &  3.87 &   --  \\
 CSIMon-000660  & 1.36 &  1.40 &  1.86 &  5.25 &  5.25  & 21.4$\pm$2.5 &  1.16$\pm$0.44 & 68$\pm$22               &  7.64 &  7.50 \\
 CSIMon-000774  & 3.10 &  1.83 &  2.49 &  3.46 &   --   & 30.7$\pm$1.3 &  0.84$\pm$0.28 & 57$\substack{+33\\-23}$ &  4.73 &   --  \\
 CSIMon-000811  & 1.08 &  0.91 &  1.97 &   --  &  7.88  & 13.5$\pm$1.7 &  1.07$\pm$0.35 & 64$\pm$26               &   --  &  8.19 \\
 CSIMon-000824  & 1.96 &  1.48 &  2.23 &  7.05 &   --   & 14.8$\pm$1.4 &  0.92$\pm$0.40 & 68$\substack{+22\\-37}$ &  7.90 &   --  \\
 CSIMon-000928  & 0.58 &  0.63 &  1.65 &  9.92 &   --   & 20.1$\pm$2.8 &  2.38$\pm$1.05 & 90                      & 10.09 &   --  \\
 CSIMon-001054  & 0.63 &  0.36 &  2.17 &   --  &  4.08  & 20.4$\pm$2.0 &  0.76$\pm$0.28 & 49$\substack{+41\\-28}$ &   --  &  3.52 \\
 CSIMon-001131  & 0.55 &  0.36 &  2.01 &   --  &  5.18  &  8.8$\pm$1.7 &  0.45$\pm$0.21 & 27$\pm$14               &  4.44 &  4.46 \\
 CSIMon-001140  & 1.10 &  1.31 &  1.67 &  3.87 &  3.90  & 19.8$\pm$2.3 &  0.91$\pm$0.34 & 66$\substack{+24\\-35}$ &  6.79 &  6.83 \\
 CSIMon-001167  & 0.56 &  0.30 &  2.23 &   --  &  8.78  & 11.3$\pm$2.1 &  0.88$\pm$0.42 & 62$\substack{+28\\-35}$ &  5.18 &  5.37 \\
 CSIMon-001296  & 0.91 &  0.69 &  1.99 &   --  &  9.75  & 11.0$\pm$1.6 &  1.07$\pm$0.48 & 63$\pm$27               &   --  &  8.52 \\
 CSIMon-001308  & 0.53 &  0.63 &  1.59 &  6.45 &  6.68  &  9.9$\pm$2.0 &  0.82$\pm$0.37 & 55$\substack{+35\\-28}$ &  7.86 &  8.04 \\
 CSIMon-014132$\dagger$  & 0.37 & 0.28 & 1.88 & -- & 4.48 &  --        &      --        & --                      &   --  &  3.98 \\
\hline
\end{tabular}
\tablefoot{
$P1$ and $P2$ are the Keplerian periods found from the CoRoT light curves in 2008 and 2011, respectively, 
while $R_{w,1}$ and $R_{w,2}$ are the co-rotation radii found using $P1$ and $P2$, respectively. 
The stars identified with a $\dagger$ either have no observed spectrum, or the S/N is too low to determine 
$v\sin i$, and so we cannot calculate their inclinations.
}
\end{center}
\end{table*}

\paragraph{Co-rotation radius}

A stellar rotation period independent of CoRoT light curves was measured for only one star in our 
sample, Mon-250, by determining the variability of the radial velocity of photospheric lines in 
the FLAMES spectra (Fig. \ref{fig:veilingmon250}). This method was unsuccessful for the other stars, 
since their radial velocities did not present significant variability within our errors. The 
spectroscopic period found for Mon-250 was $P_{rot}=8.6 \pm 0.5$ days, a value consistent with the 
periods found from its CoRoT light curves of $P_{kep}=8.3\pm1.0$ days in 2008 and $P_{kep}=8.9\pm0.5$ 
days in 2011. The same period of 8.6 days was found by Sousa et al. (in prep.) in both the H$\alpha$ and the 
HeI 6678\AA \space lines. This indicates that the structure responsible for the occultations in the CoRoT light 
curves is located at or near the co-rotation radius. 

This was found to be true for the star AA Tau as well, when \citet{bouvier07} 
measured its rotational period using the radial velocity of the HeI 5876\AA \space emission line and 
photospheric lines and found a value consistent with their photometric period. Therefore we find 
it reasonable to assume that the inner disk warp is located at the disk's co-rotation radius for 
the other AA Tau-like systems. 
Since we do not have an accurate measure of stellar rotation periods for the stars in our sample 
besides Mon-250, we use the periods determined from their CoRoT light curves in Equation 2 
to calculate their inclinations. 

\begin{figure}[t]
\centering
\includegraphics[width=5cm]{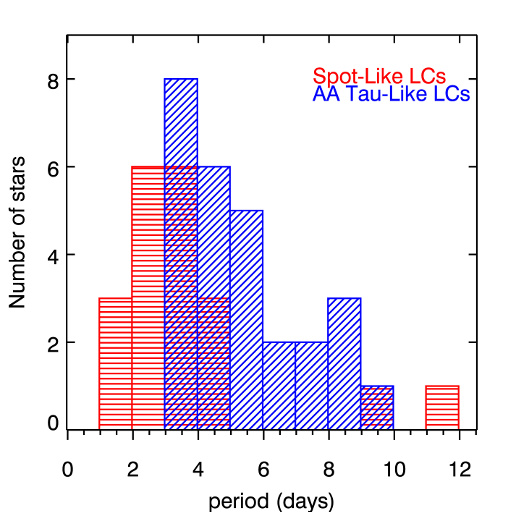}
\caption{Period histogram of classical T Tauri stars in NGC 2264 whose CoRoT light curves were 
classified as AA Tau-like (blue) or spot-like (red). The periods of spot-like light curves represent 
stellar rotational periods, while those of AA Tau-like light curves represent Keplerian rotation periods 
at the warp location. The periods of spot-like light curves were taken from the 2011 data, 
while AA Tau-like light curves from both 2008 and 2011 were used.
}\label{fig:per}
\end{figure}

In order to further investigate whether these Keplerian periods can also represent the stars' 
rotation periods, we made a period histogram of all of the CTTS in NGC 2264 with CoRoT light curves 
classified as spot-like or AA Tau-like (see Fig. \ref{fig:per}). 
The spot-like light curves are due to rotational modulation of stable 
configurations of spots on the stellar surface, and therefore these values represent the stars' 
rotational periods. The values from AA Tau-like light curves represent the Keplerian rotation 
periods of the inner disk material that occults the stellar photosphere. We see that, though the 
distributions are not identical, the values for AA Tau-like light curves fall within the region 
where rotational periods are found. 

\paragraph{Inclination}

We calculate $\sin i$ of each star using Equation \ref{eq:sini} and taking the period found from the 
CoRoT light curve as $P_{rot}$, except in the case of Mon-250, for which the spectroscopic period 
was used. For those stars observed to be periodic in both epochs, but for which a slightly different 
period was found in 2008 and 2011, the value with the smallest error was used, i.e., the one whose 
periodogram presented a peak with smaller width at half maximum and higher intensity. The results 
are shown in Table \ref{table:param}. 

It is important to note that the inclination is a very difficult parameter to determine precisely, 
especially when dealing with high inclinations. In these cases a relatively small error in 
$\sin i$ is translated into a large error in $i$. Unfortunately our errors in $\sin i$ themselves are 
relatively large to begin with, therefore our values for $i$ are very imprecise. 

\citet{hartmann01} and \citet{rebull02} discuss the difficulties in obtaining precise values of 
luminosity, temperature, and radius for TTSs because of their strong variabilities, spectrum veiling, surface 
spots, and uncertainties in calculating the distance. The former showed that typical errors in radius 
determination are on the order of 20\%, while the latter show errors of typically 12\% for TTSs in the
spectral range K5-M2. In this study we consider errors of 20\% in radius.

Errors in period and $v\sin i$ vary from one star to another. For periods, we adopted 
the half width at half maximum of the peak in each periodogram as the uncertainty, which is typically 
around 5\% to 7\%. Our errors in $v\sin i$ are typically on the order of 10\% - 12\%. This leads to 
errors around 35\% in $\sin i$, which is rather high, but consistent with other studies of 
TTS inclinations \citep[e.g., ][]{artemenko12}.

Although the actual values of inclination are imprecise, it is not difficult to distinguish between 
systems with low inclinations and those with medium to high inclinations, since for low inclinations 
errors in $\sin i$ do not translate into such large errors in $i$ as with high inclinations. We see 
from Table \ref{table:param} that the star Mon-1131 is seen at a low inclination, and we can 
consequently rule out the proposed occultation mechanism as the primary cause for the light curve 
variations. However, it is possible that there are two structures in each hemisphere, separated 
by $180^{\circ}$, obscuring the photosphere and making it seem as if the period were half the 
true value. If this were the case its inclination would be $64^{\circ} \substack{+26\\-36}$, though 
further investigation is needed to determine its true rotation period. For the time being 
this star has been excluded from further investigation using the 
occultation model, though it is studied in \citet{stauffer14b}, where different mechanisms are 
proposed to explain its variability.

For the other stars in our sample for which an inclination can be estimated, they tend toward high 
values, at least within the uncertainties (see Table \ref{table:param}). 
There are many stars for which the value of $\sin i$ found was larger than 1. In some cases 
a reasonable interval of values within the error bars falls under 1, in which case we argue that the 
system's inclination should be anywhere between the inverse sine of the lowest value 
($\sin i - \Delta \sin i$, where $\Delta \sin i$ is the uncertainty in $\sin i$) and $90^{\circ}$. We 
then consider the inclination to be the mean value in this interval (for example, a star of  
$\sin i=1.07\pm 0.32$ has an inclination of $69^{\circ} \pm 21^{\circ}$, since the inverse sine of the 
lowest value,  $\sin i=1.07-0.32=0.75$, is $48^{\circ}$).

For the star Mon-928, its value of $\sin i$ is much larger than 1, even when considering the error bars. 
It is possible that the material occulting this star is located farther away from the star than its 
co-rotation radius, in which case the Keplerian period we calculated using the CoRoT light curves 
is larger than the stellar rotation period, resulting in a larger value of $\sin i$. It is also possible 
that the true period is half the value used. There is an additional, smaller, flux dip in its 2008 light 
curve located between the second and third larger flux dips (see Fig. \ref{fig:928}), which justifies 
using a period of 4.96 days, rather than 9.92 days, though we gave preference to the latter since it 
coincides with the period found in the literature of $9.48 \pm 2.34$ \citep{lamm05}. If the period 
were indeed half the value we considered, this would lead to an inclination of $66^{\circ} \pm 24^{\circ}$. 
In either case, it is probably safe to consider that this system has a high inclination. Therefore, we used 
the occultation model described in Sect. \ref{sec:introoccmodel} to find which interval of inclinations results in the 
best fit for the light curve, and consider this as the system's inclination. We did this as well for 
the three stars for which we have no measure of $v\sin i$, where we assume without proof that the systems 
are seen nearly edge-on. The new values are presented in Table \ref{table:occmodel}.

\begin{table*}[ht]
\begin{center}
\caption{Occultation model parameters found for the AA Tau-like candidates. 
}\label{table:occmodel}
\begin{tabular}{c c c c c c}
\hline
\hline
 CSIMon ID      & Epoch & $R_w (R_*)$ & $h_{max}/R_w$ & $\phi _w (^\circ)$ & Inclination $(^\circ)$\\
\hline
 CSIMon-000056  &  2008 &   9.29 & 0.24-0.25 ($\pm$0.10) & 320 ($\pm$40) - 360           & 71$\pm$6   \\
 CSIMon-000056  &  2011 &   9.45 & 0.24-0.26 ($\pm$0.10) & 290 ($\pm$70) - 360           & 71$\pm$6   \\
 CSIMon-000250  &  2008 &  11.71 & 0.23-0.28 ($\pm$0.05) & 140 ($\pm$30) - 360           & 74$\pm$3   \\
 CSIMon-000250  &  2011 &  12.28 & 0.24-0.27 ($\pm$0.05) & 270 ($\pm$30) - 360           & 74$\pm$3   \\
 CSIMon-000296  &  2011 &   6.86 & 0.21-0.31 ($\pm$0.09) & 290 ($\pm$40) - 360           & 72$\pm$5   \\
 CSIMon-000297  &  2008 &   5.95 & 0.17-0.28 ($\pm$0.04) & 180 ($\pm$30) - 360           & 75$\pm$2   \\
 CSIMon-000358  &  2011 &   6.51 & 0.17-0.19 ($\pm$0.06) & 210 ($\pm$50) - 360           & 73$\pm$4   \\
 CSIMon-000379  &  2011 &   5.89 & 0.22-0.34 ($\pm$0.12) & 110 ($\pm$30) - 270 ($\pm$90) & 70$\pm$7   \\
 CSIMon-000441  &  2008 &   3.97 & 0.17-0.23 ($\pm$0.11) & 160 ($\pm$20) - 310 ($\pm$50) & 70$\pm$7   \\
 CSIMon-000456  &  2011 &   6.92 & 0.18-0.24 ($\pm$0.05) &         360                   & 74$\pm$3   \\
 CSIMon-000498  &  2008 &   4.73 & 0.17-0.21 ($\pm$0.10) & 145 ($\pm$35) - 165 ($\pm$15) & 70$\pm$6   \\
 CSIMon-000498  &  2011 &   4.77 & 0.19-0.22 ($\pm$0.10) & 130 ($\pm$50) - 165 ($\pm$15) & 70$\pm$6   \\
 CSIMon-000654  &  2008 &   3.87 & 0.10-0.13 ($\pm$0.06) & 180 ($\pm$60) - 310 ($\pm$50) & 73$\pm$4   \\
 CSIMon-000660  &  2008 &   7.64 & 0.25-0.33 ($\pm$0.04) & 260 ($\pm$20) - 360           & 74$\pm$3   \\
 CSIMon-000660  &  2011 &   7.50 & 0.27-0.29 ($\pm$0.05) & 200 ($\pm$30) - 360           & 74$\pm$3   \\
 CSIMon-000774  &  2008 &   4.73 & 0.15-0.21 ($\pm$0.03) & 220 ($\pm$40) - 360           & 75$\pm$2   \\
 CSIMon-000811  &  2011 &   8.19 & 0.26-0.28 ($\pm$0.08) &         360                   & 72$\pm$5   \\
 CSIMon-000824  &  2008 &   7.90 & 0.24-0.27 ($\pm$0.08) & 260 ($\pm$40) - 360           & 72$\pm$5   \\
 CSIMon-000928  &  2008 &   6.36 & 0.28-0.30 ($\pm$0.10) & 120 - 300 ($\pm$60)           & 71$\pm$6   \\
 CSIMon-001054  &  2011 &   3.52 & 0.16-0.21 ($\pm$0.09) & 270 ($\pm$70) - 360           & 70$\pm$5   \\
 CSIMon-001140  &  2008 &   6.79 & 0.16-0.24 ($\pm$0.03) & 190 ($\pm$20) - 360           & 75$\pm$2   \\
 CSIMon-001140  &  2011 &   6.83 & 0.14-0.24 ($\pm$0.03) & 190 ($\pm$20) - 360           & 75$\pm$2   \\
 CSIMon-001167  &  2011 &   5.37 & 0.15-0.18 ($\pm$0.05) & 330 ($\pm$30) - 360           & 74$\pm$3   \\
 CSIMon-001296  &  2011 &   8.52 & 0.31-0.34 ($\pm$0.13) & 260 ($\pm$80) - 330 ($\pm$30) & 69$\pm$8   \\
 CSIMon-001308  &  2008 &   7.86 & 0.28-0.30 ($\pm$0.06) & 330 ($\pm$30) - 360           & 73$\pm$4   \\
 CSIMon-001308  &  2011 &   8.04 & 0.20-0.27 ($\pm$0.06) & 340 ($\pm$20) - 360           & 73$\pm$4   \\
 CSIMon-014132  &  2011 &   3.98 & 0.18-0.32 ($\pm$0.14) & 120 ($\pm$40) - 290 ($\pm$80) & 68$\pm$9   \\
\hline
\end{tabular}
\tablefoot{
$R_w$ is the radius where the warp is located, assumed to be at the co-rotation radius 
(i.e., $R_w=R_{co}$); $h_{max}$ is the maximum warp height found for each flux dip, expressed in 
stellar radii $R_*$; and $\phi _w$ is the azimuthal extension of the warp found for each flux dip. The 
inclinations shown here were constrained using the occultation model.
}
\end{center}
\end{table*}

\begin{figure}[t]
\centering
\includegraphics[width=4.4cm]{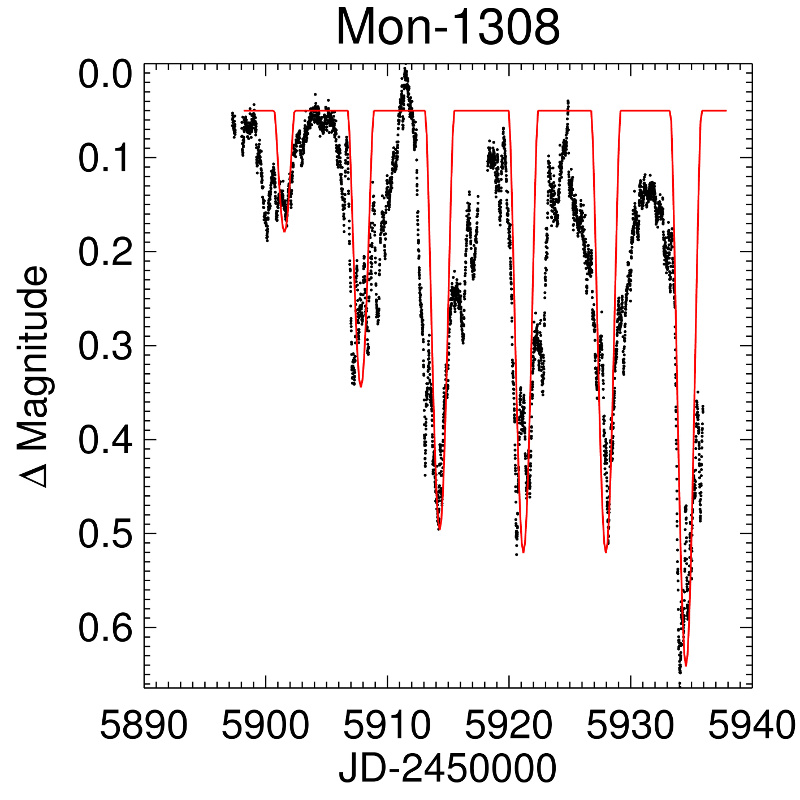}
\includegraphics[width=4.4cm]{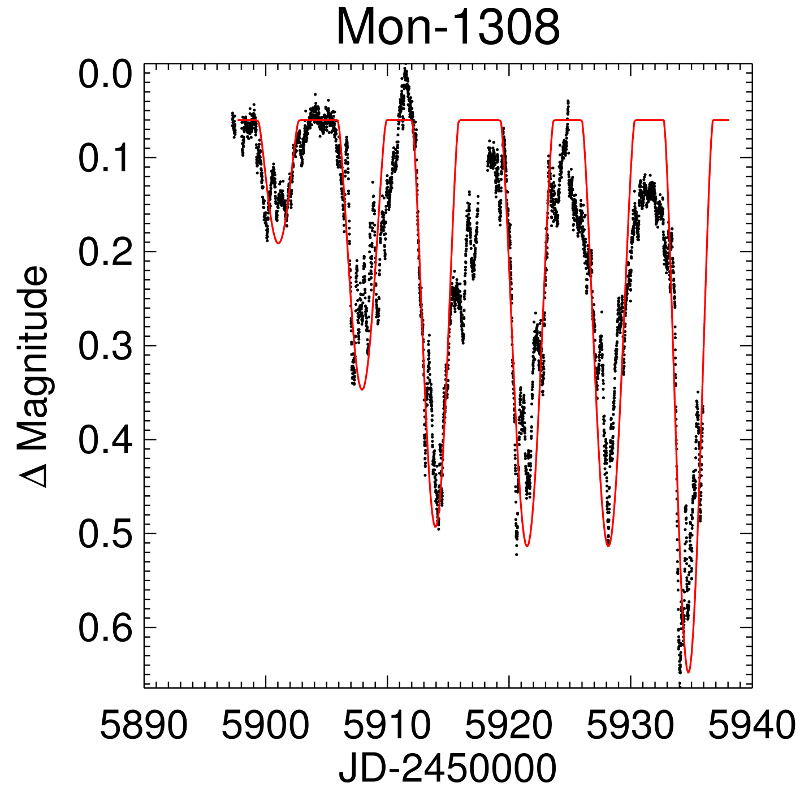}
\caption{Fitted light curves of the star Mon-1308 using an inclination of 55$^{\circ}$ (left) and 
75$^{\circ}$ (right). In both cases the values of warp maximum height and azimuthal extension were 
set to those that best fit the width and amplitude of the minima without extrapolating to unphysical 
values. 
}\label{fig:aatmodelinc}
\end{figure}

For the rest of the AA Tau-like stars, we used the occultation model in a similar fashion as a fine tuning for 
the inclination. We noticed that in many cases we were not able to successfully reproduce the observed 
widths in flux dips using the whole range of values of $i$ initially calculated, unless we used other 
parameters that were unphysical (such as azimuthal extension $\phi_w > 360^{\circ}$). Therefore, we took all 
values of $i$ within the error bars and chose only the interval that could successfully reproduce the 
widths of the observed light curves with physically possible values of $\phi_w$. This restricted the 
inclinations much more than the original calculation, resulting in much smaller error bars, as can be seen 
in Table \ref{table:occmodel}. Fig. \ref{fig:aatmodelinc} shows an example of the best 
fits for a light curve using the originally calculated value of inclination 
(with $\phi_w \leq 360^{\circ}$) and the new value presented in Table \ref{table:occmodel}. It is easy 
to see that this value reproduces the widths of the flux dips much more closely. 

\paragraph{Warp parameters and fitted light curves}

\begin{figure*}[t]
\centering
\includegraphics[width=4.4cm]{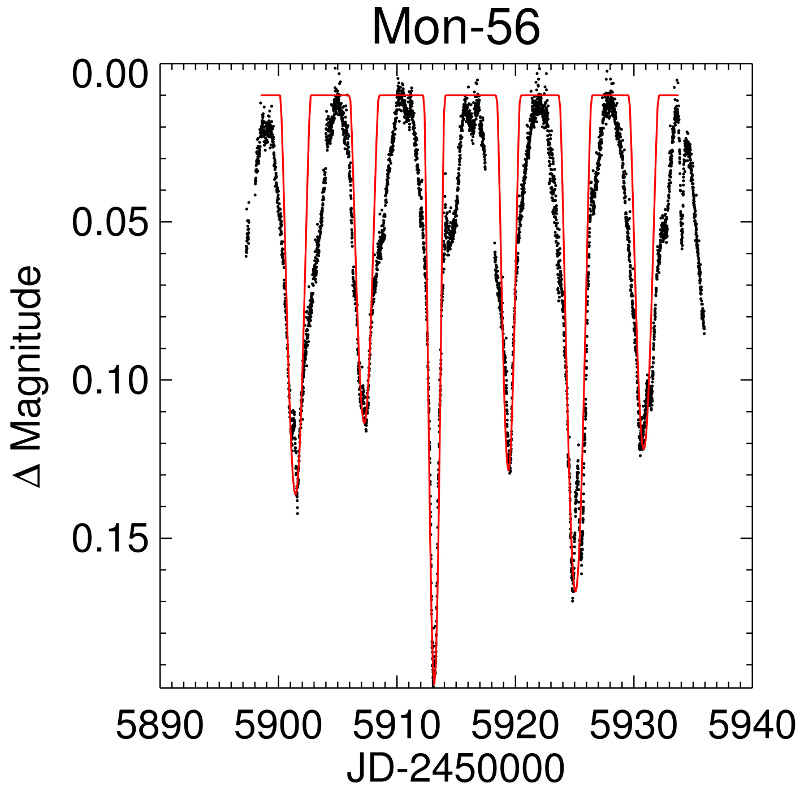}
\includegraphics[width=4.4cm]{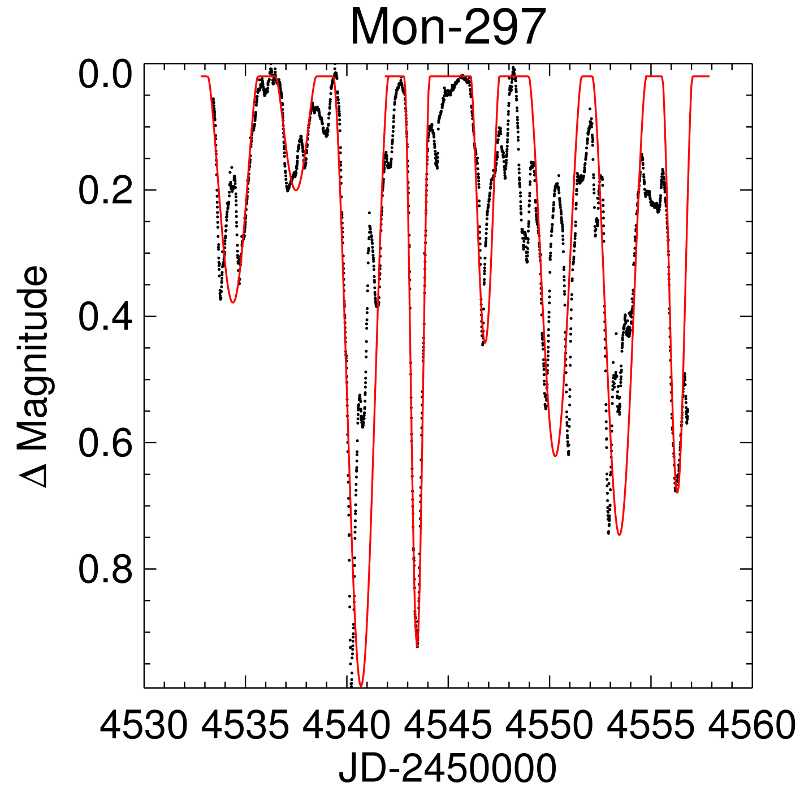}
\includegraphics[width=4.4cm]{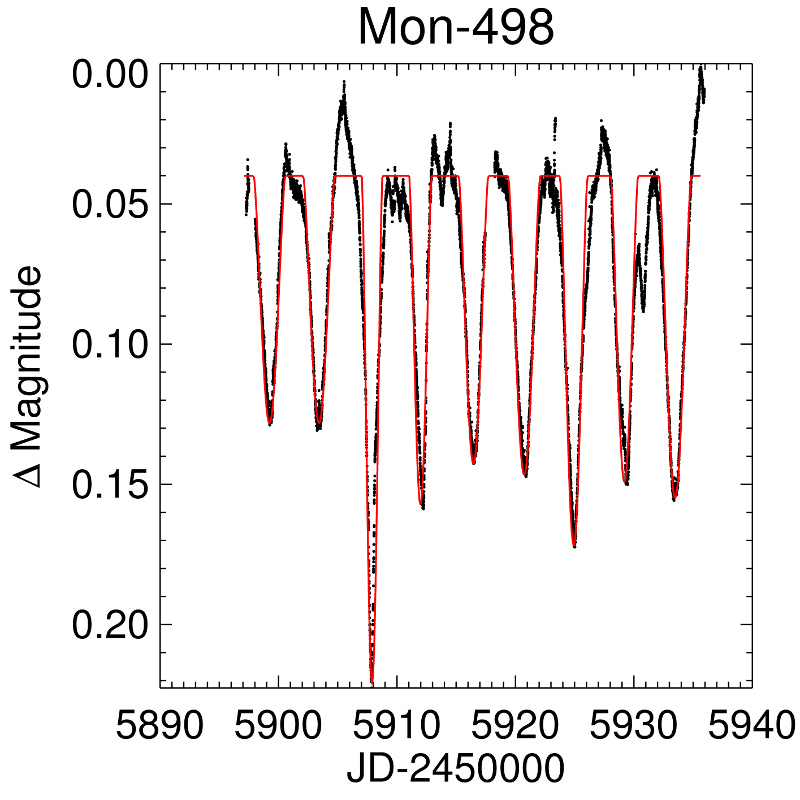}
\includegraphics[width=4.4cm]{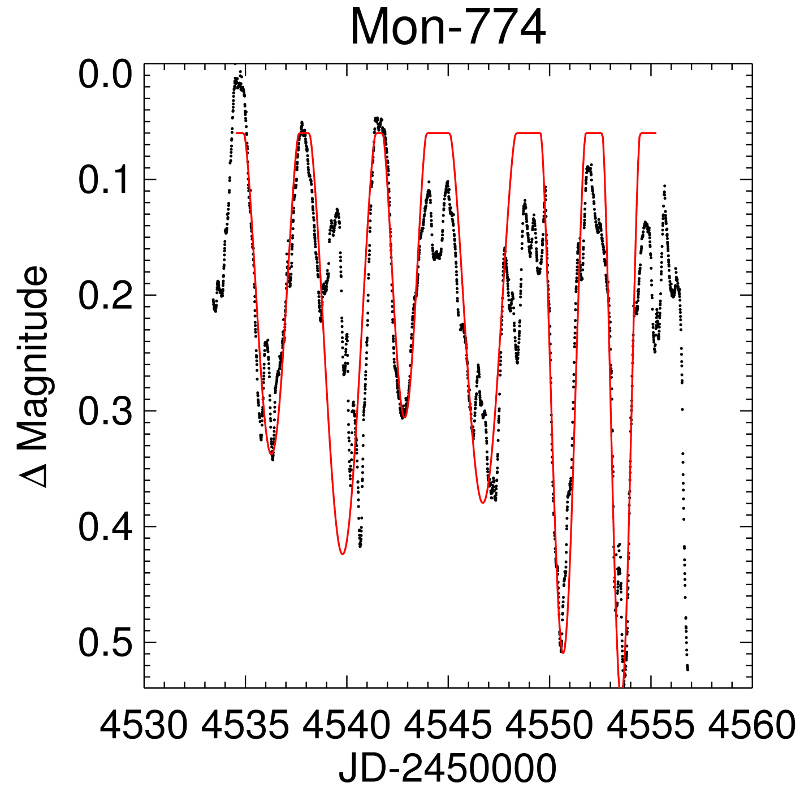}
\includegraphics[width=4.4cm]{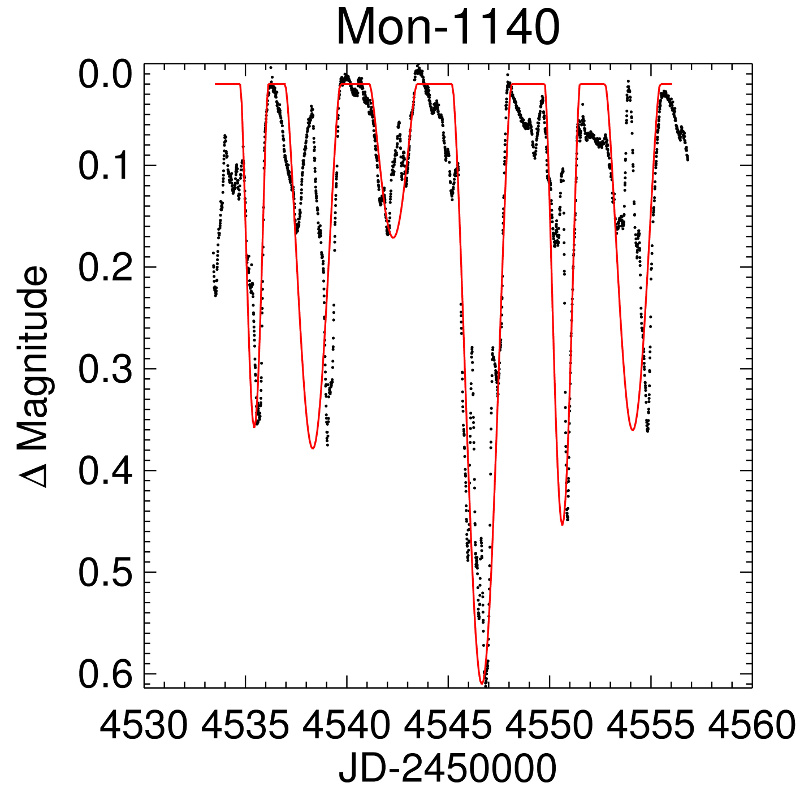}
\includegraphics[width=4.4cm]{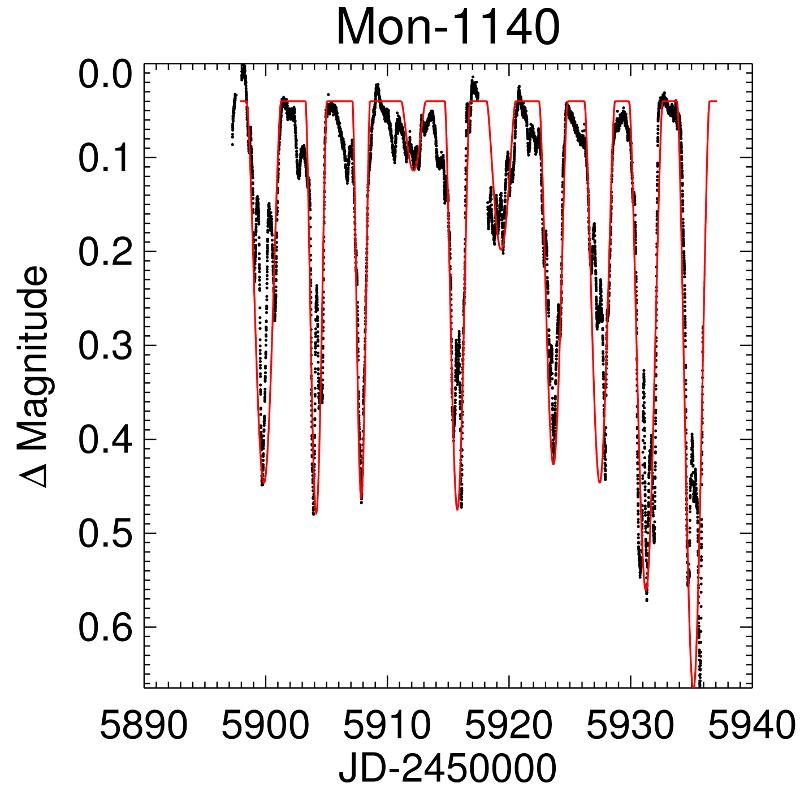}
\includegraphics[width=4.4cm]{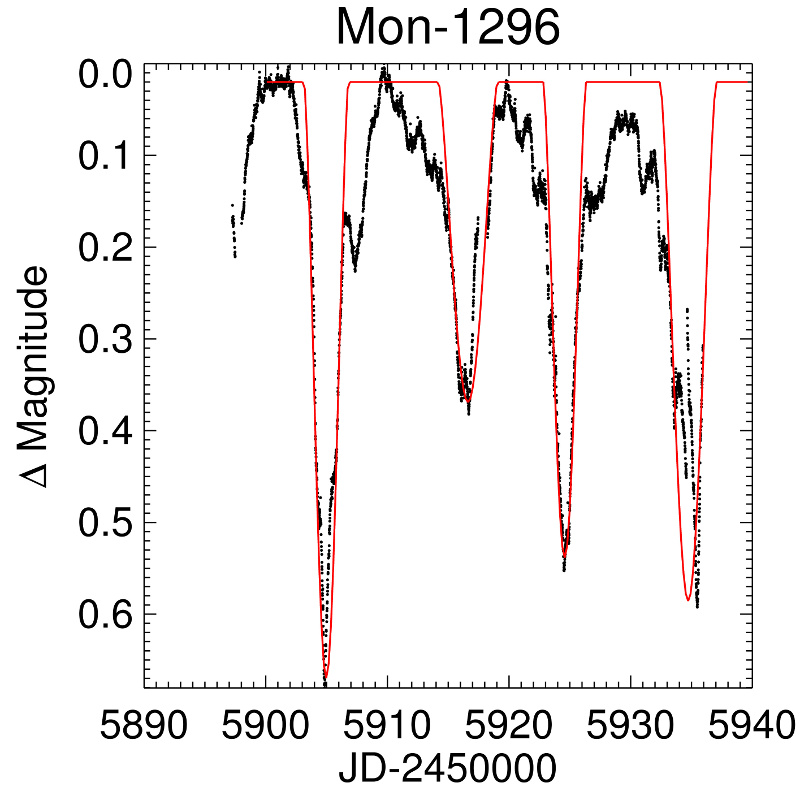}
\includegraphics[width=4.4cm]{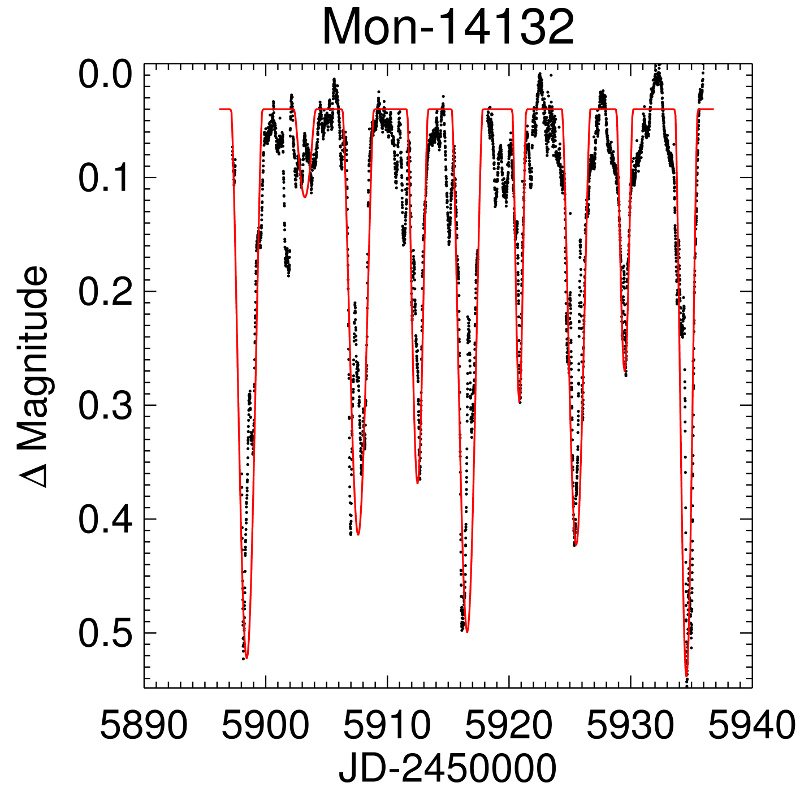}
\caption{Examples of simulated light curves of AA Tau-like stars (red) plotted over 
CoRoT light curves (black). 
}\label{fig:aatmodel}
\end{figure*}

The occultation model was used interactively to simulate the CoRoT light curves using the values given in Table 
\ref{table:param} for stellar mass, radius, and co-rotation radius, with the warp maximum height and azimuthal 
extension as free parameters, and inclination as a semi-free parameter, constrained to values within the 
error bars presented in Table \ref{table:param}. Each minimum of each light curve was fitted independently 
to account for slight changes in the warp from one rotation cycle to another. Some examples of the synthetic 
light curves reproduced by the occultation model are shown in Fig. \ref{fig:aatmodel} (synthetic light 
curves of all other stars are shown in the online material, Fig. \ref{fig:app3}). 
Inclinations were determined first, as described above, by verifying for 
which values we were able to reproduce the width of each minima without having to resort to unphysical values 
of azimuthal extension. We used a maximum inclination of $77^{\circ}$, since for inclinations larger than this 
the outer flared disk would overshadow the inner disk region\footnote{This effect may not be as significant for 
sources that have anemic disks, since they may possess flatter outer disks. However, this would only be relevant 
for the star Mon-314, which has aperiodic photometric behavior and therefore is not analyzed using the 
occultation model at any rate.}, making it impossible to observe the stellar photosphere \citep{bertout00}. 
Error bars in warp maximum height and azimuthal extension were calculated by modeling 
the minima with the value of inclination fixed at its lowest and highest possible values, and finding the 
best fit in each case.  

Table \ref{table:occmodel} shows the final values obtained for the model parameters. Since they vary from 
one rotation cycle to another, we present the intervals of the values found for each light curve. Most of 
the values found for the warp's maximum height and azimuthal extension are similar 
to those found for AA Tau. The average ratio between the warp's maximum height and the radius at which 
it is located is $h_{max}/R_w=0.23$, though individual values range from 0.10 to 0.34 (for comparison we 
note that the value used for AA Tau in \citet{bouvier99} was of $h_{max}/R_w=0.30$). 
The average value of the warp's azimuthal extension is 
$\phi_c=300^{\circ}$, but is often a full $360^{\circ}$, the same as AA Tau, and can be as low as $110^{\circ}$. 
Inclinations generally need to be greater than $\sim 65^{\circ}$ for the occultation model to successfully 
reproduce the widths of the minima, but in some cases may be as low as $60^{\circ}$ for relatively large warp 
heights and shallow flux dips. 

We see that the warp maximum height should vary between rotation cycles by on average 11\%, and at times by 
up to 57\%, to account for the variations in amplitude of the light curves. The 
azimuthal extension of the warp should vary by on average 17\%, and up to 65\%, between rotation cycles to 
account for the variations in width of the minima present in the CoRoT light curves. If the photometric 
variability of these stars is truly due to occultations by an inner disk warp caused by the interaction 
between the magnetosphere and the inner disk region, as we propose, then these values show how dynamic this 
interaction is, as has been predicted by MHD models.

\subsection{Color-magnitude diagrams and color variation}\label{sec:cmd}

\begin{figure*}[thbp]
\centering
\includegraphics[width=6cm]{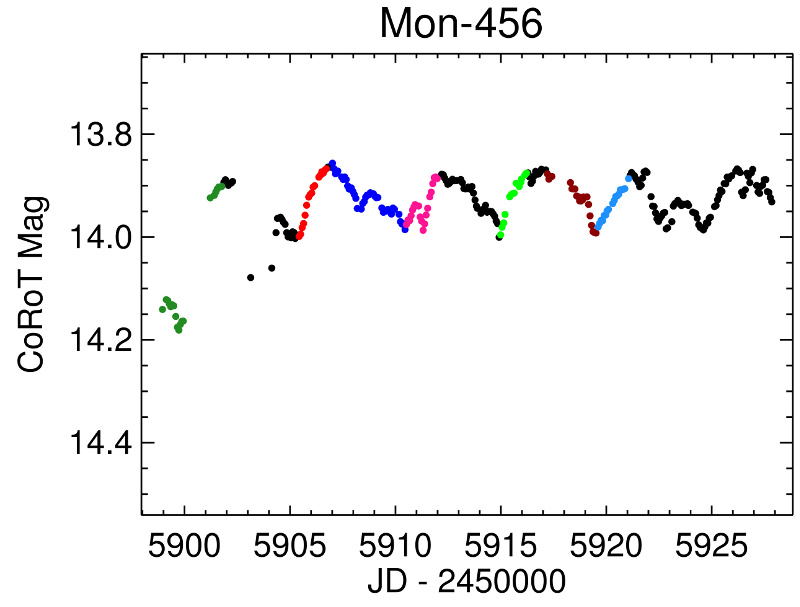}
\includegraphics[width=6cm]{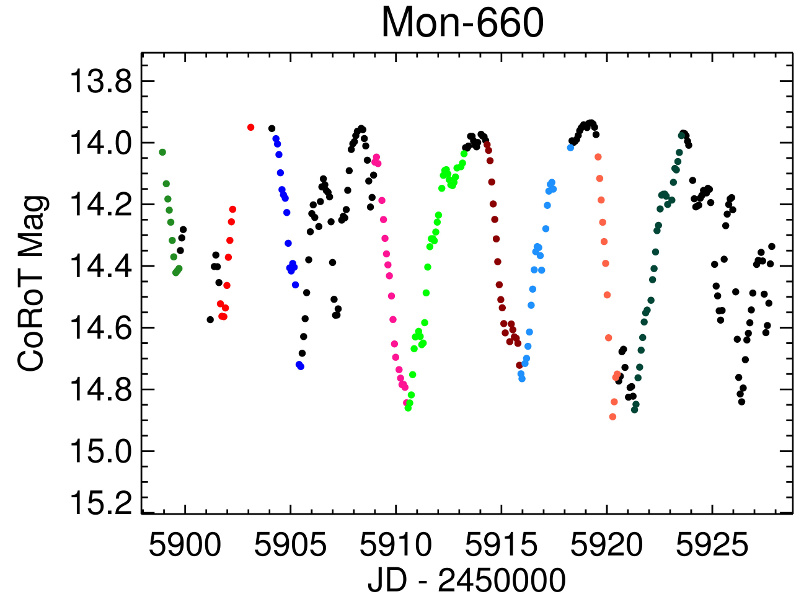}
\includegraphics[width=6cm]{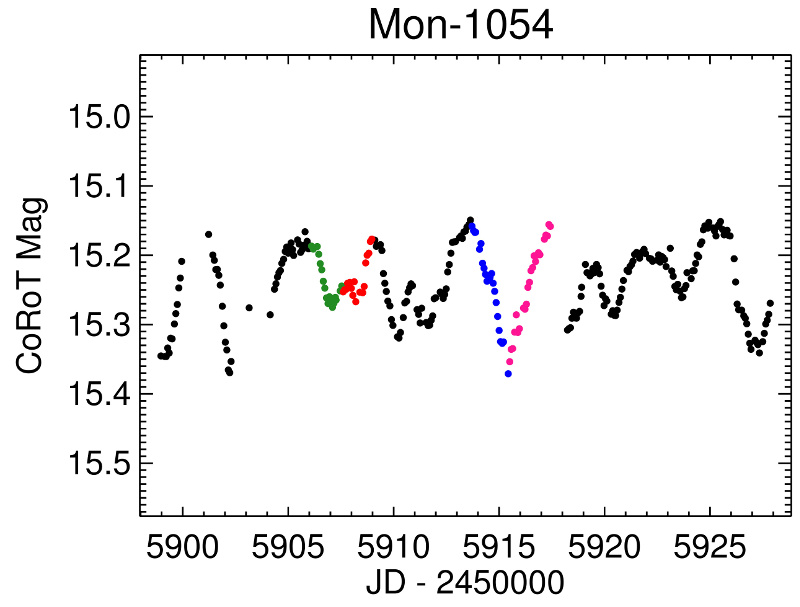}
\includegraphics[width=6cm]{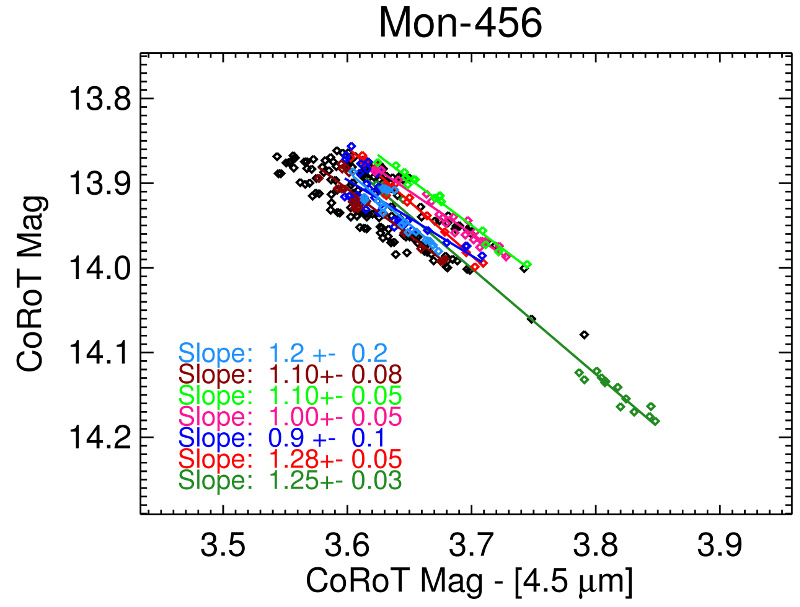}
\includegraphics[width=6cm]{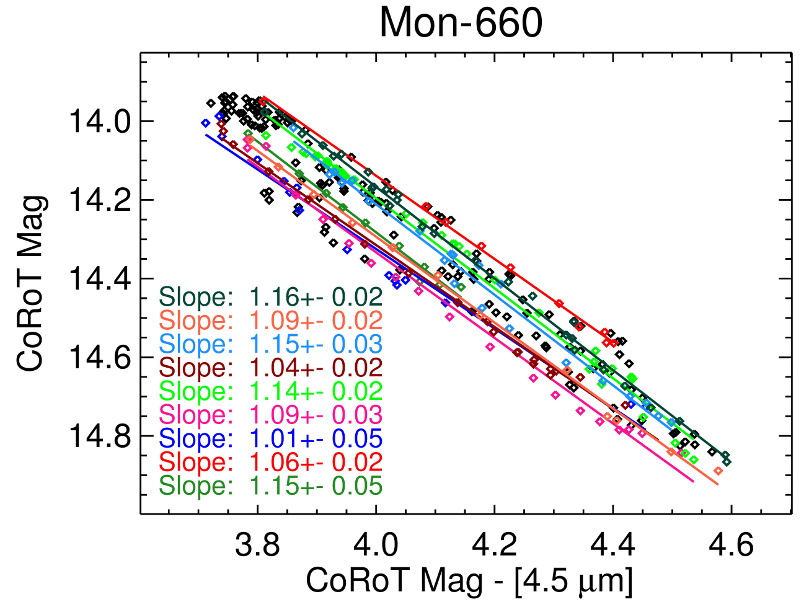}
\includegraphics[width=6cm]{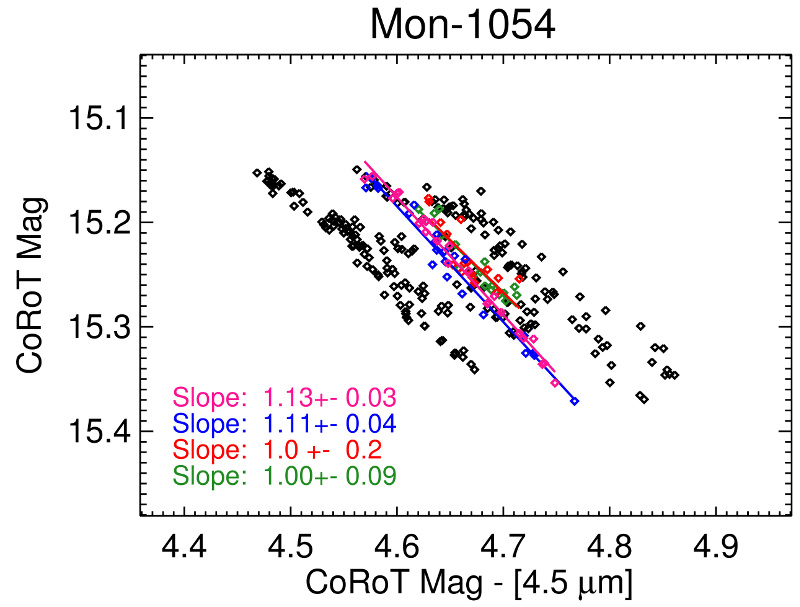}
\caption{Top: CoRoT light curve of the stars Mon-456, Mon-660, and Mon-1054. Bottom: CoRoT vs. 
CoRoT-IRAC$_{4.5 \mu \mathrm{m}}$ color magnitude diagrams of the same stars. The slopes were calculated 
for various parts of the light curves where CoRoT and \textit{Spitzer} IRAC data best correlate. Different 
colors show different parts of the light curves for which slopes were calculated.
}\label{fig:colmag}
\end{figure*}

We constructed color magnitude diagrams of the stars in our sample (both those that present 
AA Tau-like light curves and those that present aperiodic extinction dominated light curves), to search 
for trends characteristic of extinction. We used data taken simultaneously with CoRoT and \textit{Spitzer}, 
plotting CoRoT magnitude against CoRoT magnitude minus \textit{Spitzer} IRAC$_{4.5 \mu \mathrm{m}}$ magnitude. 
It is not possible to use this method to clearly distinguish if the photometric variability is due to extinction 
or a configuration of spots on the surface, but if we assume that 
extinction is the main source of the flux dips present in both the CoRoT and IRAC light curves, then we 
can estimate extinction laws from the slopes of the distributions (Fig. \ref{fig:colmag}). 

In order to be sure that we are analyzing parts of the light curve where the same phenomenon is responsible 
for the variability in the optical and infrared bands, we calculated slopes using linear regression only for the 
parts of the data where the CoRoT and IRAC light curves correlate reasonably well. We also separated these parts 
in order to calculate slopes for different flux dips and different parts of the flux dips, as shown in Fig. 
\ref{fig:colmag}, in order to verify if different extinction laws could be found for different parts of 
the disk warp. We find that entrances into and exits from the occultations show the same slopes, within 
the uncertainties. In one case, Mon-456 (see left panels of Fig. \ref{fig:colmag}), slightly different 
slopes were found during different flux dips. This might indicate a small variability in the grain distribution 
of the inner disk warp during rotation cycles, or it could be due to other factors, such as variable disk emission, 
interfering with the IRAC light curves. 

We were able to estimate extinction laws for ten of the stars in our sample using this method, finding values  
between those consistent with ISM ($A_{4.5 \mu \mathrm{m}}/A_R \sim 0.05$) and up to 
$A_{4.5 \mu \mathrm{m}}/A_R \sim 0.29 \pm 0.04$. Slopes of values 
less than 1 were not considered, since extinction laws derived would be unphysical. In these cases, other 
factors, such as variable disk emission in $4.5 \mu \mathrm{m}$ or a configuration of hot or cold spots on the 
stellar surface, could be strongly influencing the distribution of points in the color magnitude diagrams. It 
is possible that variable disk emission is also influencing our derived extinction laws, causing us to find 
lower slopes and therefore lower extinction laws than were estimated in Sect. \ref{sec:optir}. More accurate estimates 
will require a more extensive study of these and other color magnitude diagrams, which will be treated in a 
future paper.

To check for color variation in the periodic and quasi-periodic stars in our sample, we compared CoRoT 
and I-band light curves with $u-r$ for the stars whose light curves were classified as AA Tau-like or 
AA Tau candidates in 2011. We plotted these light curves in phase, using the periods determined 
from the CoRoT data (Table \ref{table:uxori2}). The same initial date was taken for all three plots for easy 
comparison. Though the CFHT data were not taken simultaneously with CoRot, by folding the light curves in 
phase, using the same initial date and period, we can see how the color varies during the flux dips. For 
most stars, we see a reddening at phase 0.5 during the eclipses, as is to be expected from an extinction 
event. 

For four stars (Mon-250, Mon-358, Mon-456, and Mon-1054), we can observe a bluing at phase 0.5 for one or 
two rotation cycles, while the other cycles either maintain relatively constant color or become slightly 
redder. This effect was observed in AA Tau \citep{bouvier03}, which showed little color variation except 
during some minima, when it became bluer. This was attributed to the appearance of part of the hot spot 
associated with the accretion shocks on the stellar surface, during eclipses from the inner disk warp. 
This shows very strong evidence that these four stars are undergoing the same physical processes as AA Tau 
did during its stable accretion phase. 

\begin{figure*}[t!]
\centering
\includegraphics[width=11cm]{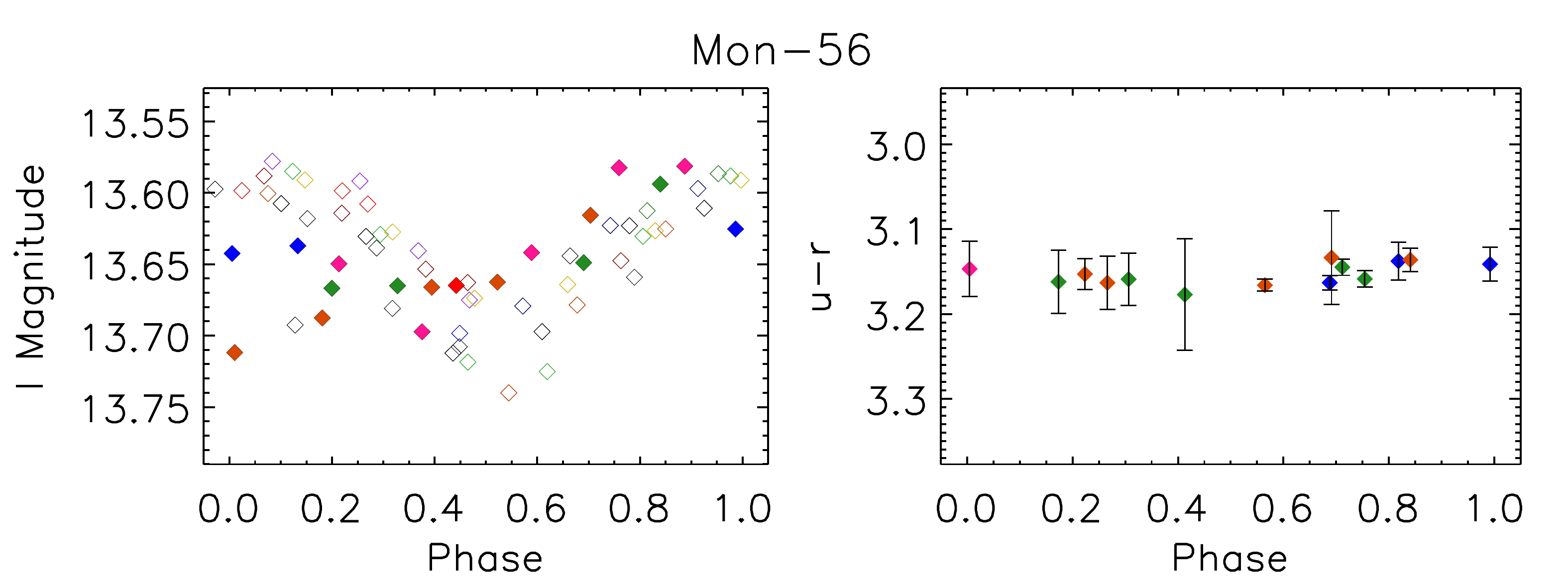}
\includegraphics[width=11cm]{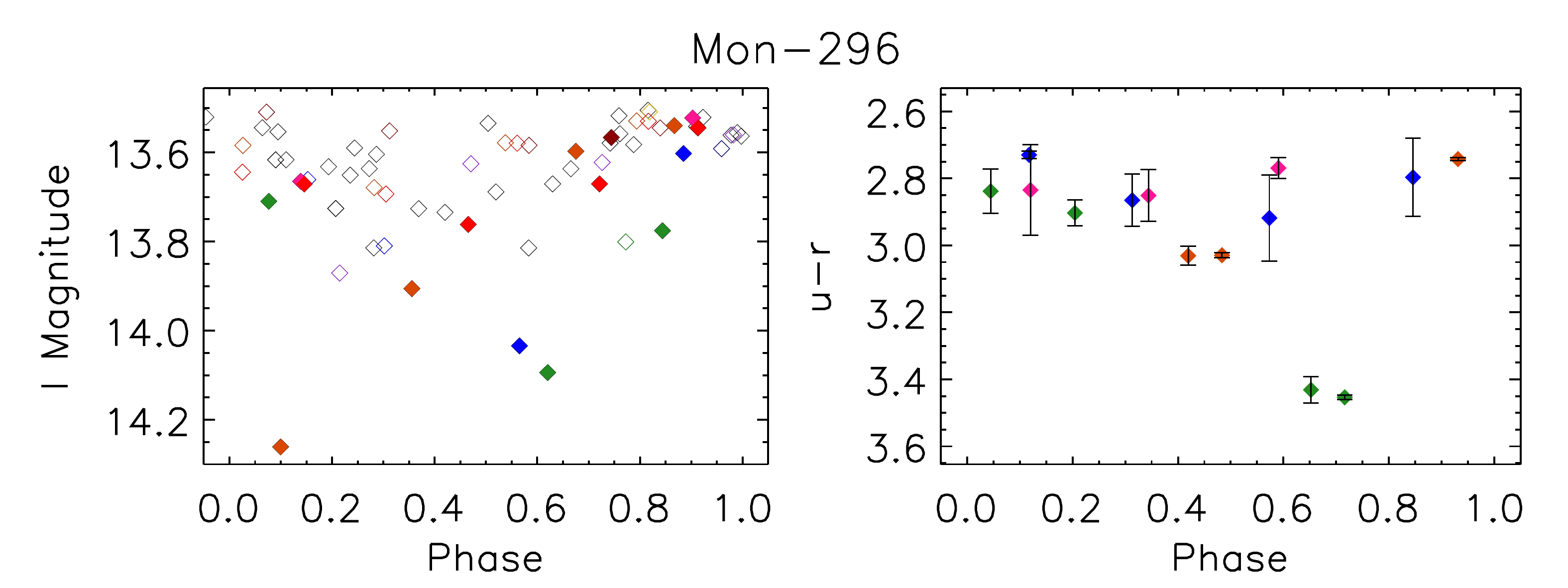}
\includegraphics[width=11cm]{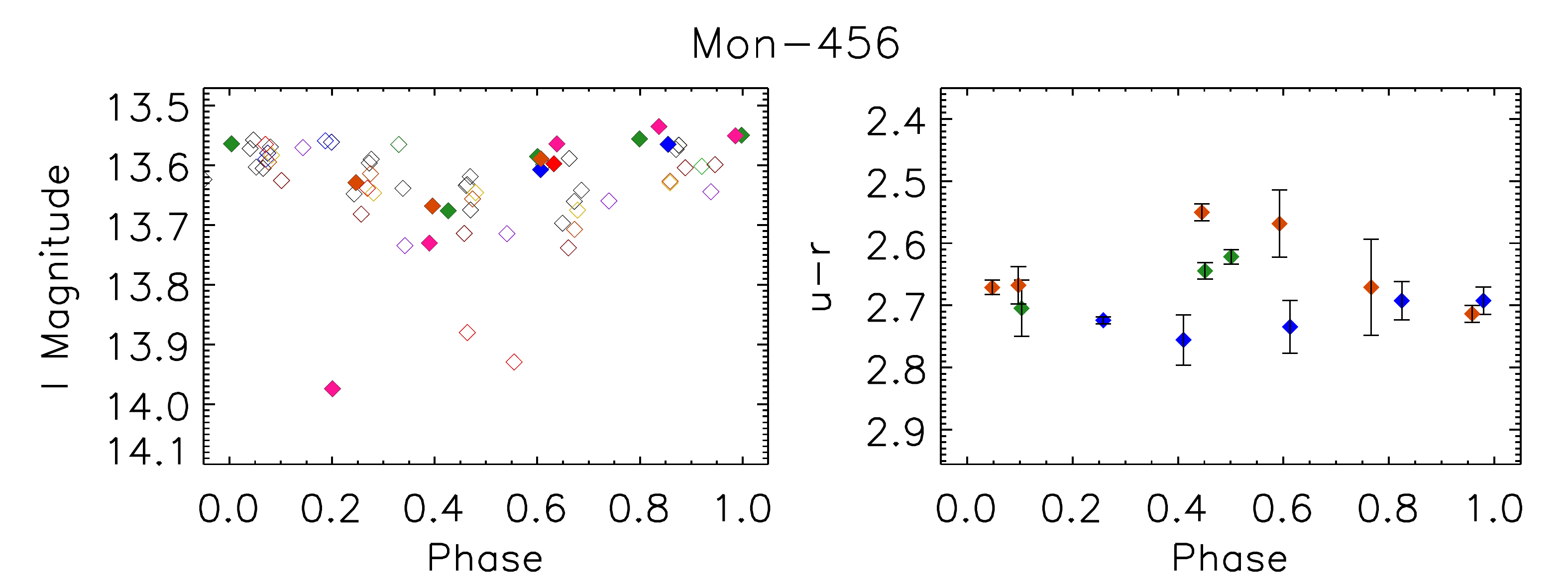}
\includegraphics[width=11cm]{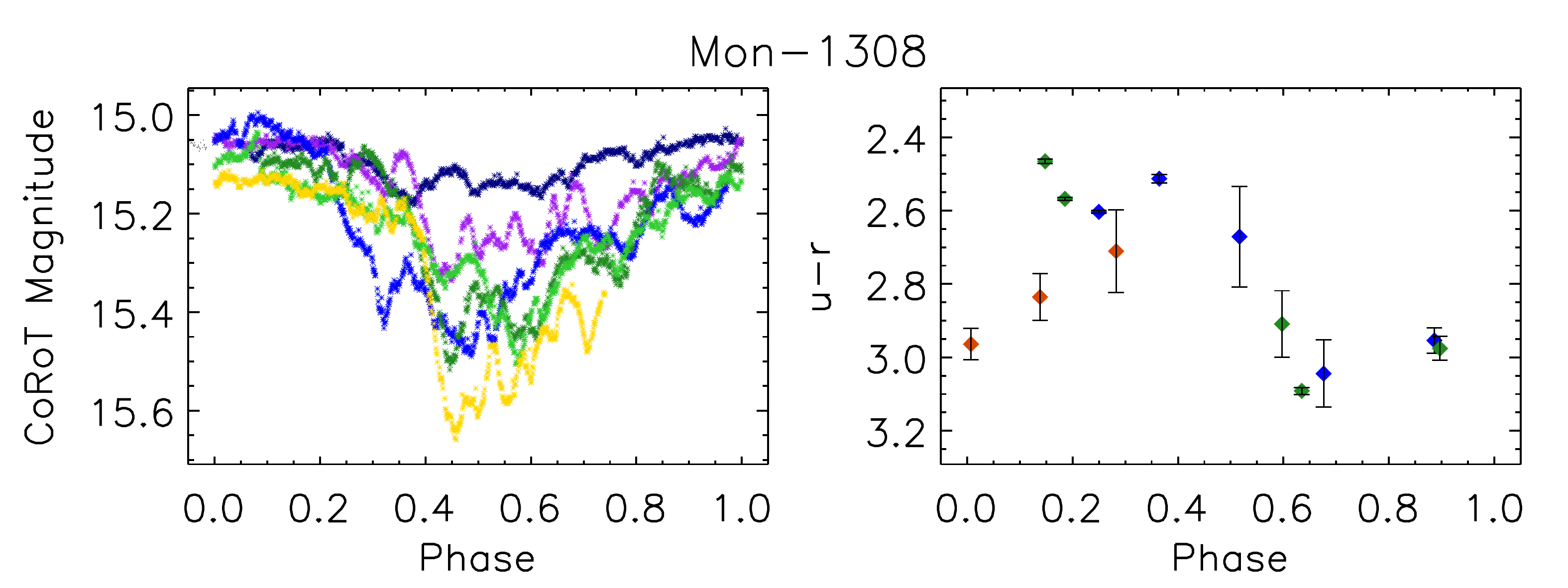}
\caption{Examples of I-band and u-r light curves folded in phase. The colors used for u-r represent the 
same rotation cycle as the filled diamonds of the same color in I. Mon-56 shows no measurable  
variability in $u-r$; Mon-296 shows reddening events in $u-r$ during its eclipses in the I-band; 
Mon-456 shows a slight reddening during one of its I-band minima (blue diamonds) and bluing 
events during the other I-band minima (orange and green diamonds); and Mon-1308 shows a bluing, 
followed by a reddening event during its minima in the CoRoT light curve. It was not possible to identify flux 
dips in the phase-folded I-band light curve of Mon-1308 because of insufficient cadence, so the phase-folded CoRoT 
light curve is shown instead. The same period and initial date were used for CoRoT and u-r, though in 
this case colors do not represent the same phase in both plots. 
}\label{fig:colorphase}
\end{figure*}

Three stars undergo an interesting phenomenon where they become slightly bluer during the beginning of 
the eclipses and then become redder. This could also be evidence that the hot spot associated with the 
accretion shock is partially showing, but in this case it is soon occulted by the inner disk warp. 
These stars are Mon-1140, Mon-1308, and Mon-1167. For two stars, Mon-56 and Mon-14132, we see little or 
no change in color. Fig. \ref{fig:colorphase} shows some of these light curves and $u-r$ color diagrams 
folded in phase. The others are shown in the online material, Fig. \ref{fig:app4}.

\subsection{Other characteristics of AA Tau-like stars}\label{sec:other}

\begin{figure}[t]
\centering
\includegraphics[width=3.5cm]{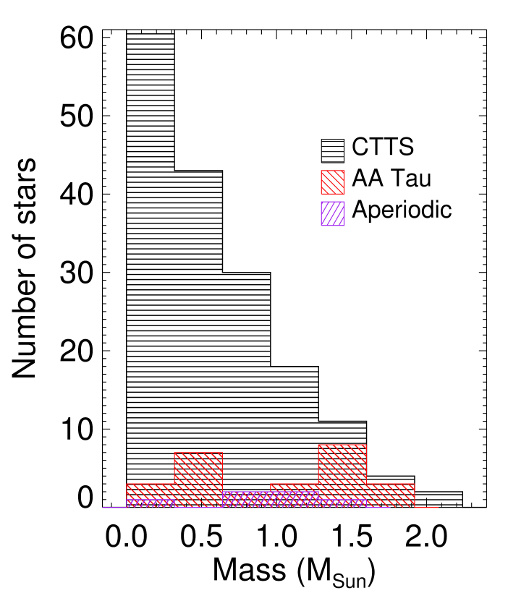}
\includegraphics[width=3.7cm]{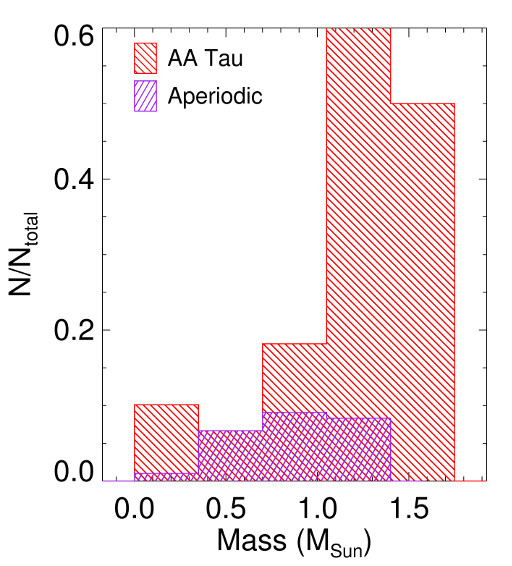}
\includegraphics[width=4.0cm]{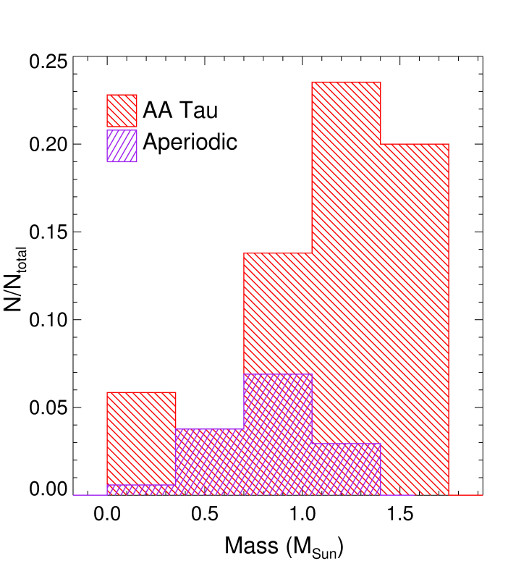}
\caption{Top left: mass histogram of all CTTS with CoRoT light curves (black), stars with AA Tau-like light 
curves (red), and stars whose CoRoT light curves are dominated by aperiodic extinction events (purple).
Top right: number of AA Tau-like stars (red) and aperiodic extinction stars (purple), divided by the total 
number of CTTS observed by CoRoT in that mass bin. Stars of both CoRoT observing runs were used, and 
any star that presented AA Tau-like photometric behavior in either run was included in the AA Tau category.
Bottom: number of AA Tau-like stars (red) and aperiodic extinction stars (purple), divided by the total 
number of CTTS in the sample of \citet{venuti14}, including those that were not observed by CoRoT.
}\label{fig:mass}
\end{figure}

Fig. \ref{fig:mass}, top left panel, shows a mass histogram of AA Tau-like stars (red) and stars 
with aperiodic extinction dominated light curves (purple), plotted over a histogram of all CTTS in 
our study of NGC 2264 that present CoRoT light curves from either observing run. 
While the CTTS show a tendency toward lower masses, the number of stars declining rapidly 
for larger masses, the AA Tau-like stars are equally present in the whole range of masses 
considered. Therefore, though there are AA Tau-like stars of all masses in the CTTS range, the 
ratio of these stars to the total number of stars in that mass bin increases with increasing mass, 
meaning they are relatively more common among higher mass stars than lower mass ones (top right 
diagram in Fig. \ref{fig:mass}). Stars that present AA Tau-like photometric behavior in 
either 2008 or 2011 were included in the AA Tau category.

Inclination is a major factor in whether we observe AA Tau-like events or not; therefore, we 
know that the observed AA Tau-like stars are only a fraction of the total number of stars that 
present an inner disk warp. For stars of $M \lesssim 0.7 M_{\odot}$, fewer than 10\% are observed as 
AA Tau-like. If we consider that it is possible to view the occultations at angles between $\sim$ 
60$^{\circ}$ and 77$^{\circ}$, as was shown using the occultation model, then only at most 50\% of 
stars in this mass range possess an inner disk warp with two major accretion funnels. For stars of 
$M \gtrsim 0.7 M_{\odot}$, about 35\% show AA Tau-like behavior, meaning that if we assume random 
inclinations, all of these stars likely possess an inner disk warp during some part of their 
evolution.

The fraction of stars that present AA Tau-like photometric behavior is quite 
large as we approach even higher masses ($M\gtrsim1.0 M_{\odot}$), much larger than we would 
expect if all CTTS observed by CoRoT possessed an inner disk warp, given that when we consider 
random inclinations we should only see $\sim 20\%$ of AA Tau-like systems. This is 
possibly due to a selection bias, since the CoRoT targets are not selected randomly, but 
based on specific criteria, which included many systems with known periodicity, and in 2011 
included systems previously classified as AA Tau-like. Therefore, to compare with a less biased 
sample, we plotted a histogram of the number of AA Tau-like systems observed by CoRoT compared 
to the total number of CTTS observed by the CFHT MegaCam \citep[presented in][]{venuti14}, 
including those that were and were not observed by CoRoT. This plot, shown in the bottom panel of 
Fig. \ref{fig:mass}, may also have a biased tendency, since it is possible that some of the 
systems not observed by CoRoT also present AA Tau-like variability that would be unaccounted for. 
Even so, we see that slightly over 20\% of stars with mass $M \gtrsim 1.0 M_{\odot}$ 
present this type of variability. Extrapolating this amount to the systems where the inclinations
prevent us from observing this tendency, we still conclude that nearly all of the systems with 
mass $1.0 M_{\odot} \lesssim M \lesssim 2.0 M_{\odot}$ and most ($\sim 75\%$) of the systems 
with mass $0.7 M_{\odot} \lesssim M \lesssim 1.0 M_{\odot}$ should present an inner disk warp, 
while for systems of $M \lesssim 0.7 M_{\odot}$ this fraction would be of only $\sim 25\%$.

This may be due to different stellar magnetic field configurations. The dipolar component of the magnetic field 
is responsible for the interaction with the inner disk, and a stronger dipolar component leads 
to a larger truncation radius. As was discussed in Sect. \ref{sec:stab}, a larger truncation radius tends  
to favor a stable accretion regime. \citet{gregory12} found that the magnetic field configuration of 
a star depends strongly on its internal structure, and that for intermediate to high mass T Tauri 
Stars ($0.5 M_{\odot} \lesssim M_* \lesssim 2.0 M_{\odot}$), those that are fully convective, or have 
as yet only developed a small radiative core, tend to have strong dipoles. At the estimated age of 
2 - 3 Myr of NGC 2264, \citet{gregory12} show that stars in this mass range have not yet developed 
a significant radiative core. They should therefore possess strong dipolar components in their magnetic 
fields. 

There is still very little information available on magnetic fields of T Tauri stars of smaller mass 
($M_* \lesssim 0.5 M_{\odot}$), because of observational constraints. At least one T Tauri star in this 
mass range, V2247 Oph, has had its magnetic field topology measured and was found to have a weak 
dipolar component \citep{donati10v2247}. Therefore, it is possible that among fully convective stars, a 
significant difference in magnetic field configurations may exist among these two mass ranges. 
\citet{morin10} showed that among main sequence stars, those of very low mass show magnetic 
field topologies that differ considerably from fully convective stars of somewhat higher mass. 
While the latter generally possess magnetic fields with strong dipoles, main sequence stars of very 
low mass ($M_* \lesssim 0.2 M_{\odot}$) show a number of different magnetic topologies, ranging 
from very strong, axisymmetric, nearly dipolar fields to weaker fields with a strong non-axisymmetric 
component, and a strong toroidal component. This supports the idea that a different magnetic field 
regime may dominate among stars of lower mass.

However, our results alone are insufficient to conclude that a significant difference exists among 
magnetic field configurations in these two mass ranges. There are other factors that may influence 
our higher detection of AA Tau-like systems among those of somewhat higher mass. For instance, 
mass accretion rates tend to decrease toward lower masses \citep{venuti14}, which would increase 
the truncation radius at this mass range. This would favor a stable accretion mechanism, but could 
also lead to the disk warp being located farther away from the co-rotation 
radius than we have assumed. If they are located significantly farther away, we would need higher 
inclinations to observe the eclipses in these systems, making it more difficult to do so.

It is also possible that this effect is simply the result of an observational bias. Stars of lower mass 
are generally fainter than those of higher mass, and the signal from the CoRoT observations could 
be too low among stars below a certain mass limit to allow an accurate classification of AA 
Tau-like light curves.

\section{Discussion}\label{sec:discuss}

The occultation model initially proposed for the star AA Tau appears to be adequate at 
reproducing the general aspects of the flux dips of the stars listed in Table 
\ref{table:occmodel}. It is capable of reproducing their widths and amplitudes using values 
of inclination, warp maximum height, and azimuthal extension that are consistent with those 
found for AA Tau itself, with only a few stars showing values of warp height and azimuthal 
extension that are significantly lower. 

We note that for stars whose light curves are classified as AA Tau-like in both CoRoT 
observing runs (2008 and 2011), the values found for warp maximum height and azimuthal extension 
in one epoch are always within the same range as those found in the other. This means that, although 
the occulting structure should be variable on a timescale of days to weeks because of the 
dynamic interactions, it should be stable enough to persist on a timescale of a few years.
We also note that, for these stars, the periods found in each observing run are very similar to 
each other. Most are equal within their uncertainties. In all of these cases the radius where 
the occulting structure should be located differs by less than 4\% from one epoch to the other, 
and most by less than 2\%, which is within their respective uncertainties. It is possible 
that the structures move somewhat in radius over timescales of a few years, but they should do 
so by very little. 

Though it can reproduce the widths and amplitudes of light curve minima, the occultation model 
is unable to account for the exact shape of the flux dips, which are much more complex than 
the nearly Gaussian shapes that it reproduces. 
This model is based on a symmetric warp of homogeneous optical thickness, which likely does not 
reflect the reality of these systems. For  AA Tau, short term brightening episodes were seen to 
occur during the occultation events, and were attributed to inhomogeneities of reduced optical 
thickness in the occulting material \citep{bouvier99}. 
We observe these brightening episodes in many of the light curves in our sample. 
In most cases the light curve minima present significant structure, which could be 
due to these inhomogeneities, to the existence of a configuration of hot     
spots visible during the occultations, or to non-steady accretion within the funnel 
flows, which would lead to variable emission during the flux dips.  

\begin{figure}[t]
\centering
\includegraphics[width=4.4cm]{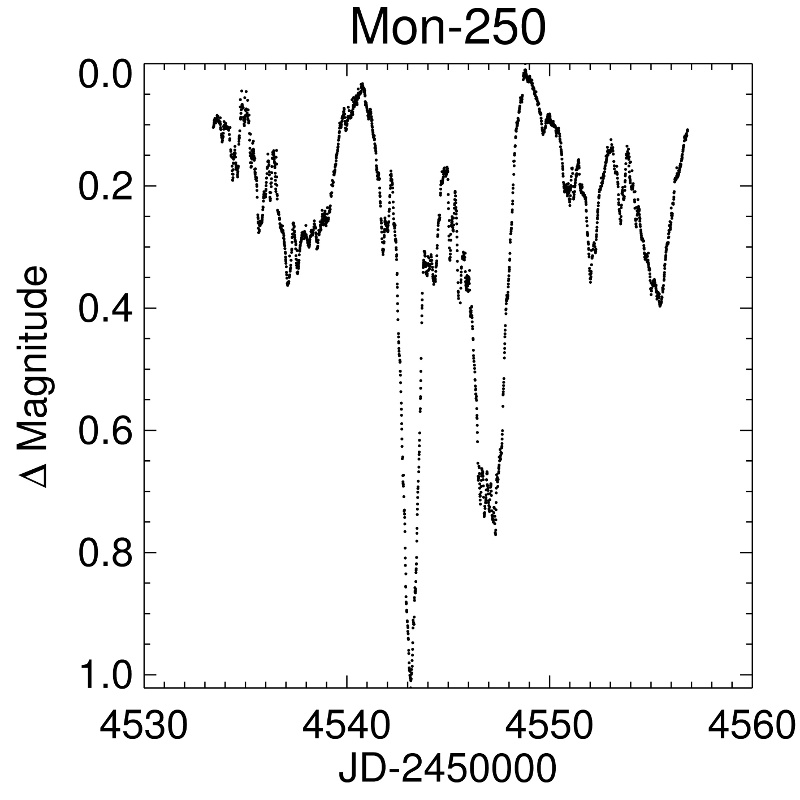}
\includegraphics[width=4.4cm]{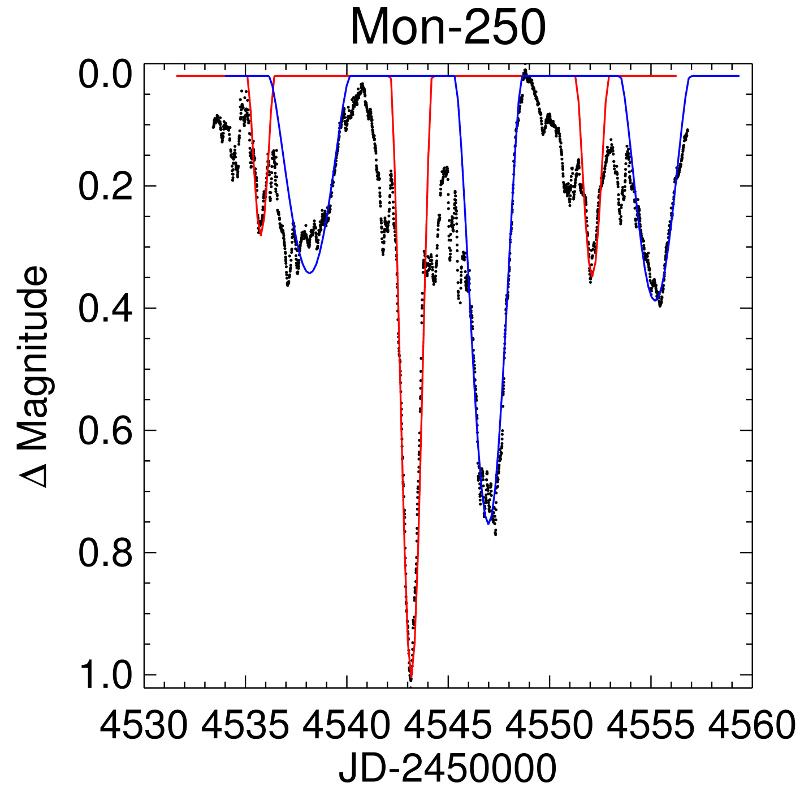}
\caption{Left: 2008 CoRoT light curve of Mon-250. Right: synthetic light curves generated with the 
occultation model for two separate accretion funnels (blue and red) overplotted on the 2008 CoRoT 
light curve of Mon-250.
}\label{fig:mon250}
\end{figure}

In some cases we observe two minima where we would expect to see one, such as in the 2008 CoRoT 
light curve of the star Mon-250 (left panel of Fig. \ref{fig:mon250}).  We know from its 2011 
CoRoT light curve and spectra that this star's rotation period is on the order of 8.6 days (Fig. 
\ref{fig:veilingmon250}); therefore, 
the 2008 light curve spans only three rotation cycles, though we see more than three flux dips. We can 
speculate that the two middle dips are part of one structure, as the last two dips would be, and that 
there is significant emission from the hot spot associated with the accretion shock superimposed onto 
the extinction from the base of the accretion funnel, or simply that the warp was very inhomogeneous. 
Unfortunately, there is no color information simultaneous with this light curve to verify these 
scenarios. However, the occultation model is incapable 
of reproducing one flux dip that would encompass both middle dips seen in the light curve using 
physically plausible parameters. We are able to reproduce each dip separately, however, if 
we consider that there are two accretion funnels, separated by $160^{\circ}$, i.e., nearly 
opposite each other (right panel of Fig. \ref{fig:mon250}).  
We propose that it is possible for the main accretion funnel to at times give way to two accretion 
funnels as a result of an increase in instabilities. 

\begin{figure}[t]
\centering
\includegraphics[width=4.4cm]{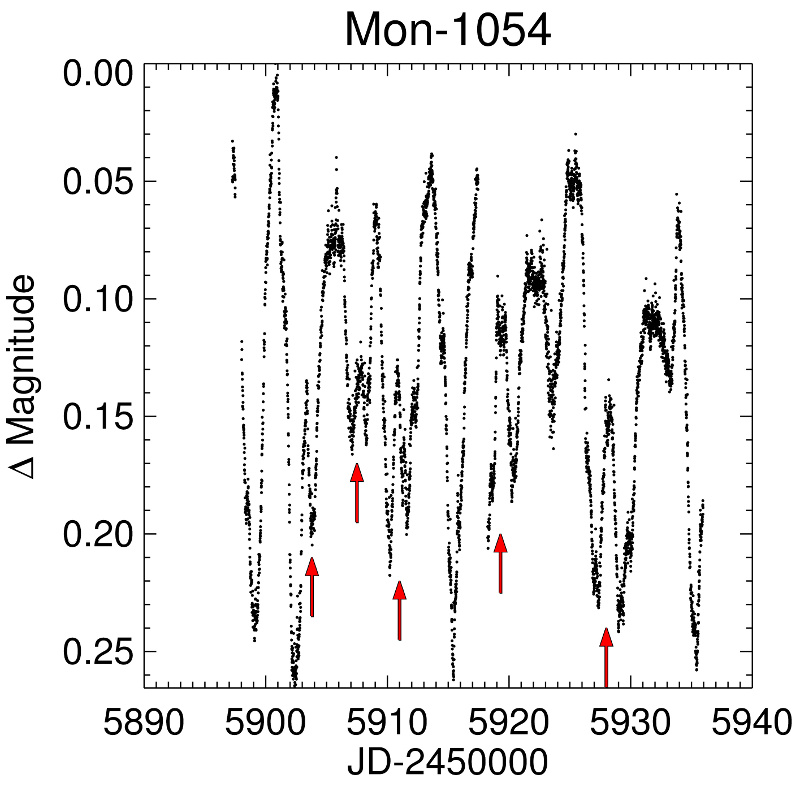}
\includegraphics[width=4.4cm]{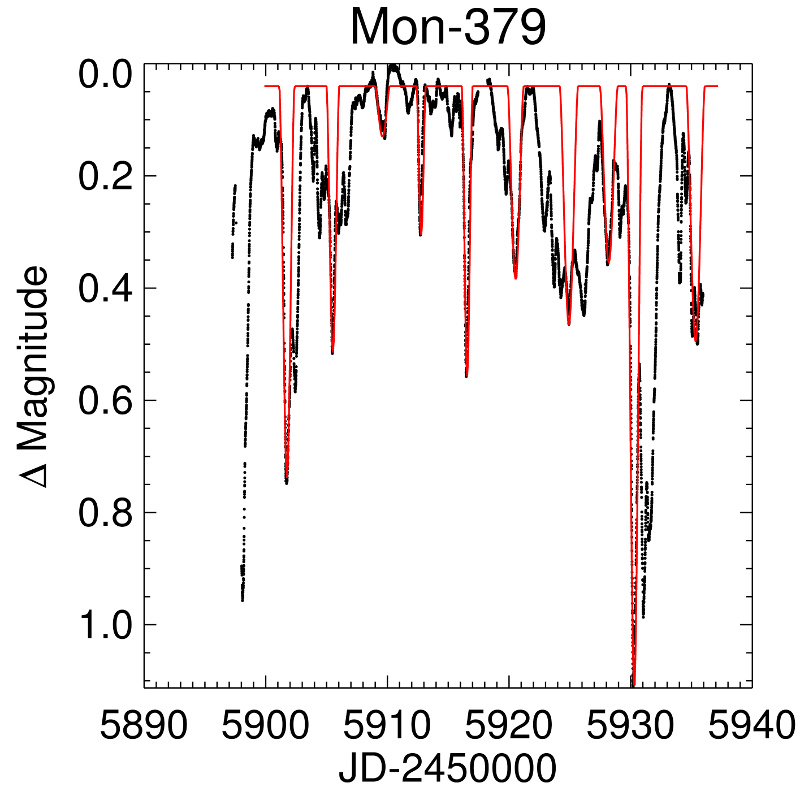}
\caption{Left: 2011 CoRoT light curve of Mon-1054. Arrows show moments when two flux dips are present 
where we would expect to see only one in a typical AA Tau-like scenario.
Right: Simulated light curve (red) overplotted on the 2011 CoRoT light curve of Mon-379 (black).
}\label{fig:mon1054}
\end{figure}

This same behavior was observed in AA Tau in 1999, when its quasi-periodic flux dips gave way to two 
flux dips per stellar rotation \citep[see Fig. 15 in ][]{bouvier07}. We may also be seeing this in 
the star Mon-1054, whose veiling features support the AA Tau-like scenario as the primary cause 
for its photometric variability (see Sect. \ref{sec:veil}). The left panel of Fig. 
\ref{fig:mon1054} shows its 2011 CoRoT light curve. Arrows point to 
moments where we observe two flux dips when we would expect to see only one in a typical AA Tau-like 
scenario. The occultation model is capable of reproducing some of these double dips with only one 
accretion funnel, assuming that another mechanism is responsible for the emission within the dip, but 
it is not able to do so for the fifth double dip indicated in the left panel of Fig. 
\ref{fig:mon1054}. In this case, only two funnels (separated by $160^{\circ}$ of 
similar height and azimuthal extension) can successfully reproduce both minima in the light curve.
This may be an indication that this star is undergoing a transition between a stable and an unstable 
accretion regime, or that it is in an intermediate state of some instability.  

This may be the case as well for the star Mon-379 during the 2011 CoRoT observations (Fig. 
\ref{fig:mon1054}, right panel), where we can observe periodic AA Tau-like dips that are easily reproduced by 
the occultation model accompanied by other extinction events. These additional extinction events may be due 
to random accretion streams characteristic of an unstable accretion regime, as was discussed in Sect. \ref{sec:stab}, 
that coexist with the main stable accretion funnel. We also note that for this particular star some of 
the dips are quite narrow and the best fit found with the occultation model has azimuthal extension 
as low as $110^{\circ}$, unlike the majority of other stars where this value is between $260^{\circ}$ 
and  $360^{\circ}$. This may also be an indication of this star's declining stability. 

The phenomenon observed in the aperiodic light curves is similar to those observed in UX Ori-type 
objects, which exhibit deep, non-periodic algol-like minima caused by extinction from circumstellar 
material in the observer's line of sight \citep[see the review by][]{grinin00}. Both our aperiodic stars 
and the UX Ori-type objects show evidence of being seen at high inclinations, which is 
favorable for circumstellar material located near the disk plane to intersect our line of sight. UX 
Ori flux dips typically present larger amplitudes and longer duration than the stars in our sample, 
probably because of larger scale disk instabilities. They are also much more common among Herbig AeBe stars, 
though a few CTTS have been classified as UXors \citep{herbst99,grinin00}. 
It is possible that some of the stars in our sample with aperiodic behavior could be considered UXors, 
though a study of color variability and polarization would be required to determine this.

\section{Conclusions}\label{sec:conc}

Classical T Tauri stars that present extinction dominated light curves, such as the quasi-periodic 
AA Tau-like behavior or aperiodic extinction events, are common in NGC 2264. Of 159 cluster members 
that show signs of accretion, 33 present one of these two behaviors, i.e., $(21\pm4)\%$. These objects 
give us insight into the phenomena that take place in the inner region of circumstellar disks,
a region that is otherwise difficult to study because of its proximity to the star and to 
instrumental limitations.

Extinction laws that include dust grains considerably larger than those found typically for the interstellar 
medium are necessary to account for the ratios between amplitudes in optical and IR light 
curves. This suggests that grain growth has occurred in these disks.

A model proposed to explain the light curve of the CTTS AA Tau, in which at the base of an 
accretion column there is an optically thick warp in the inner circumstellar disk region that 
occults the stellar photosphere periodically, was tested on 21 CTTS in NGC 2264. This 
occultation model was shown to be successful at reproducing the widths and amplitudes of AA 
Tau-like flux dips in the light curves of most of these stars, using warp parameters that 
are similar to those used for AA Tau. Typical ratios between this warp's maximum height and 
the radius at which it is located were found to be between $h_{max}/R_{w}=0.2$ and 
$h_{max}/R_{w}=0.3$. The warp's height and azimuthal extension were shown to be variable 
on a timescale of days, presenting typical variations of 10 - 20\% between rotation cycles, 
though there is strong evidence that the occulting structure can exist over timescales of a 
few years. We have shown that, when this is the case, the warp maintains similar parameters 
during that time and remains at about the same radial distance.

For the few cases where the occultation model is unable to reproduce an AA Tau-like light 
curve using a classical inner disk warp, we have proposed a scenario where more than one 
accretion column exists in each hemisphere, and dust is lifted above the disk plane 
at other locations around the disk, all of which occult the star. This can account for the 
multiple flux dips per stellar rotation seen in the light curves. In many cases, even when 
a single accretion column in the visible hemisphere is able to account for the main features 
of an AA Tau-like light curve, we observe various narrow, shallow dips alongside the deeper, 
broader flux dips attributed to occultation by the warp associated with the main 
accretion column. We attribute these traits to the existence of secondary accretion streams 
that coexist with the main accretion funnel flows. 

We ascribe AA Tau-like light curves to a stable accretion scenario, and light curves that 
present aperiodic extinction events to an unstable accretion scenario, and have shown that a star 
may suffer a transition from one to the other in the few years that separate our two CoRoT 
observing runs. For two stars with aperiodic extinction dominated light curves, we 
find evidence of hot spots associated with occultations, through an increase in veiling 
during light curve minima.
We have also shown that AA Tau-like light curves are more common among stars of intermediate 
to high mass within the CTTS range ($0.7 M_{\odot} \lesssim M_* \lesssim 2.0 M_{\odot}$) 
than among those of lower mass ($M_* \lesssim 0.7 M_{\odot}$), an aspect we believe may be 
linked to stellar magnetic field configurations. 

We have shown that all but one star in our initial sample of possible AA Tau-like systems 
(Mon-1131, which is seen at a low inclination) fit very well within the AA Tau-like scenario. 
For nine of these stars we see clear evidence of a hot spot appearing simultaneously, or 
nearly so, with the occultation, either through an increase in veiling in the spectra or through 
a bluing effect observed in the color photometry. This strongly supports the idea that the 
flux dips observed in the CoRoT light curves are due to occultation by an inner disk 
warp that is associated with stable accretion funnel flows. We can therefore state with 
reasonable confidence that these are AA Tau-like systems.

\begin{acknowledgements}
The authors thank the referee V. Grinin for his contribution to the discussion.
This work is based on data collected by the CoRoT satellite, and in part on observations made 
with the \textit{Spitzer} Space Telescope, operated by the Jet Propulsion Laboratory, California Institute 
of Technology, under a contract with NASA. Support for this work was provided by NASA through an 
award issued by JPL/Caltech. PTM, SHPA, MMG, APS and NNJF acknowledge funding support from CAPES, 
CNPq, FAPEMIG, and Cofecub. JB acknowledges funding support from Cofecub, CNES, and the grant ANR 2011 
Blanc SIMI5-6 020 01.
\end{acknowledgements}

\bibliographystyle{aa}
\bibliography{references}

\onlfig{
\begin{figure*}[h!]
\vspace*{1 cm}
\centering
\includegraphics[width=4.4cm]{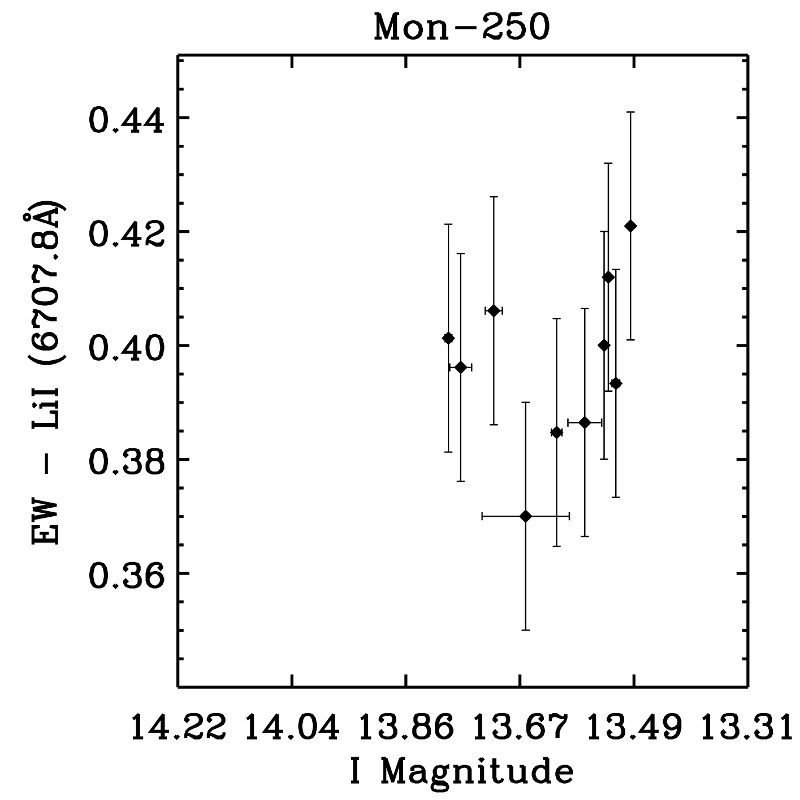}
\includegraphics[width=4.4cm]{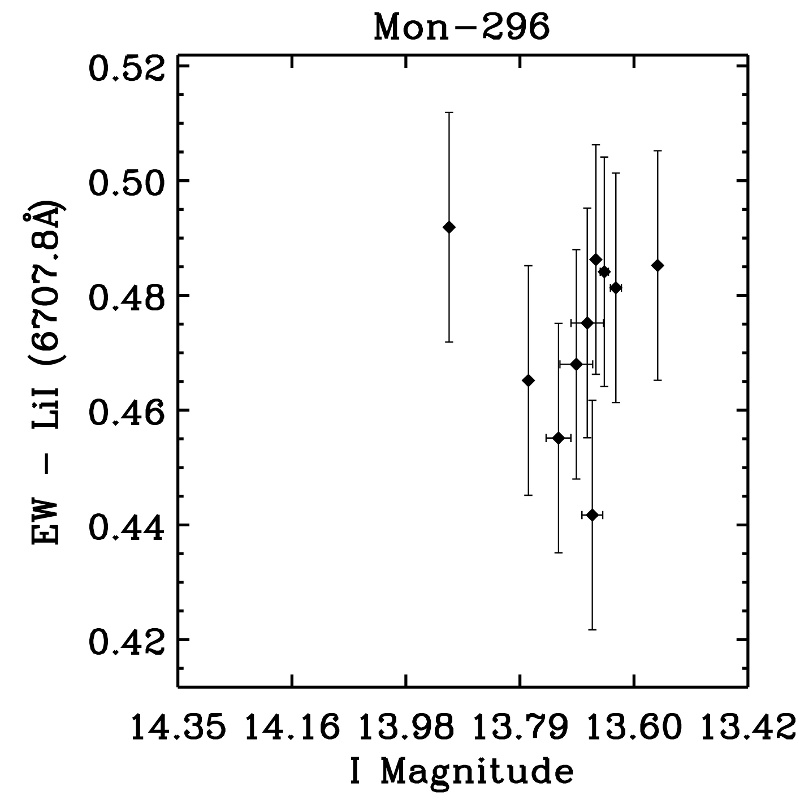}
\includegraphics[width=4.4cm]{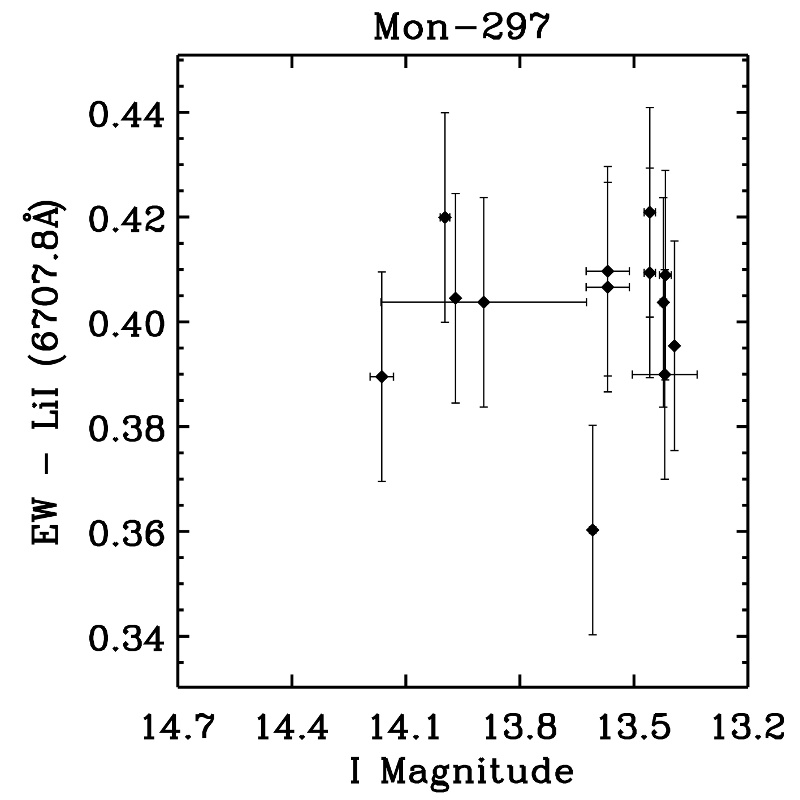}
\includegraphics[width=4.4cm]{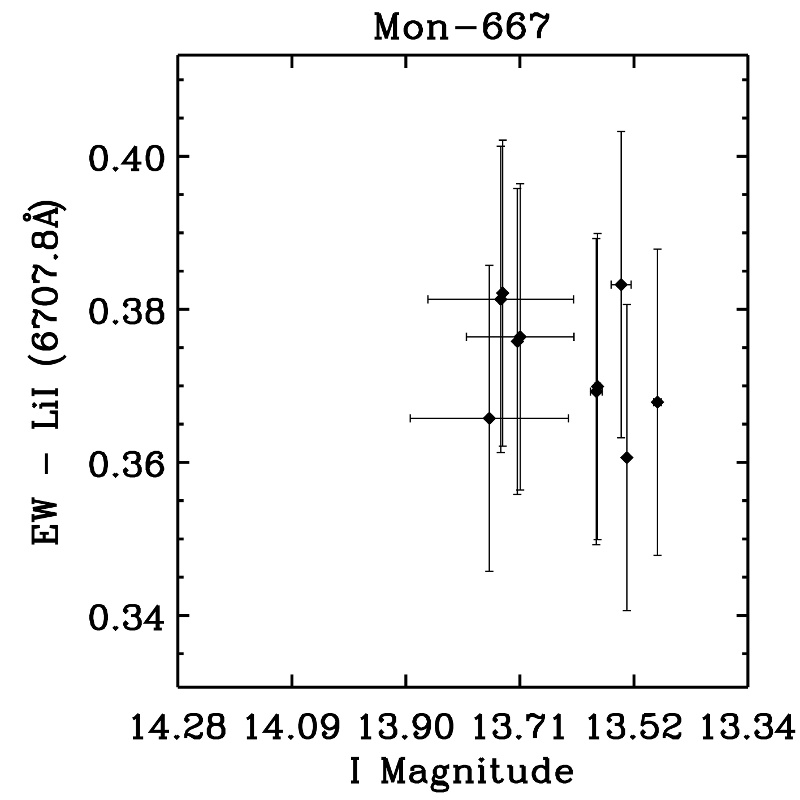}
\includegraphics[width=4.4cm]{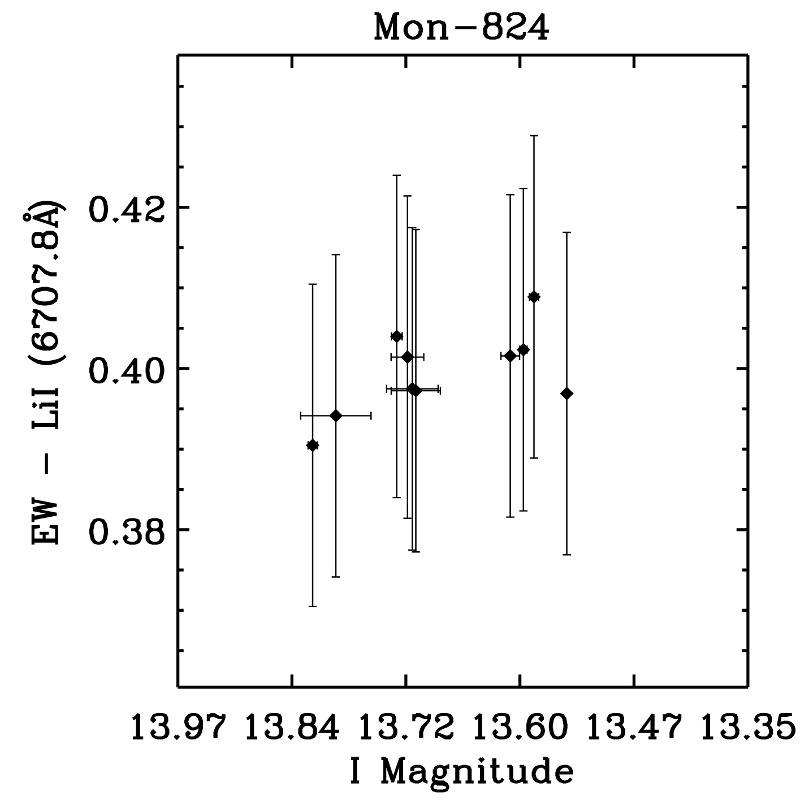}
\includegraphics[width=4.4cm]{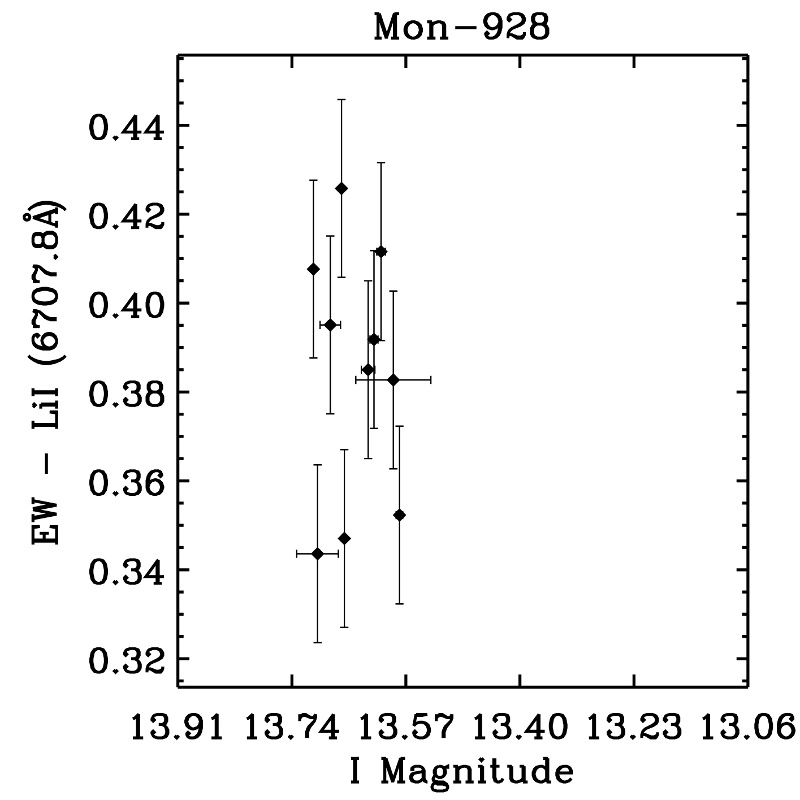}
\includegraphics[width=4.4cm]{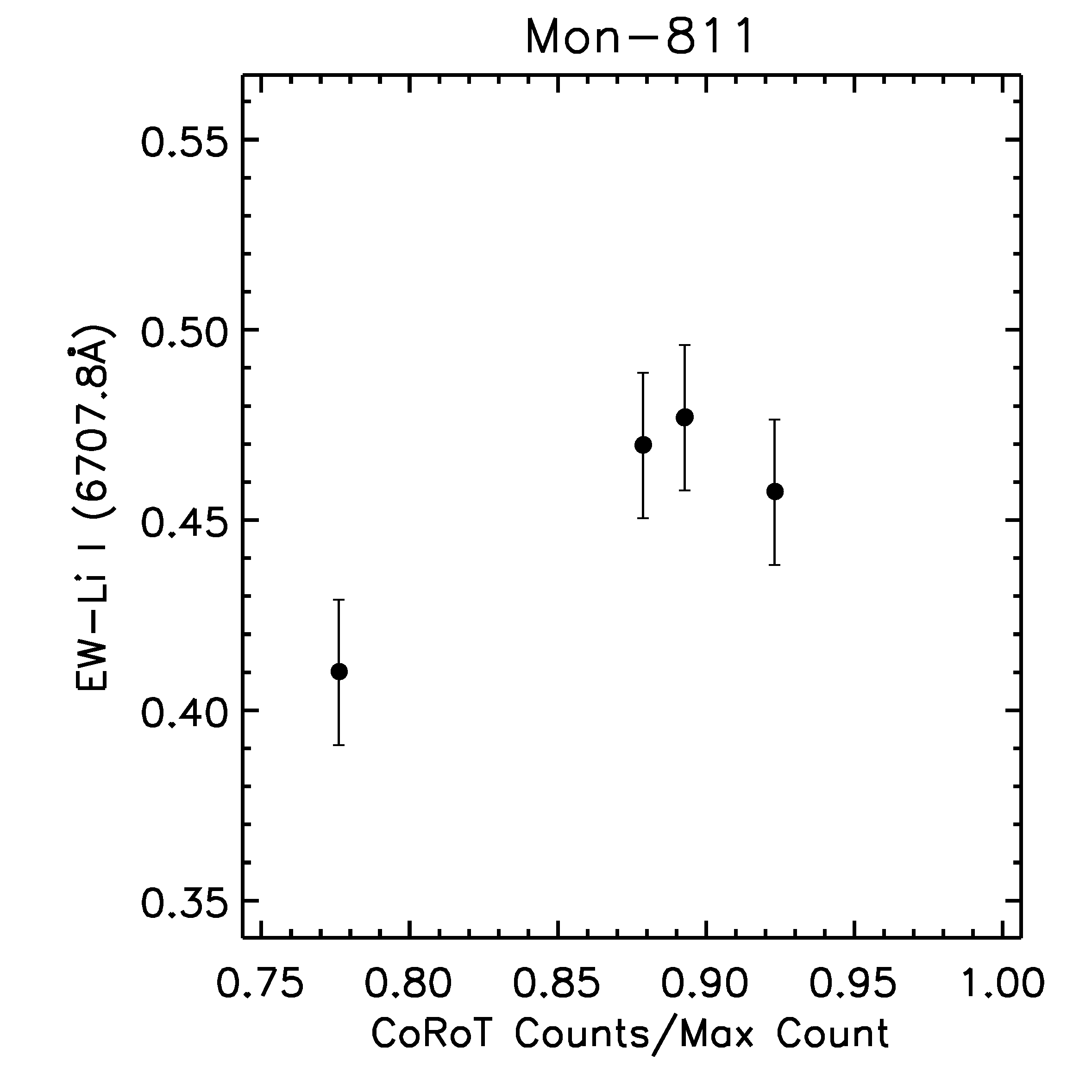}
\caption{Plots of LiIEW vs. I magnitude or CoRoT counts.
}\label{fig:app1}
\end{figure*}
}

\onlfig{
\begin{figure*}[h]
\vspace*{1 cm}
\centering
\includegraphics[width=6cm]{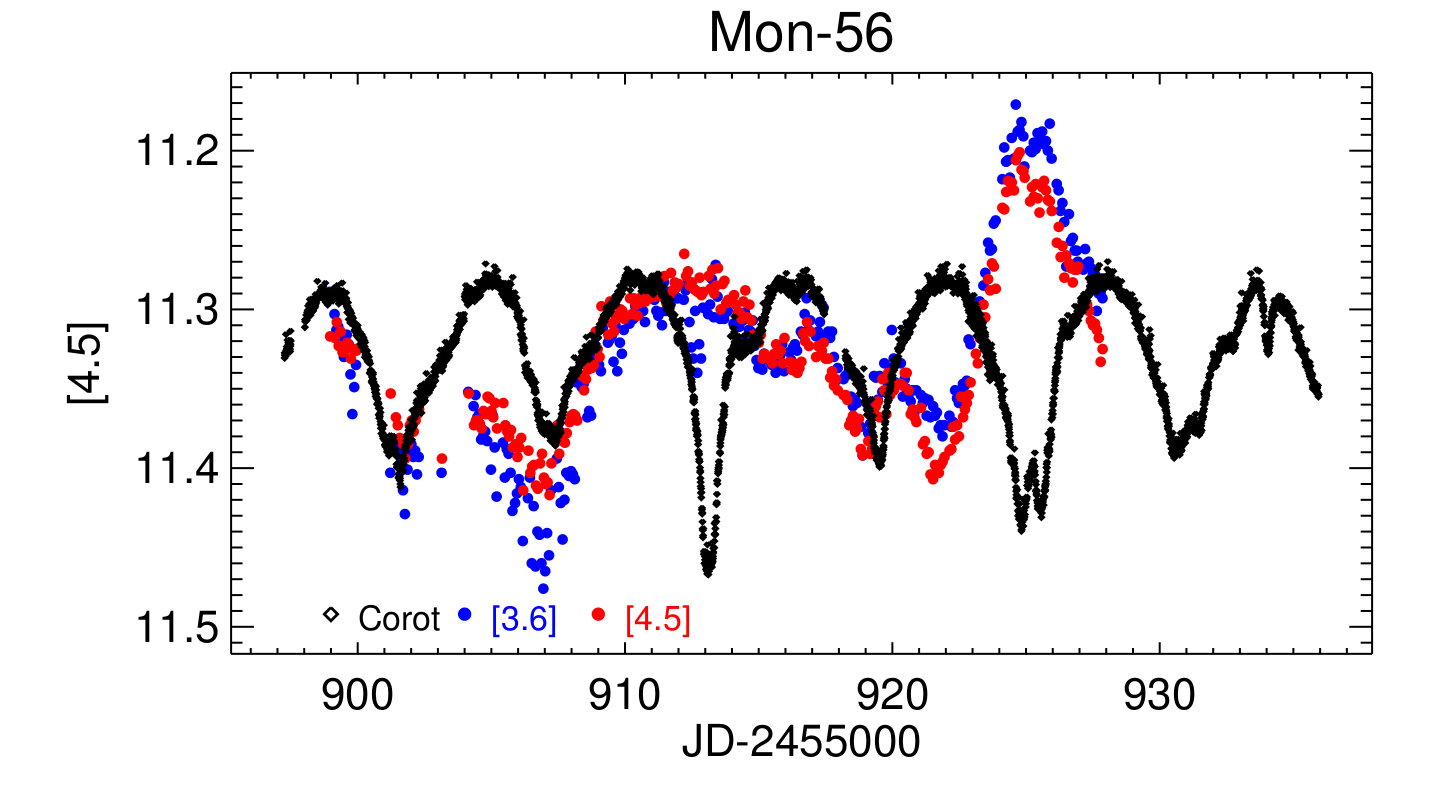}
\includegraphics[width=6cm]{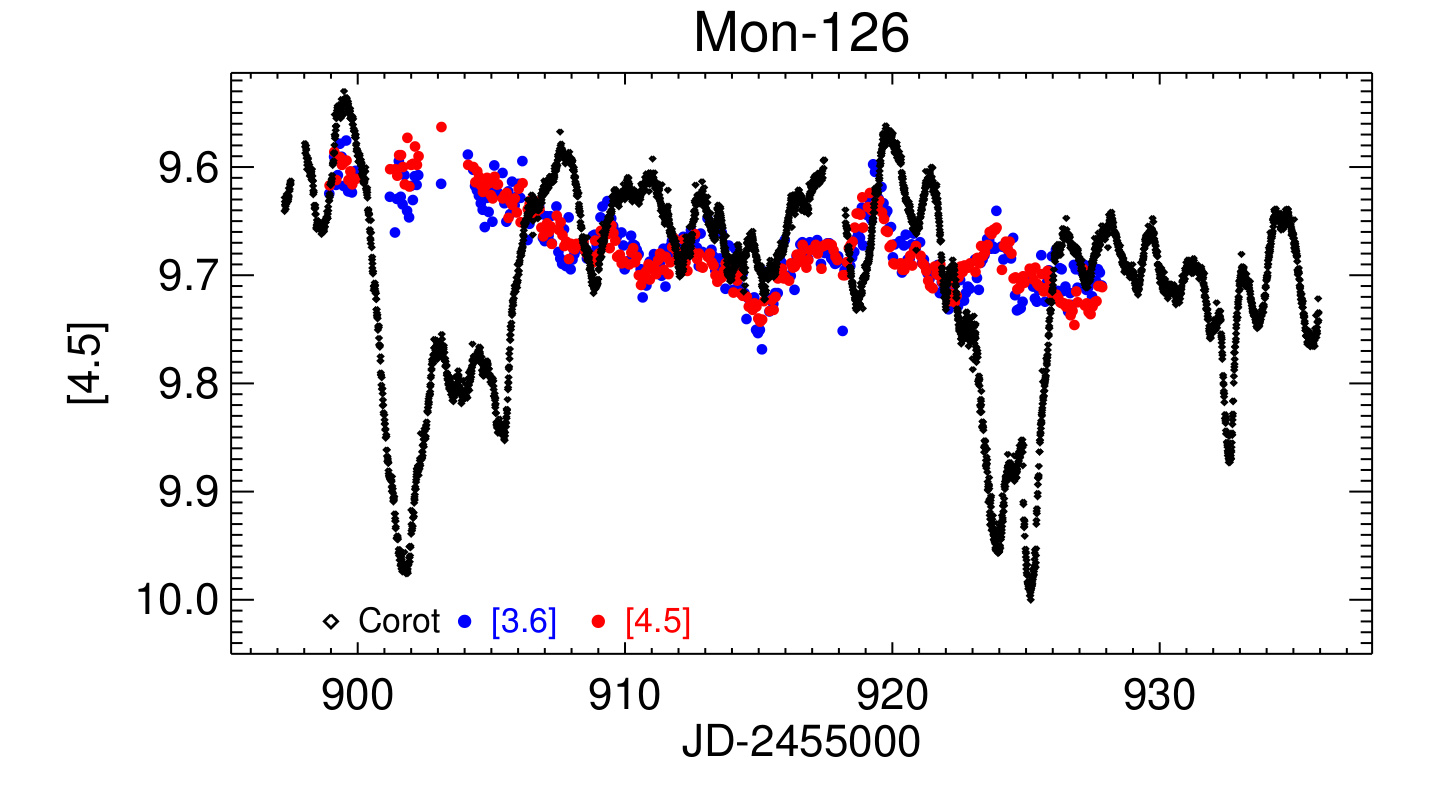}
\includegraphics[width=6cm]{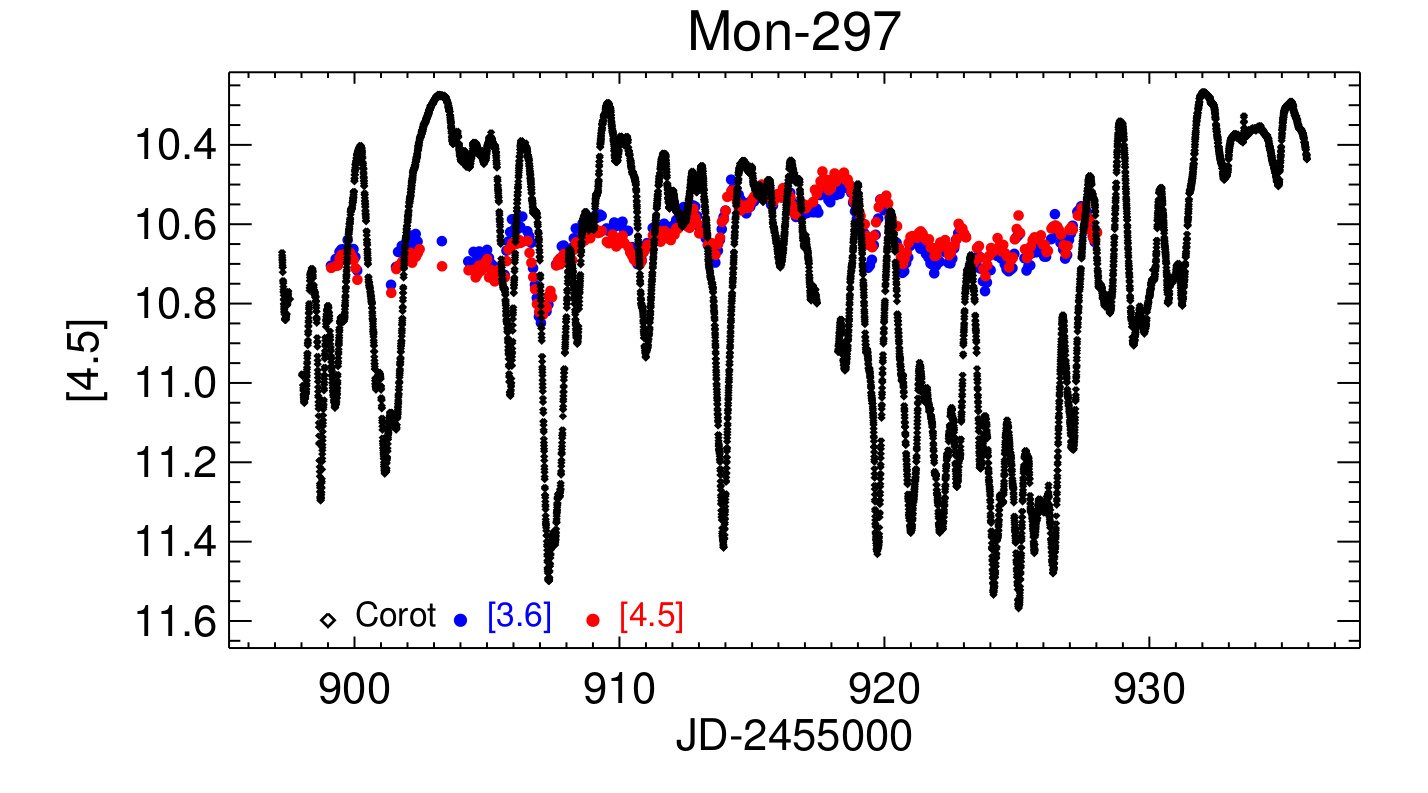}
\includegraphics[width=6cm]{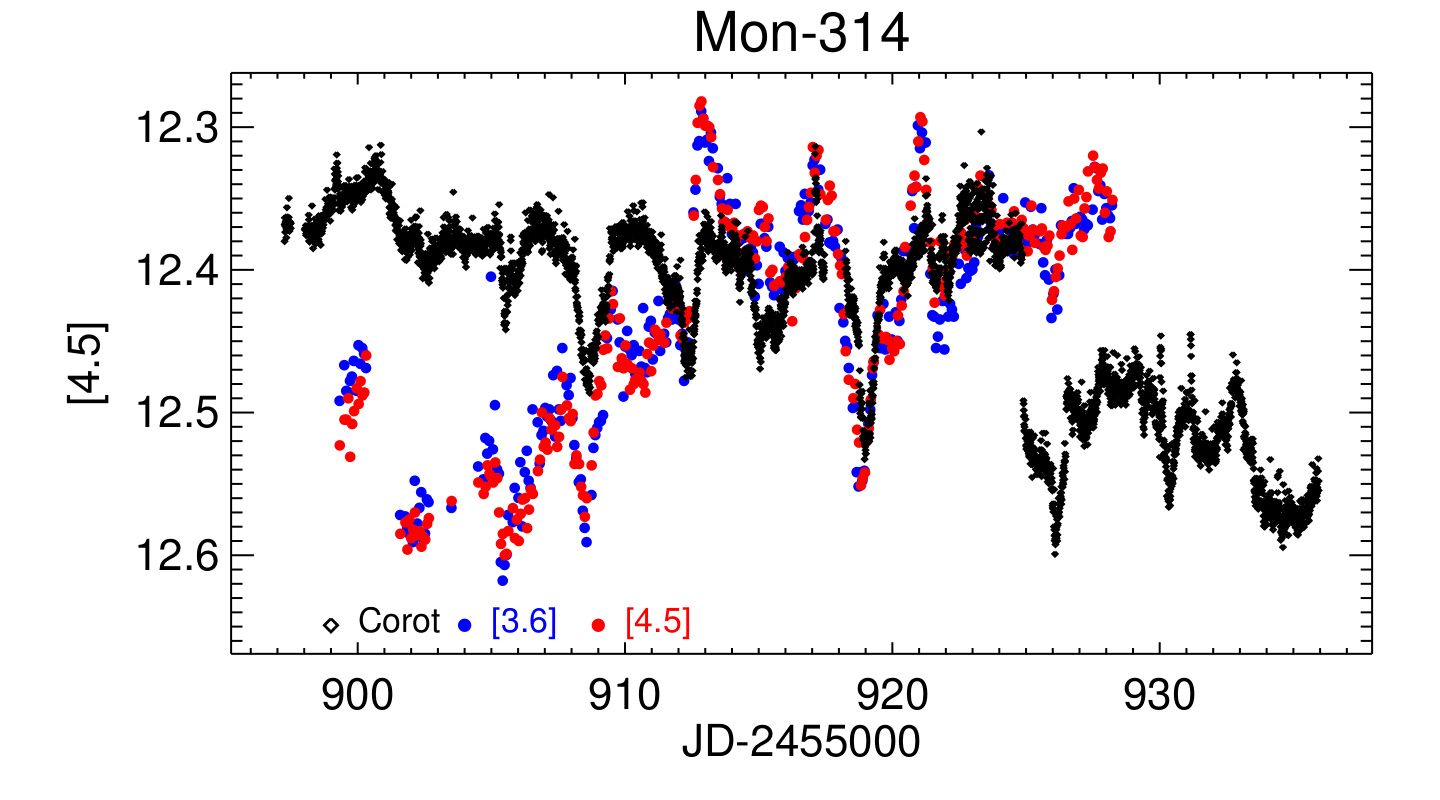}
\includegraphics[width=6cm]{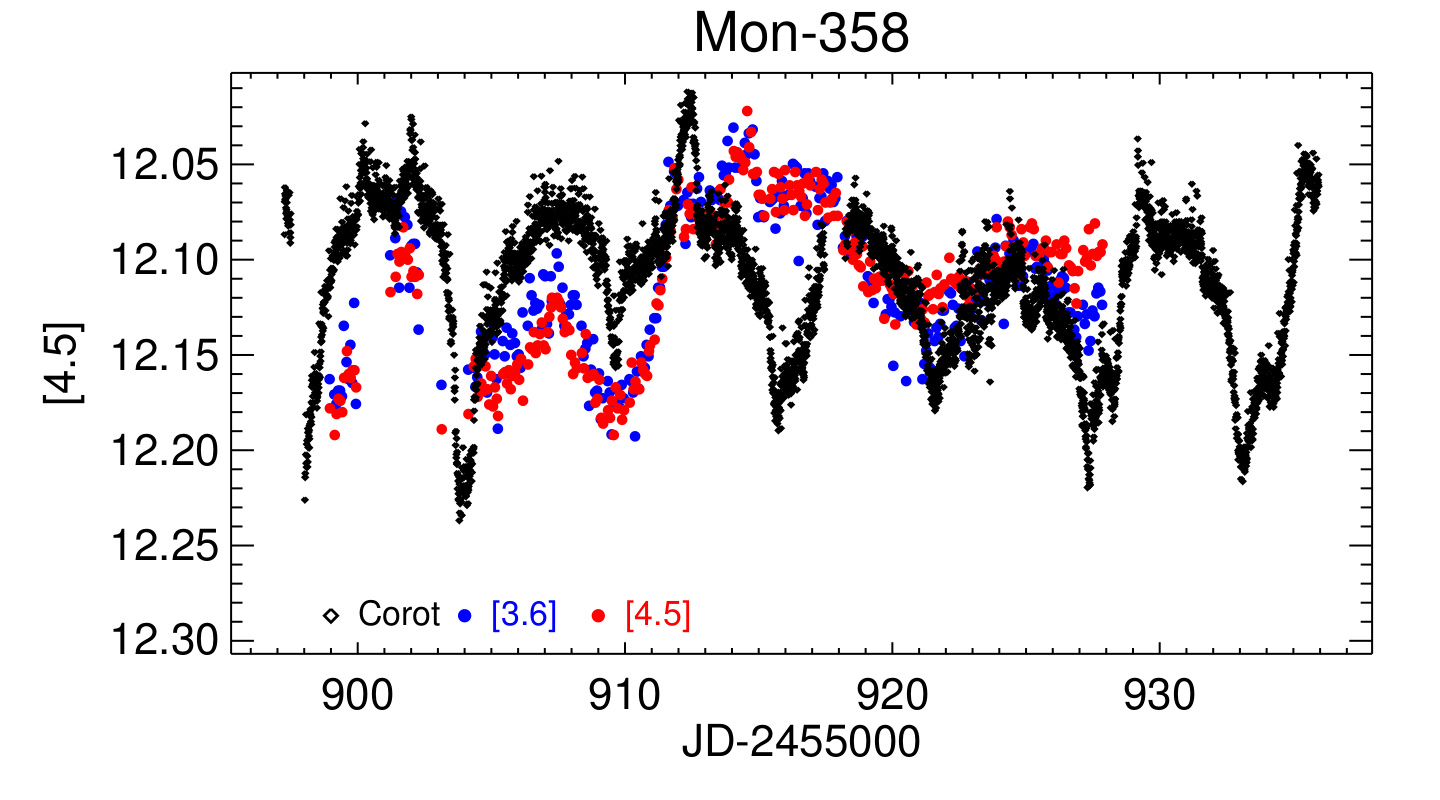}
\includegraphics[width=6cm]{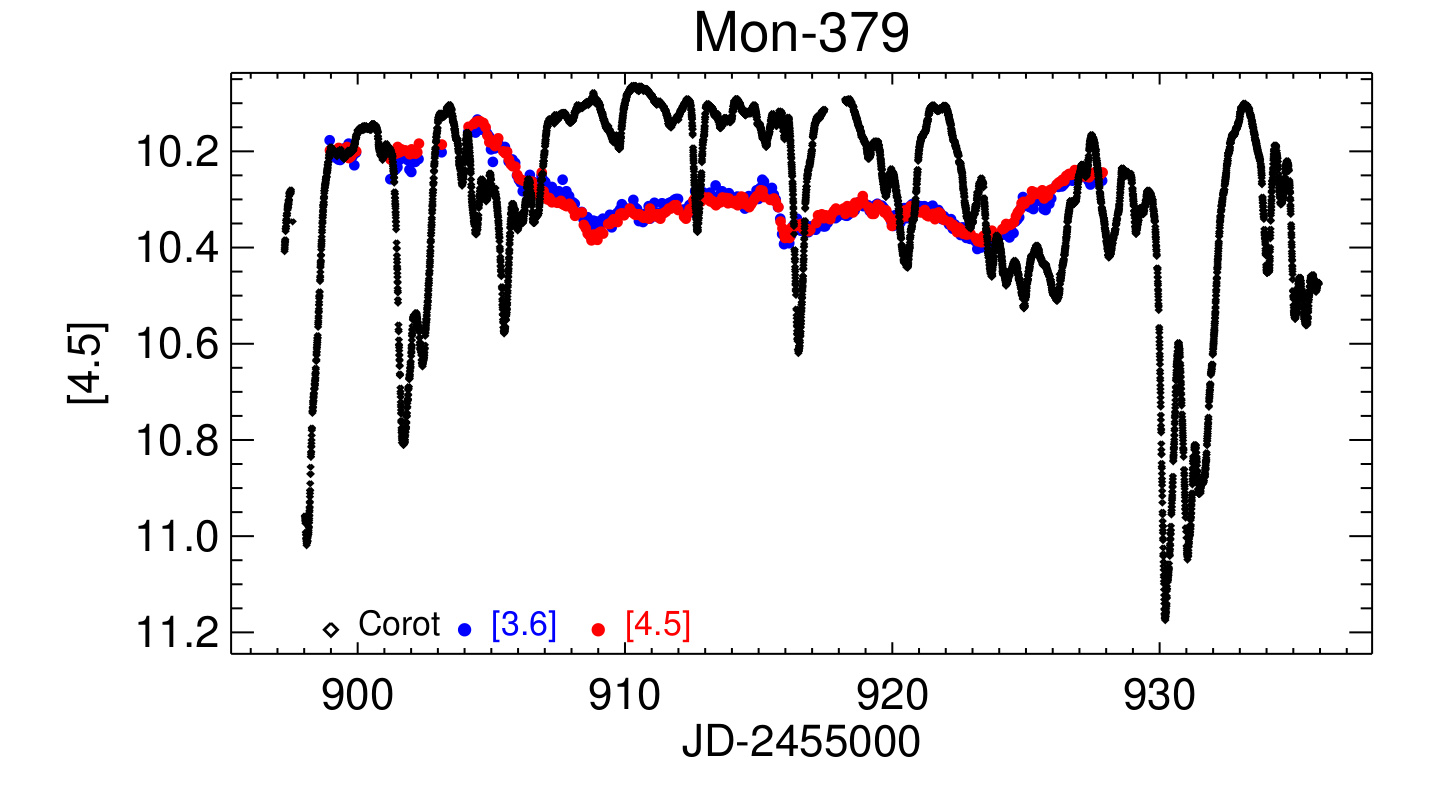}
\includegraphics[width=6cm]{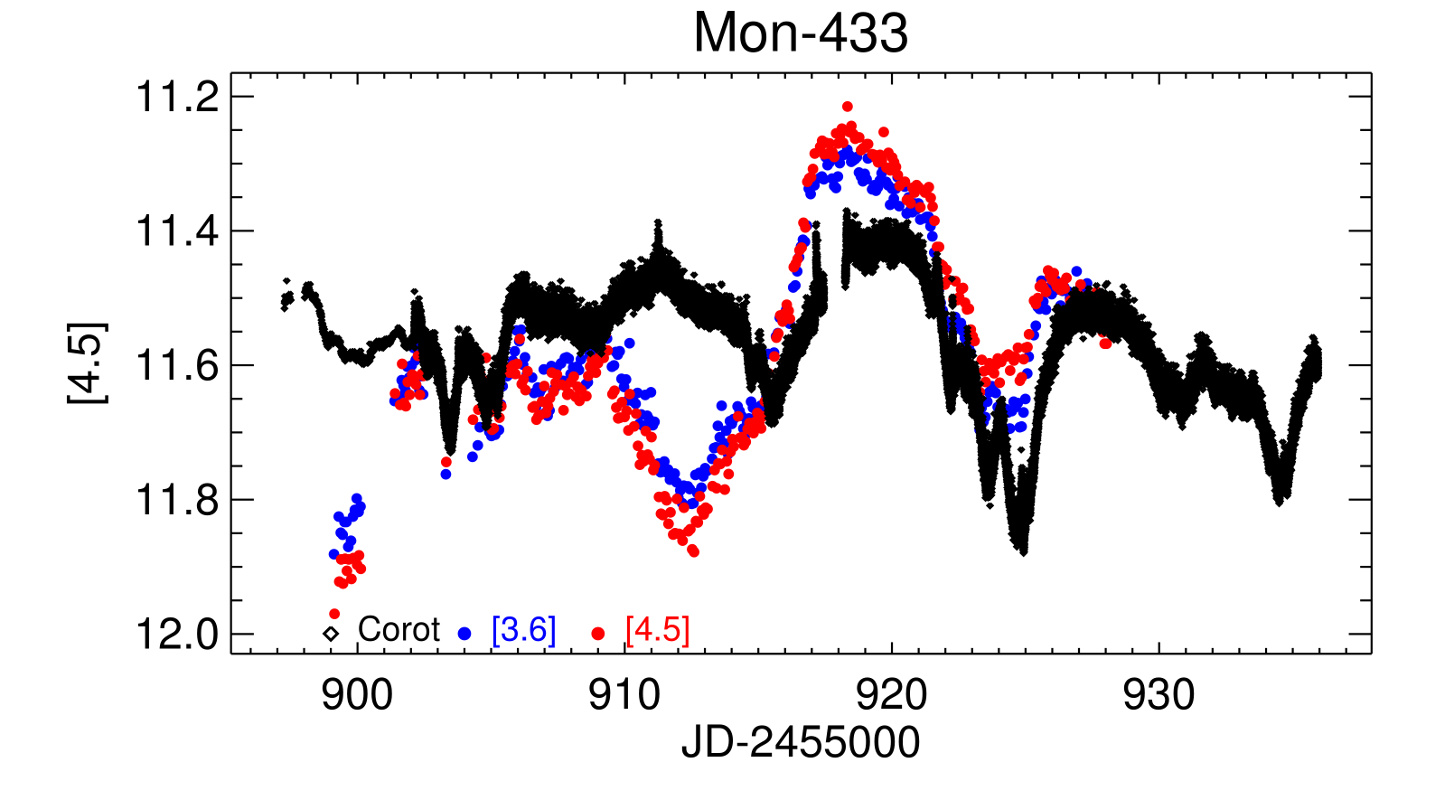}
\includegraphics[width=6cm]{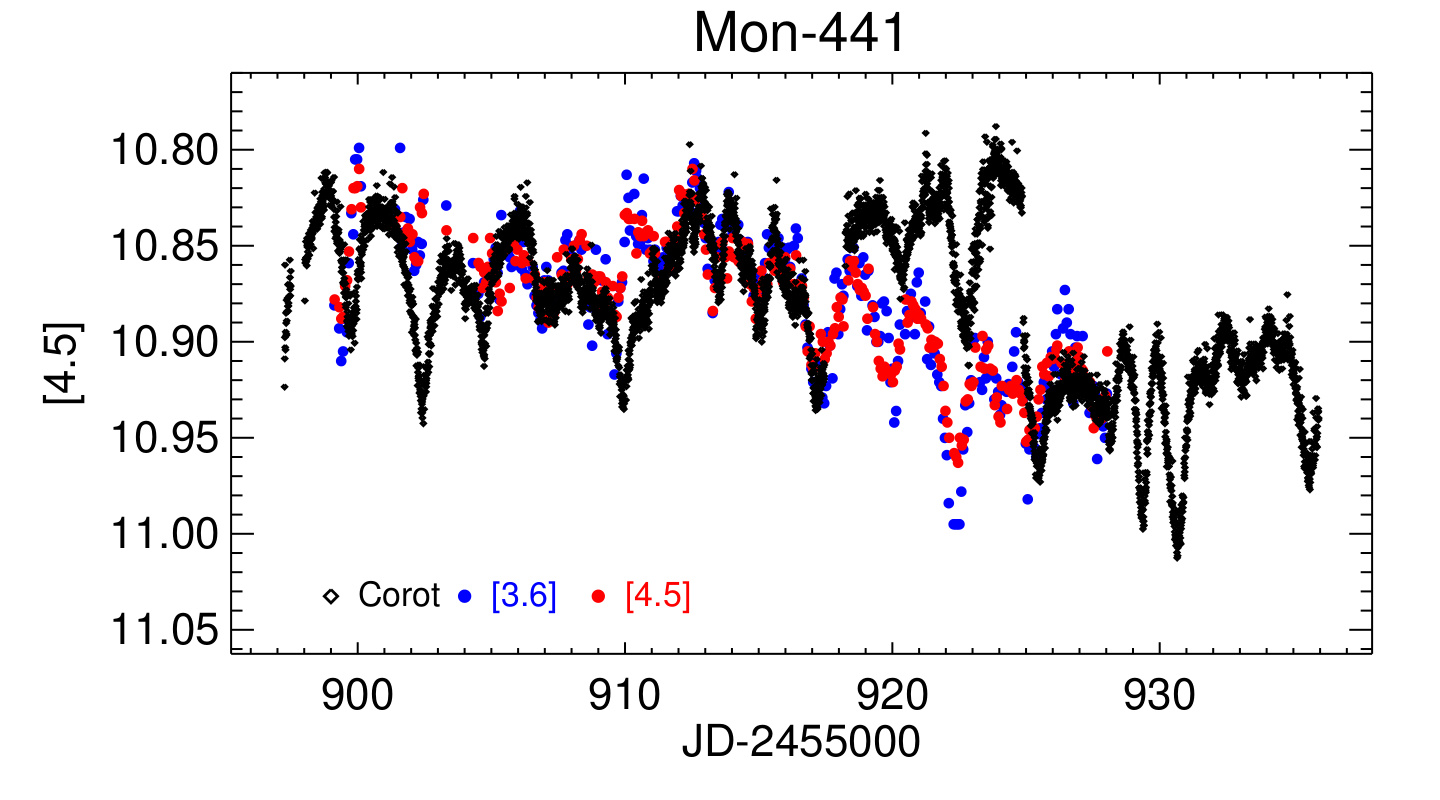}
\includegraphics[width=6cm]{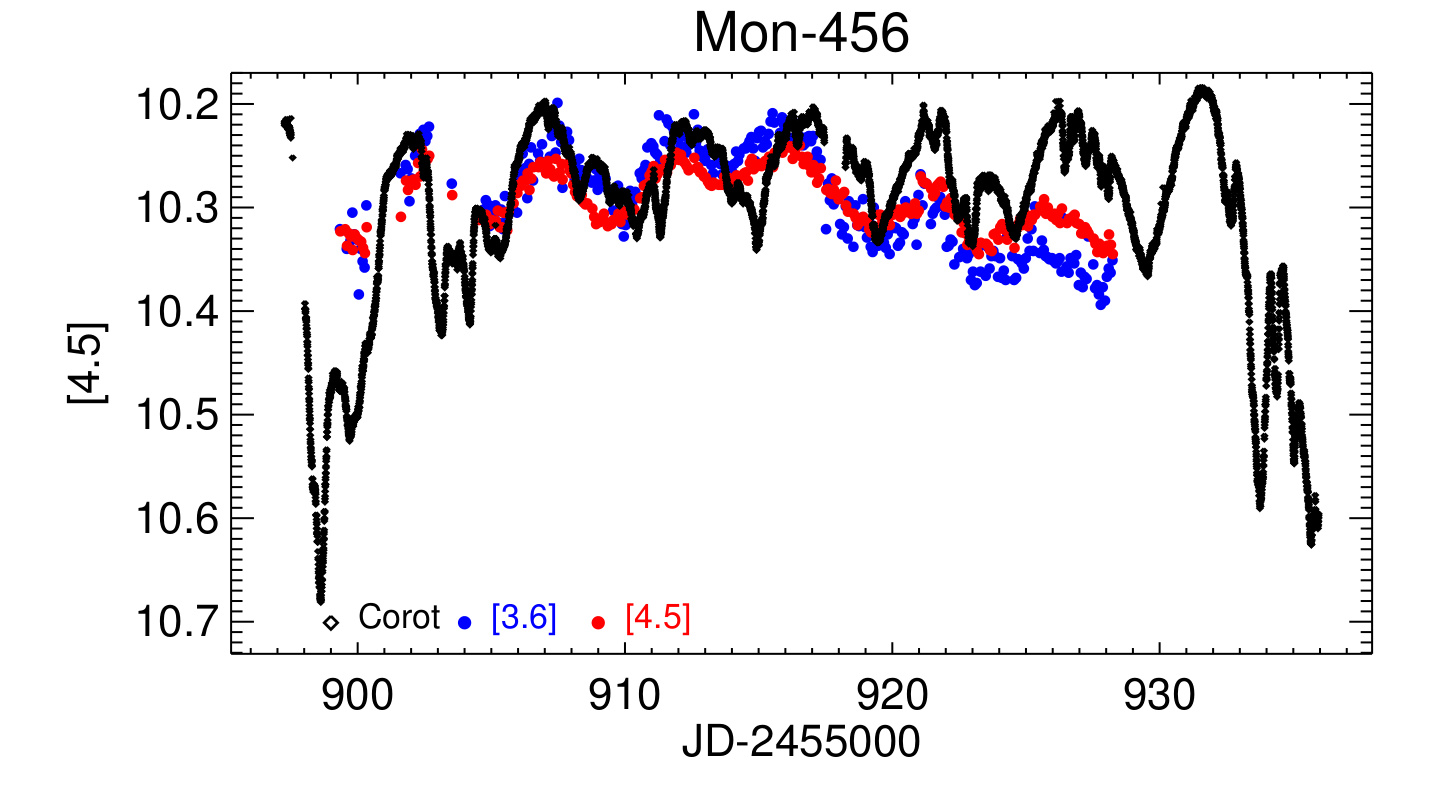}
\caption{\textit{Spitzer} IRAC $3.6 \mu m$ and $4.5 \mu \mathrm{m}$ light curves (blue and red filled
circles, respectively) overplotted on CoRoT light curves (black circles).
The CoRoT and IRAC $3.6 \mu m$ light curves were shifted in magnitude for easier comparison.
}
\label{fig:app2}
\end{figure*}

\begin{figure*}[p]
\vspace*{2 cm}
\ContinuedFloat
\centering
\includegraphics[width=6cm]{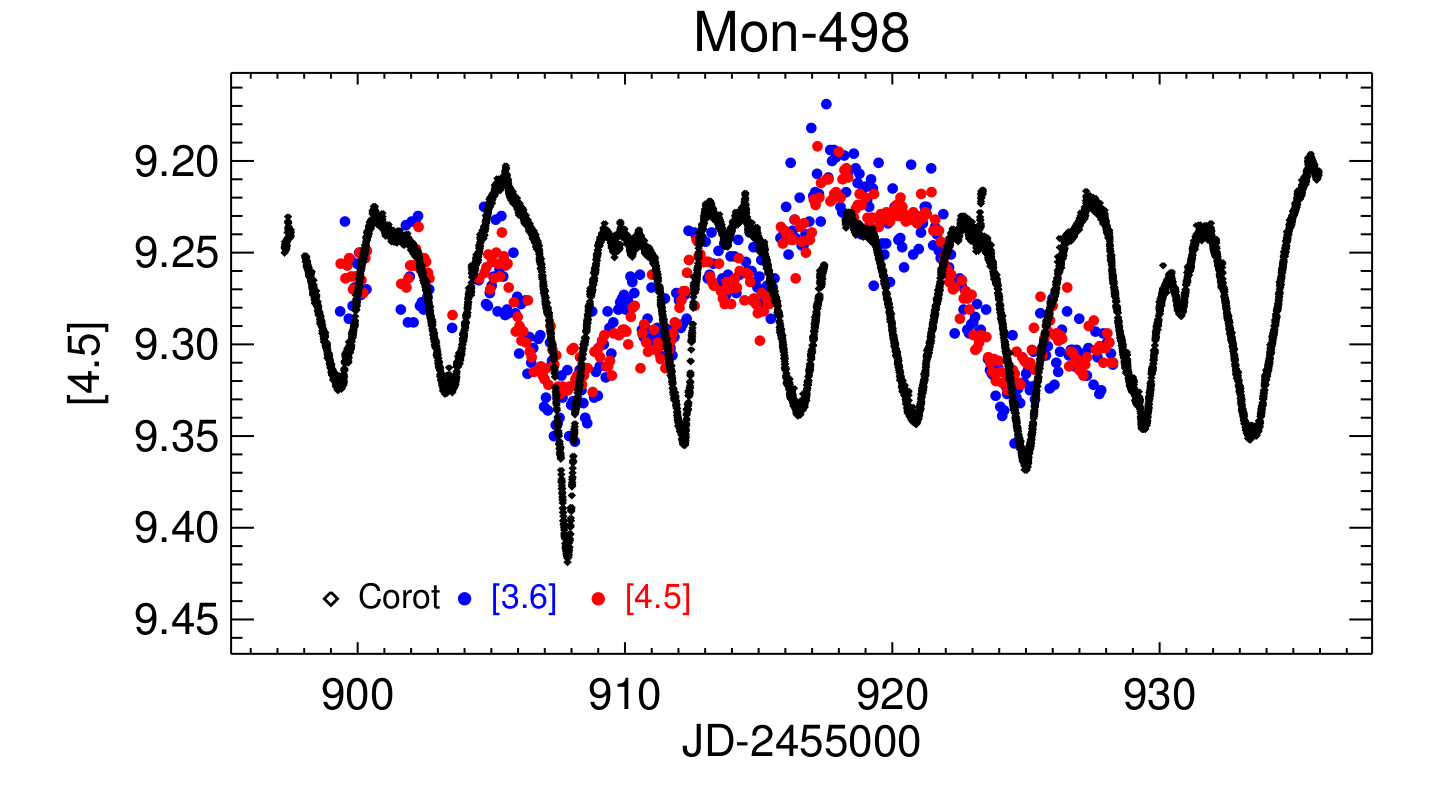}
\includegraphics[width=6cm]{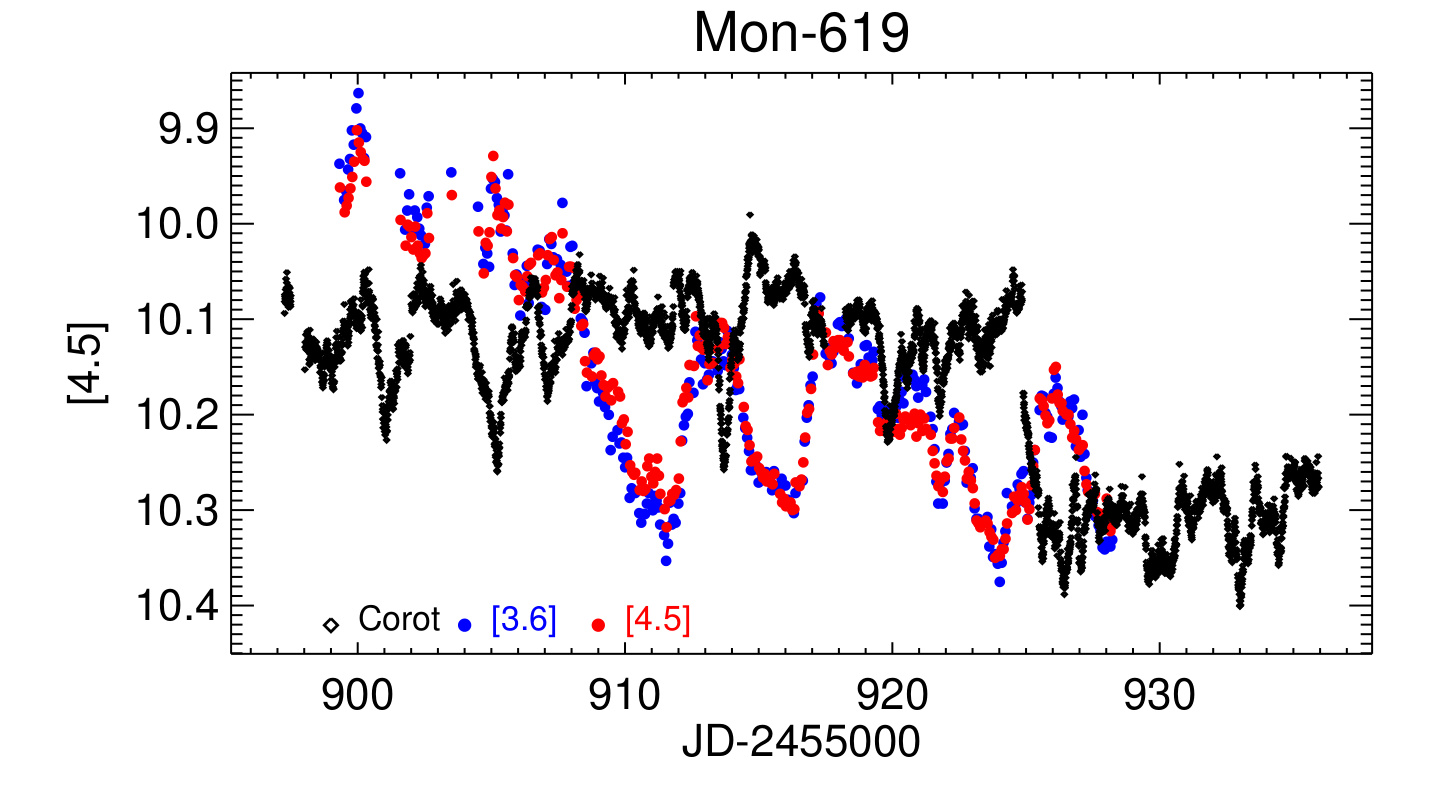}
\includegraphics[width=6cm]{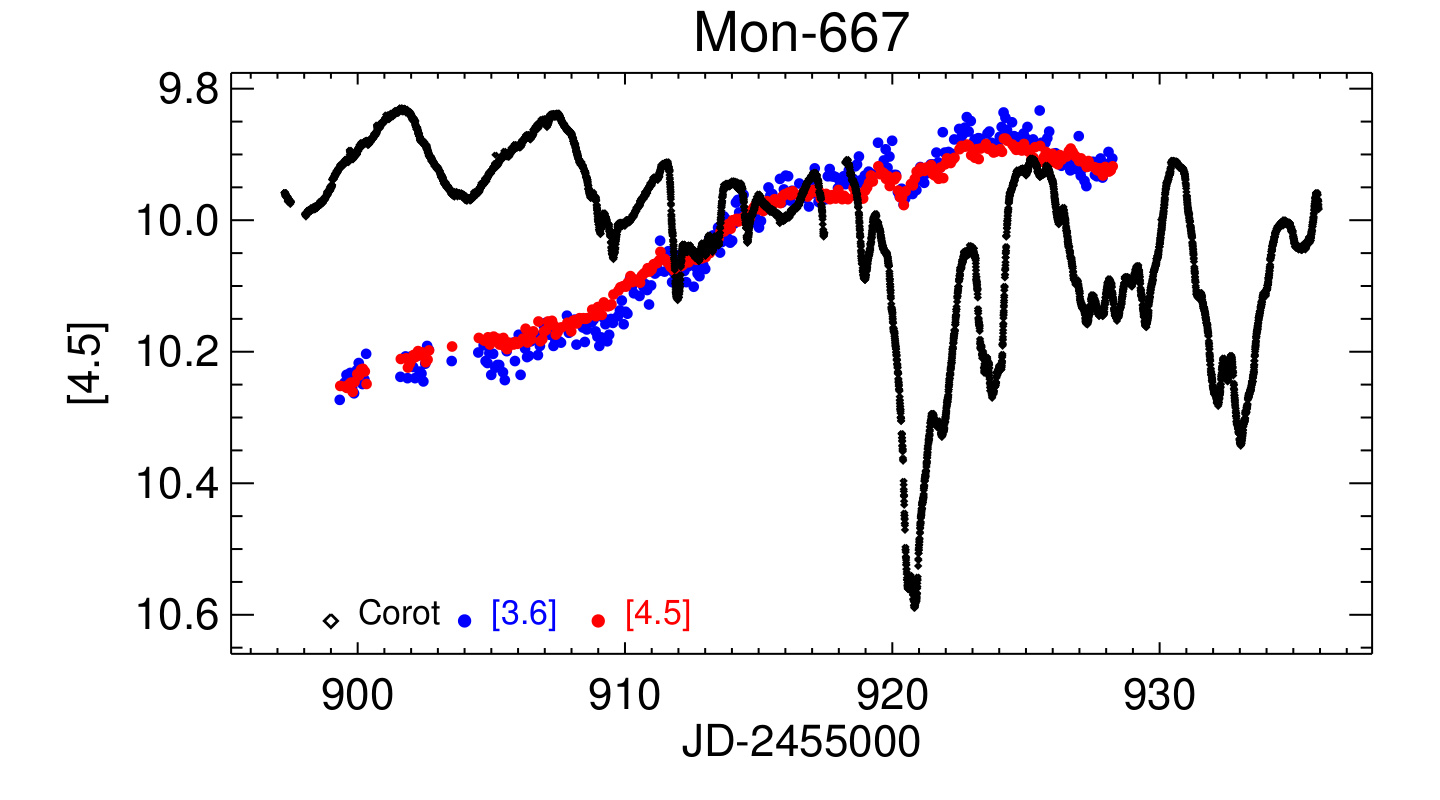}
\includegraphics[width=6cm]{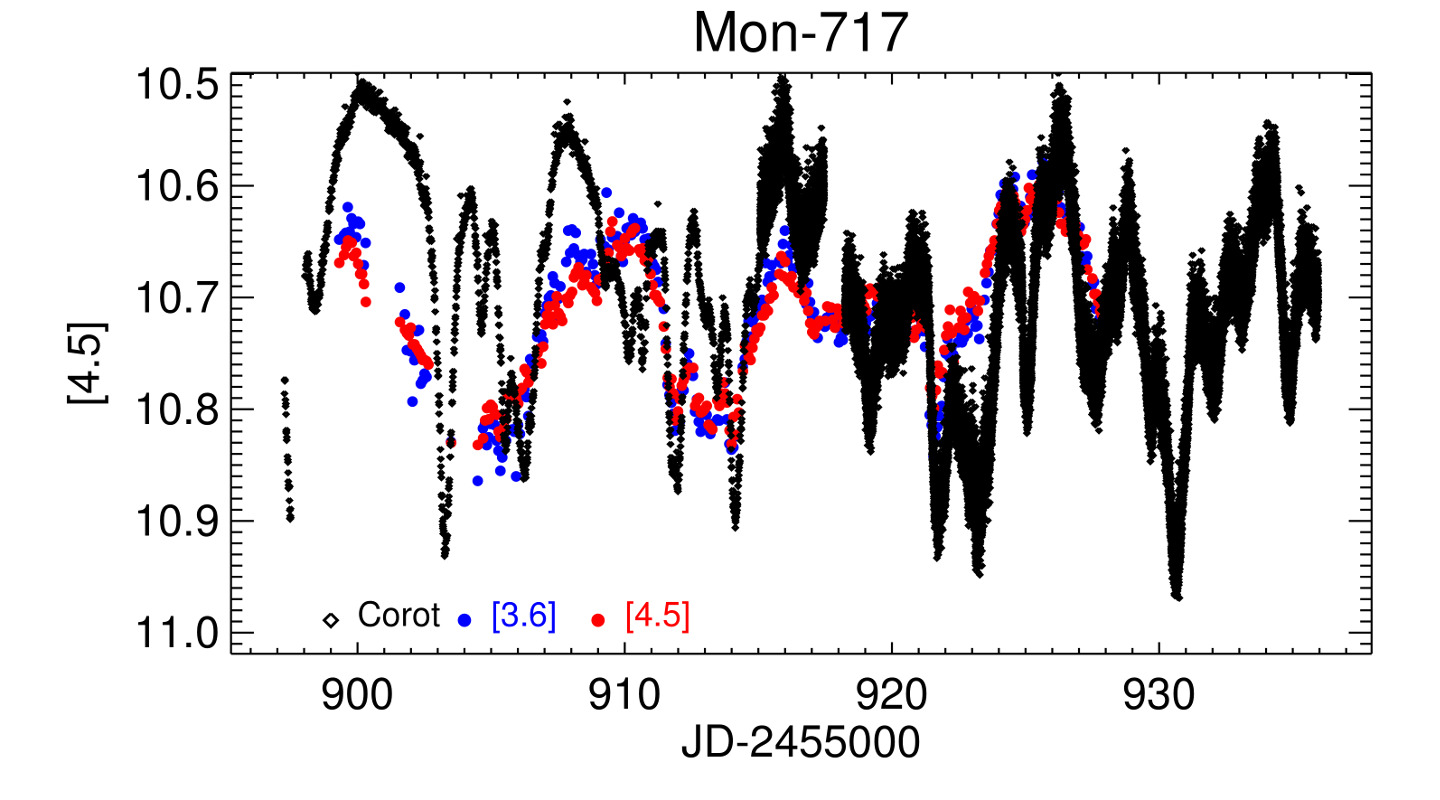}
\includegraphics[width=6cm]{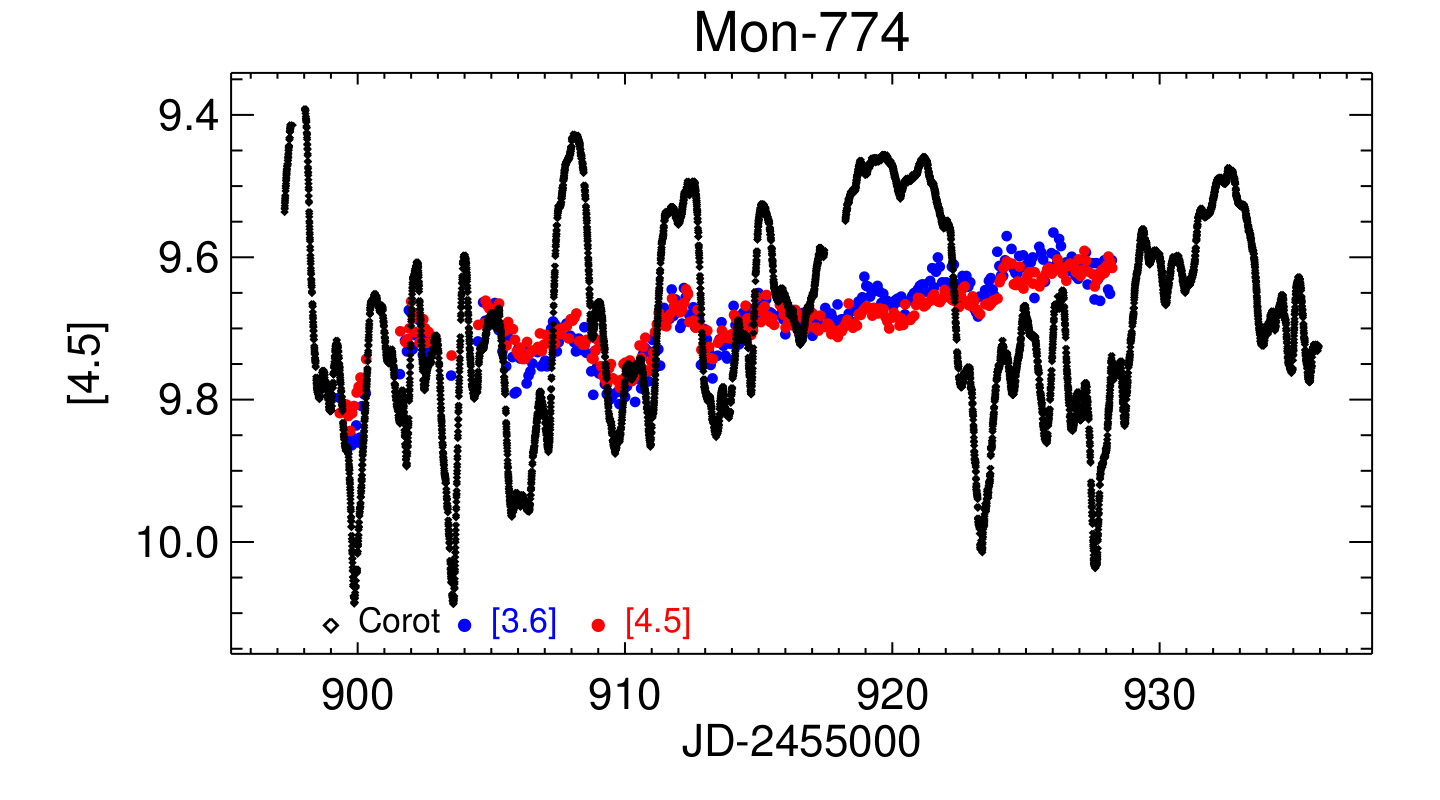}
\includegraphics[width=6cm]{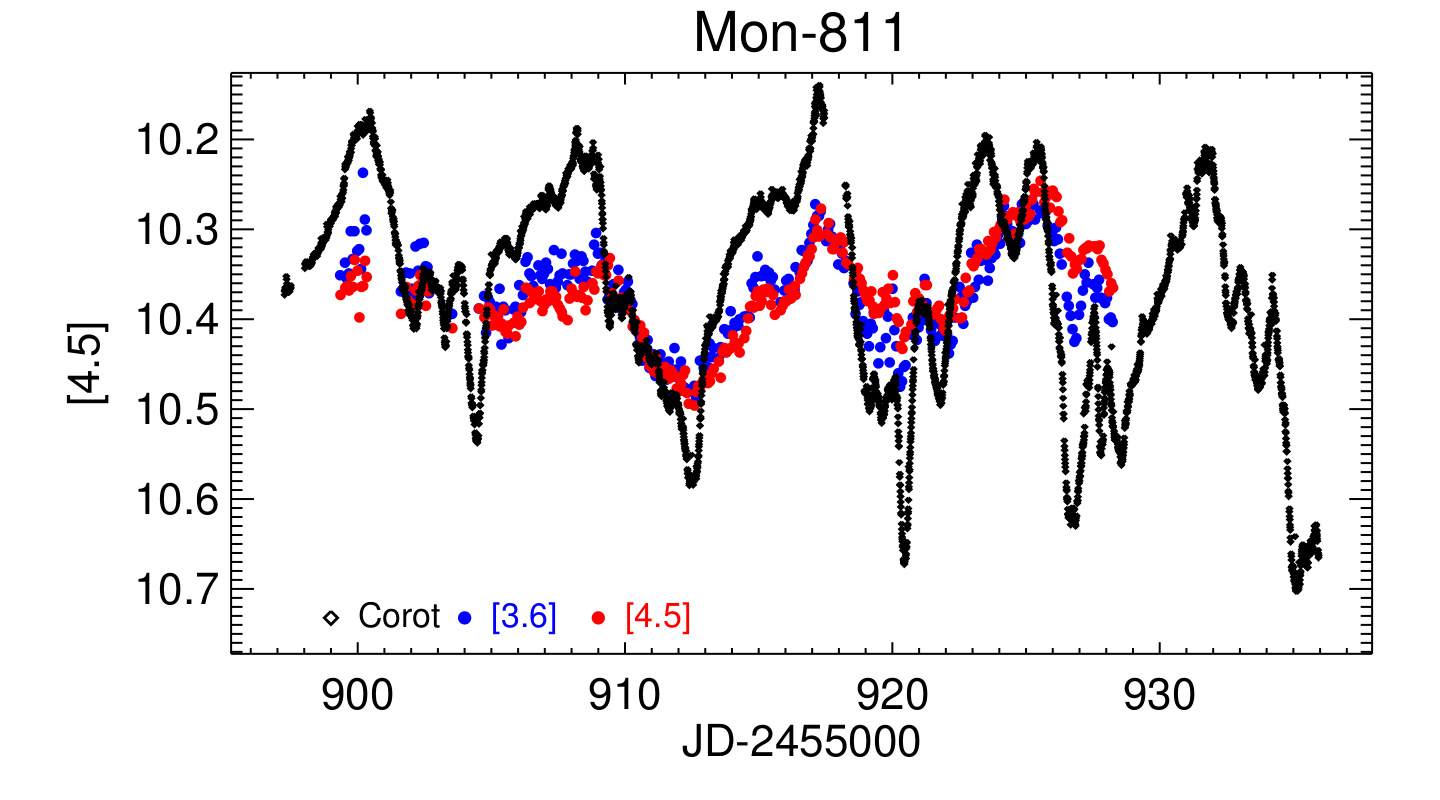}
\includegraphics[width=6cm]{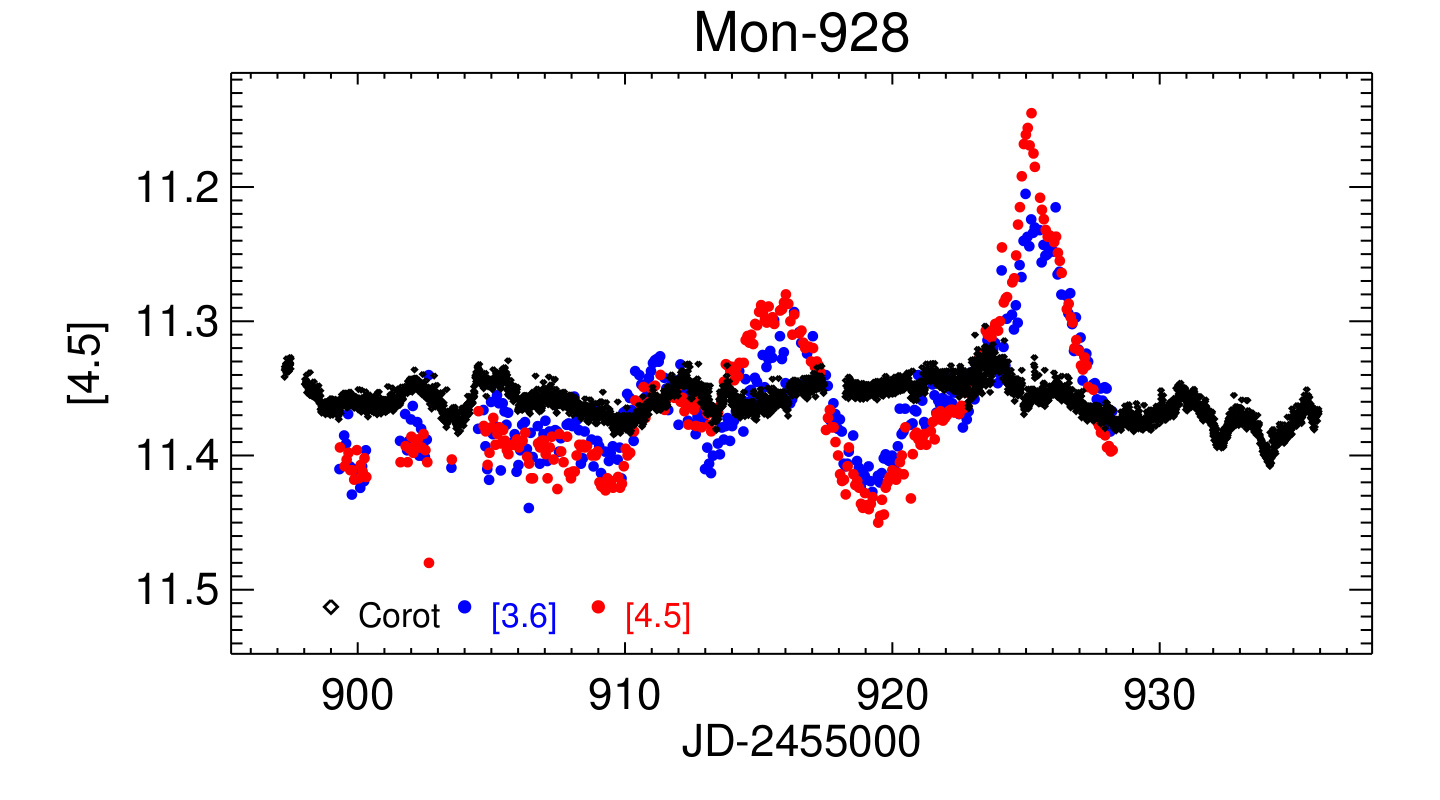}
\includegraphics[width=6cm]{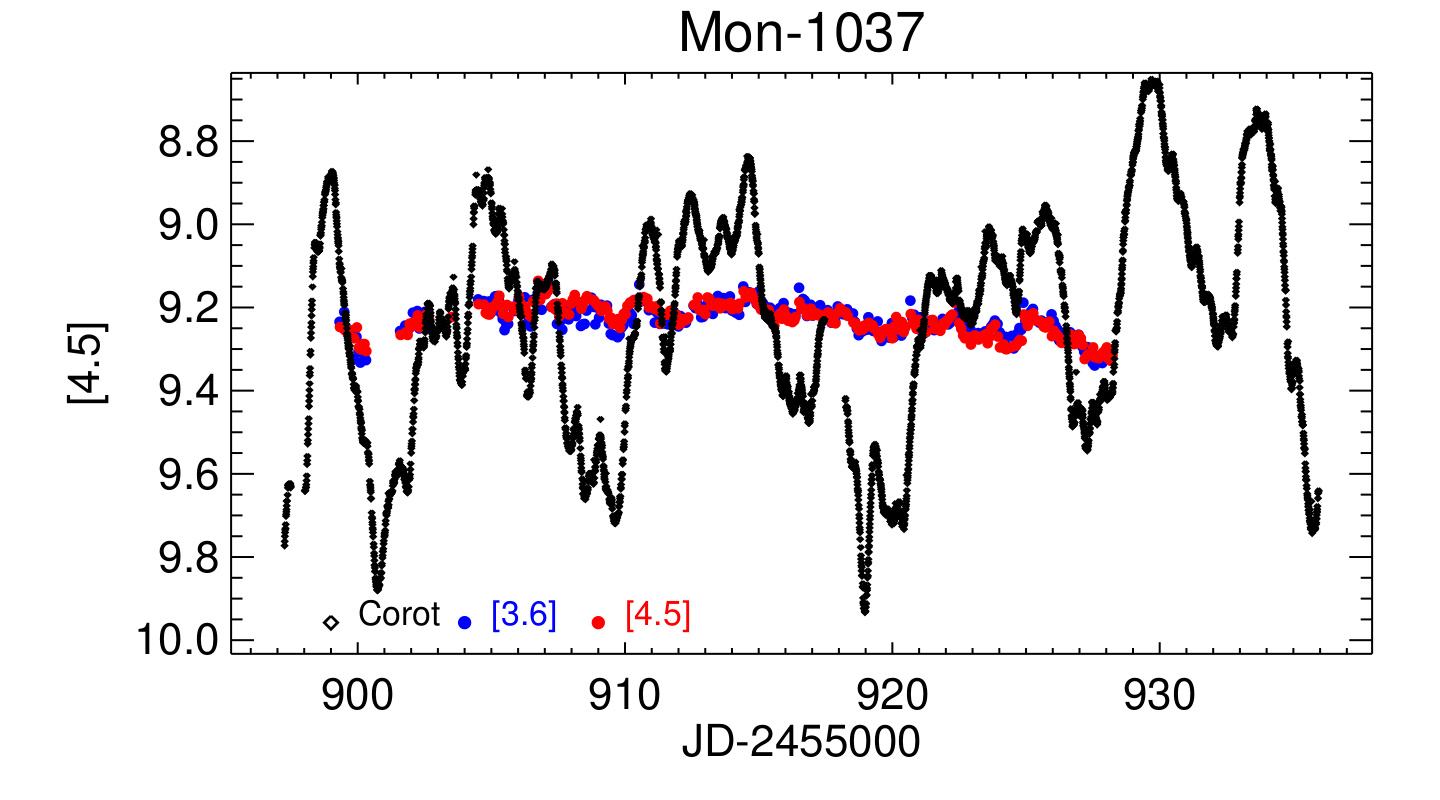}
\includegraphics[width=6cm]{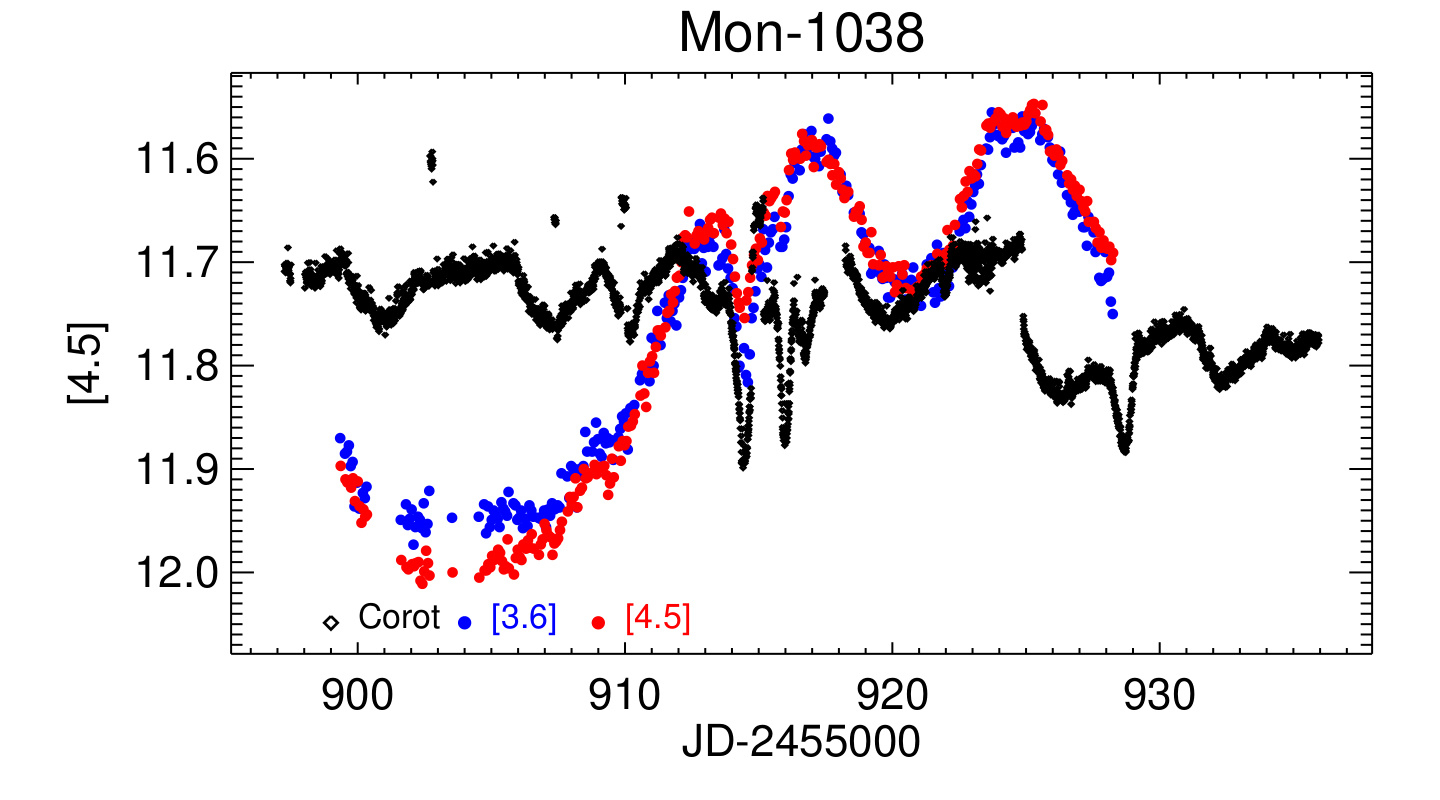}
\includegraphics[width=6cm]{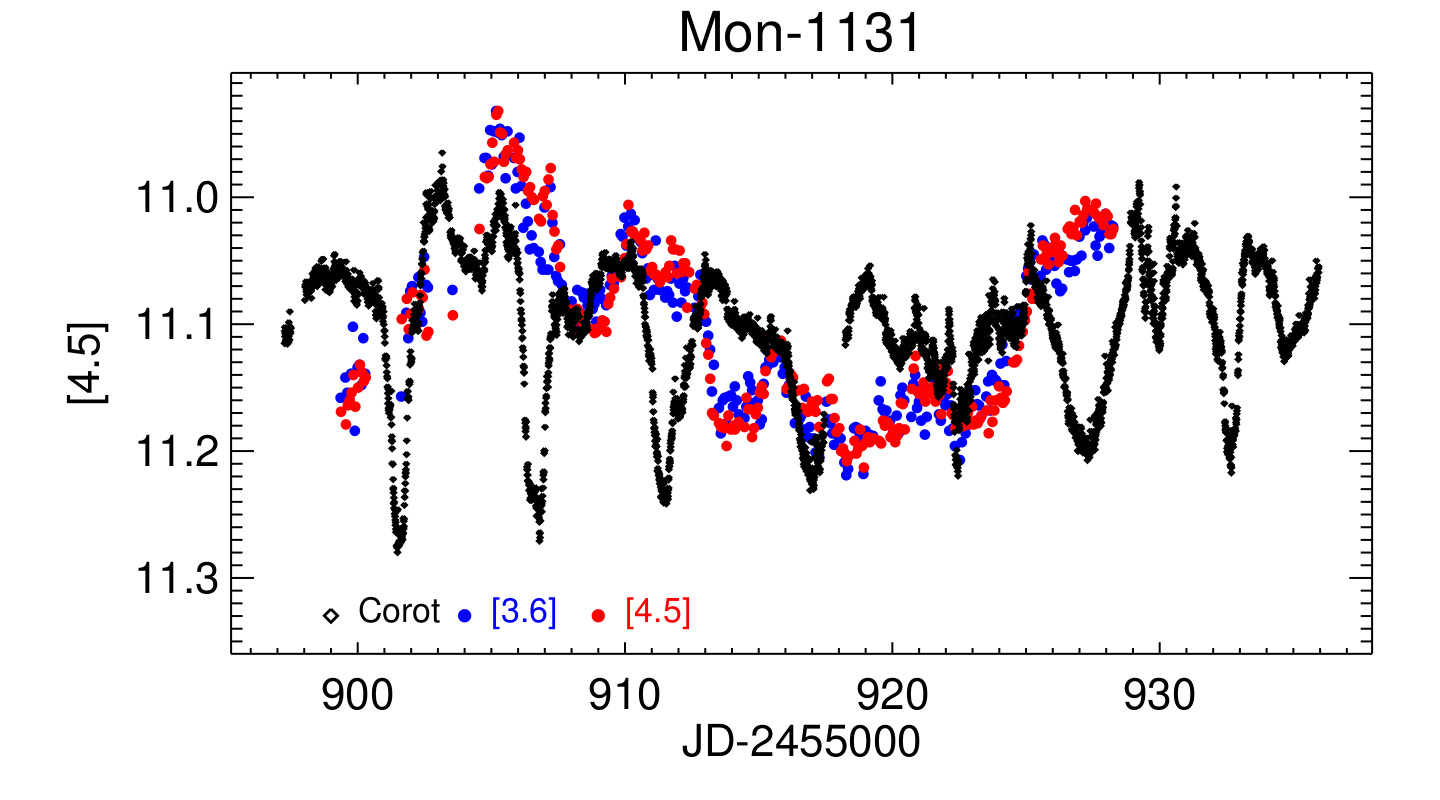}
\includegraphics[width=6cm]{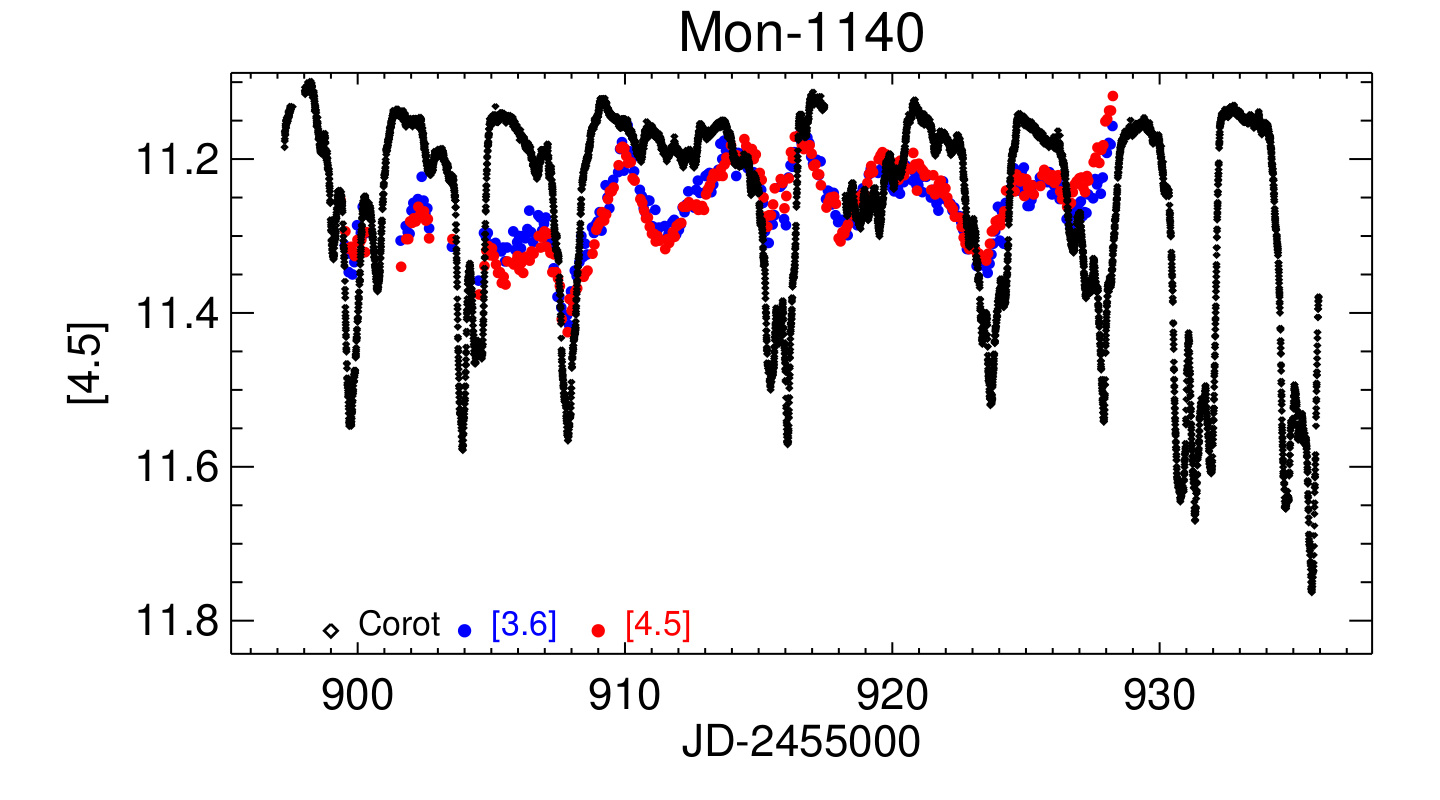}
\includegraphics[width=6cm]{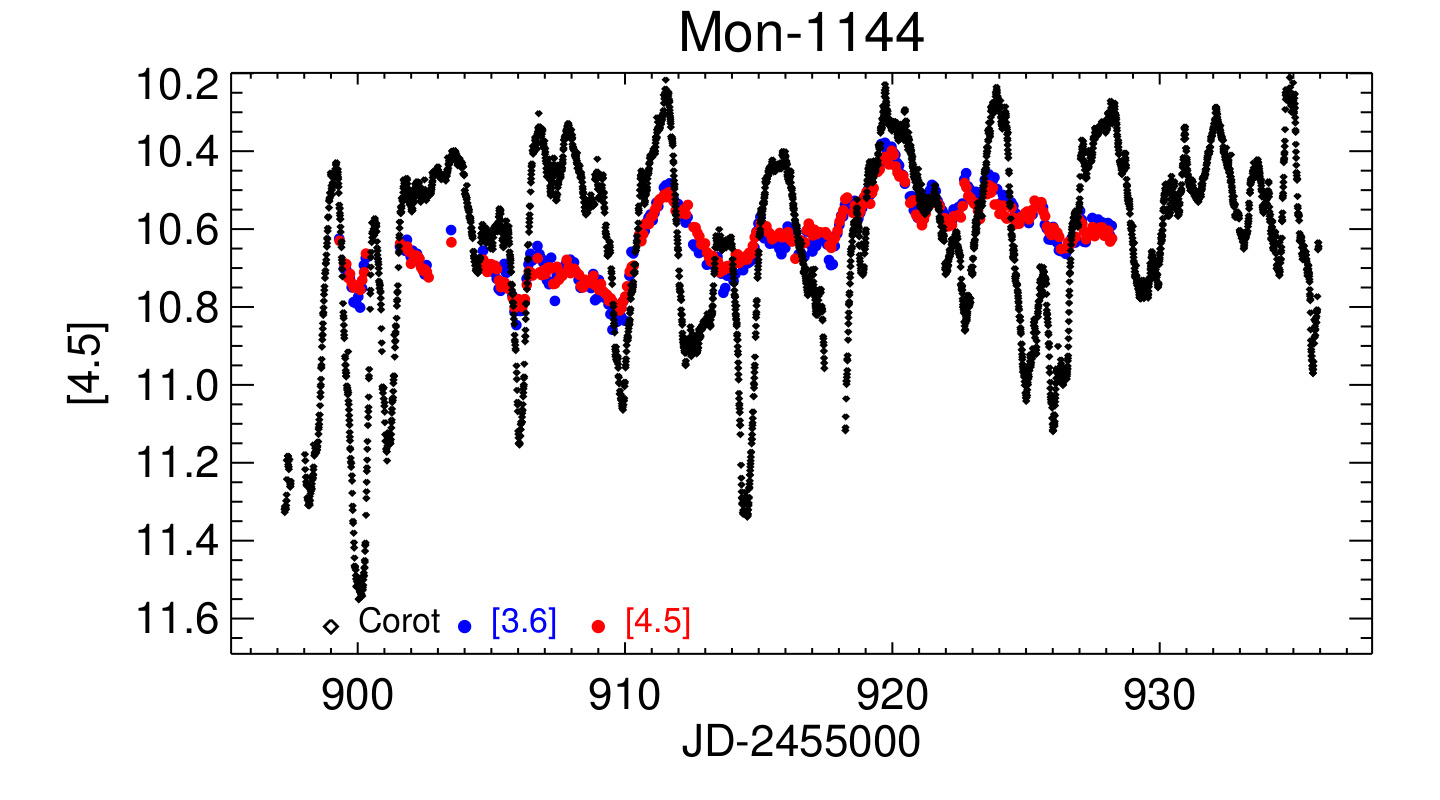}
\includegraphics[width=6cm]{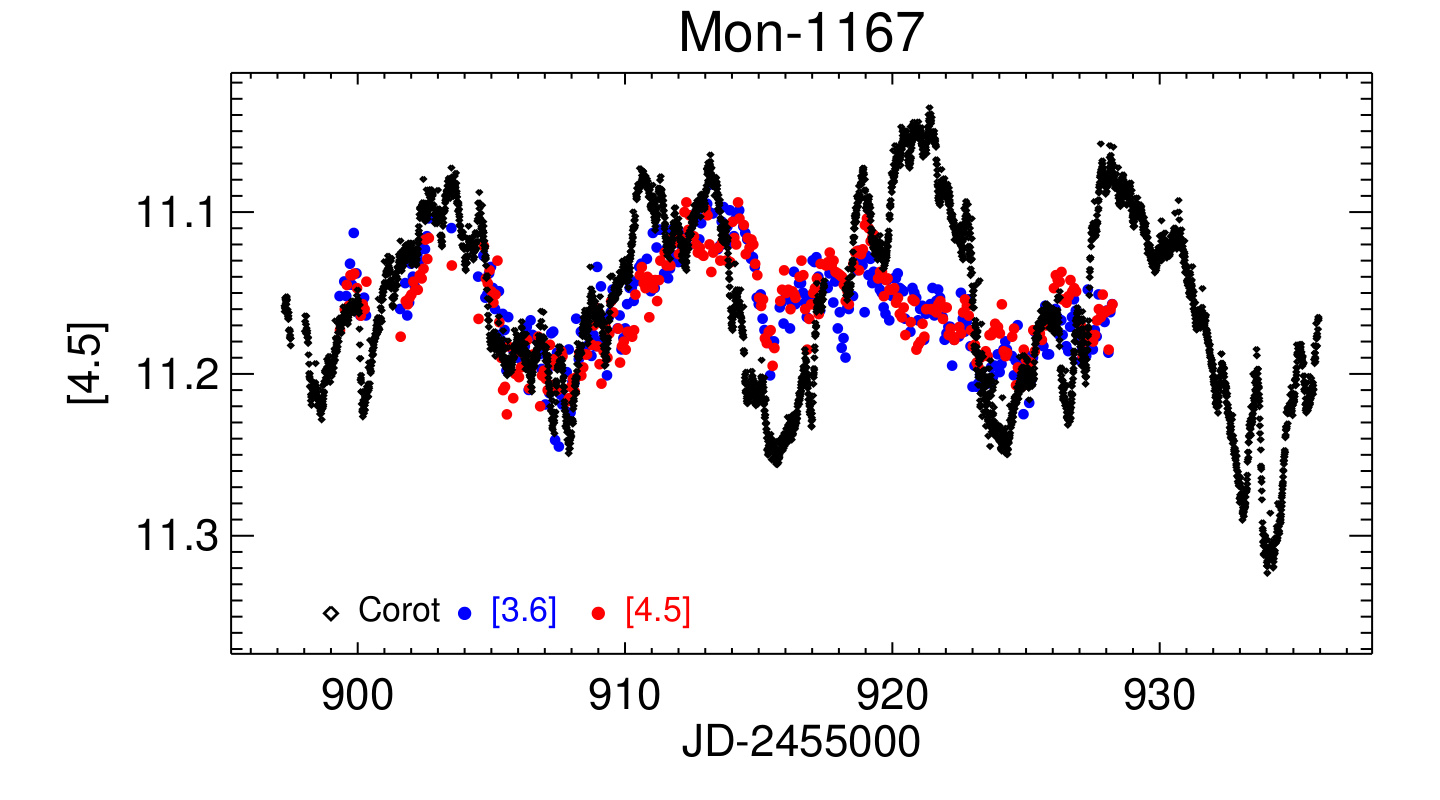}
\includegraphics[width=6cm]{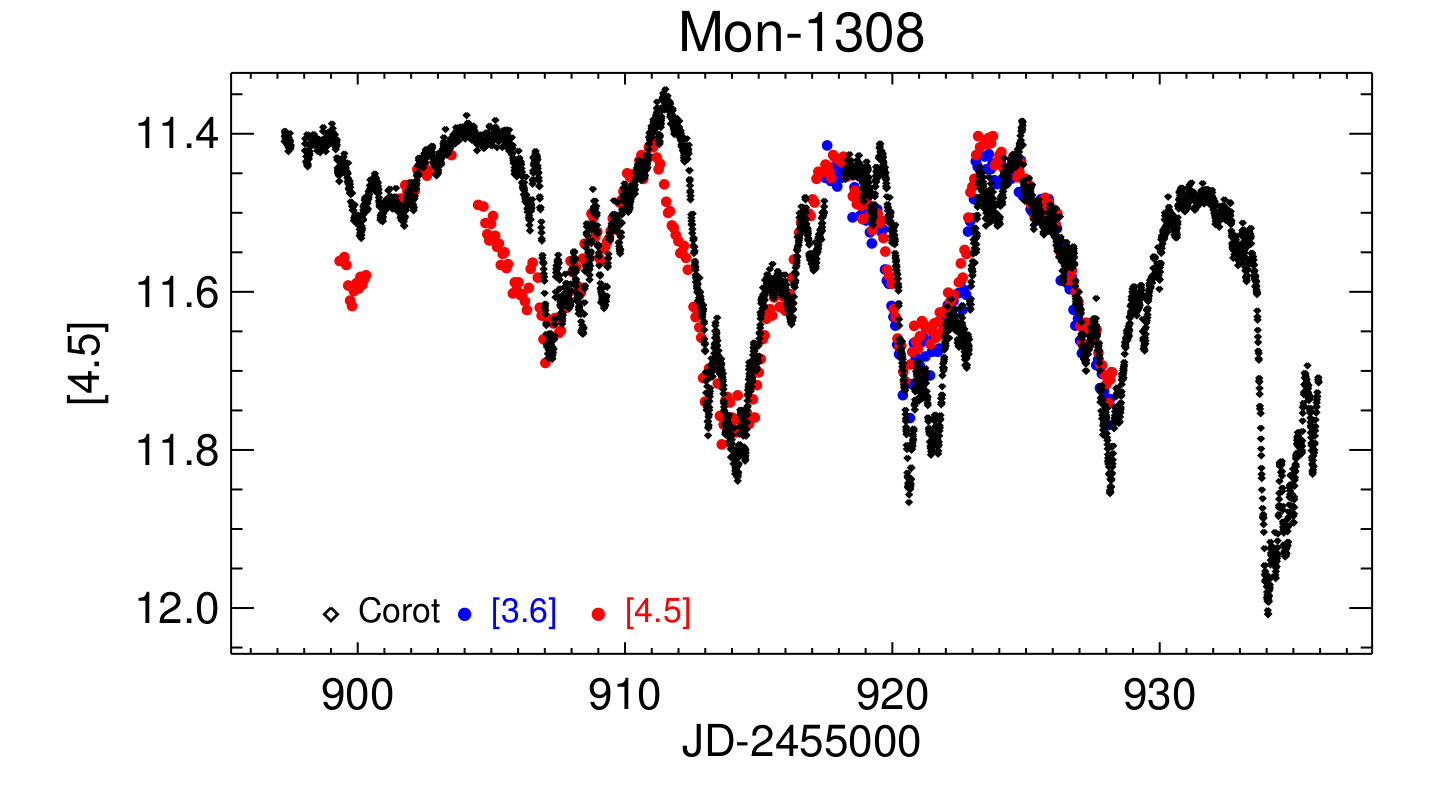}
\caption{Continued.
}
\end{figure*}
}

\onlfig{
\begin{figure*}[p]
\vspace*{1 cm}
\centering
\includegraphics[width=4.4cm]{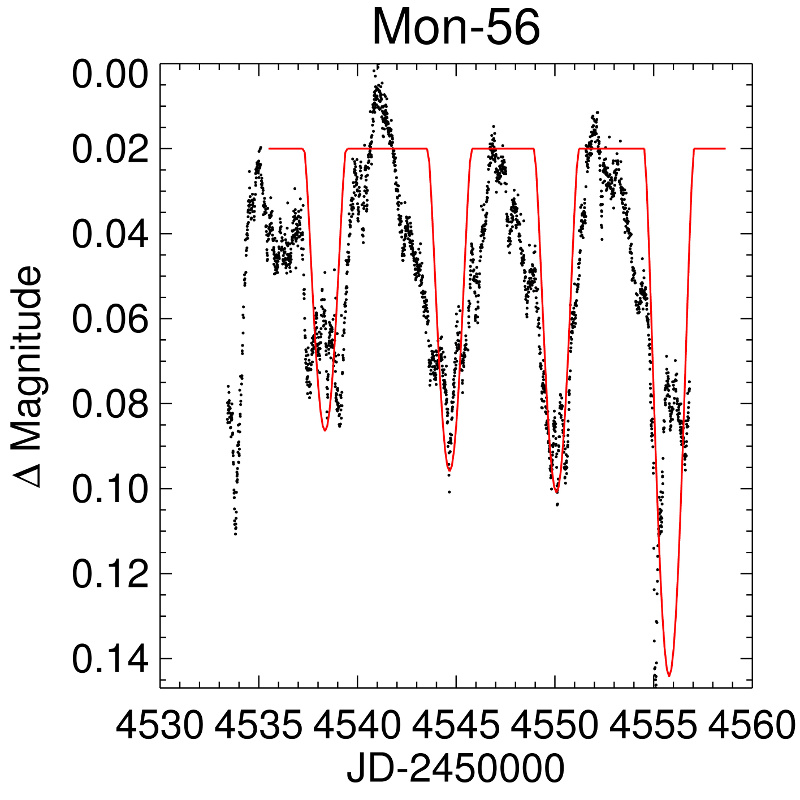}
\includegraphics[width=4.4cm]{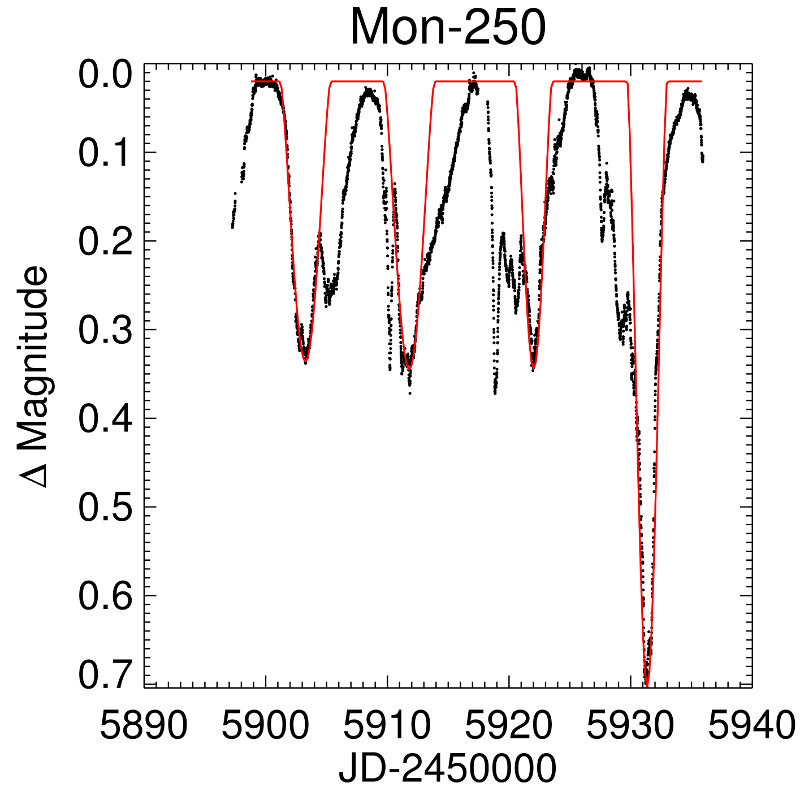}
\includegraphics[width=4.4cm]{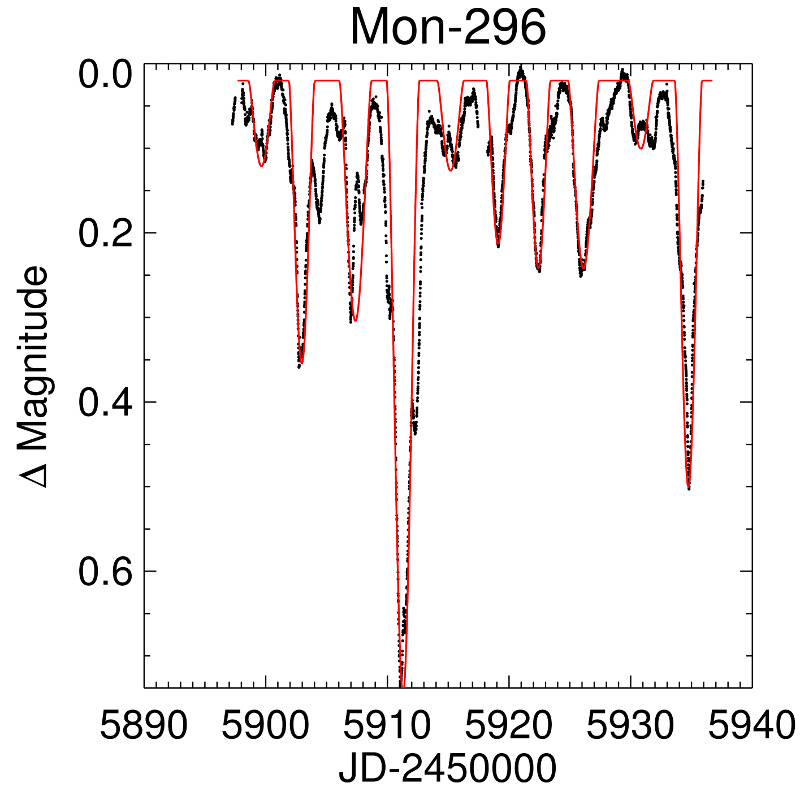}
\includegraphics[width=4.4cm]{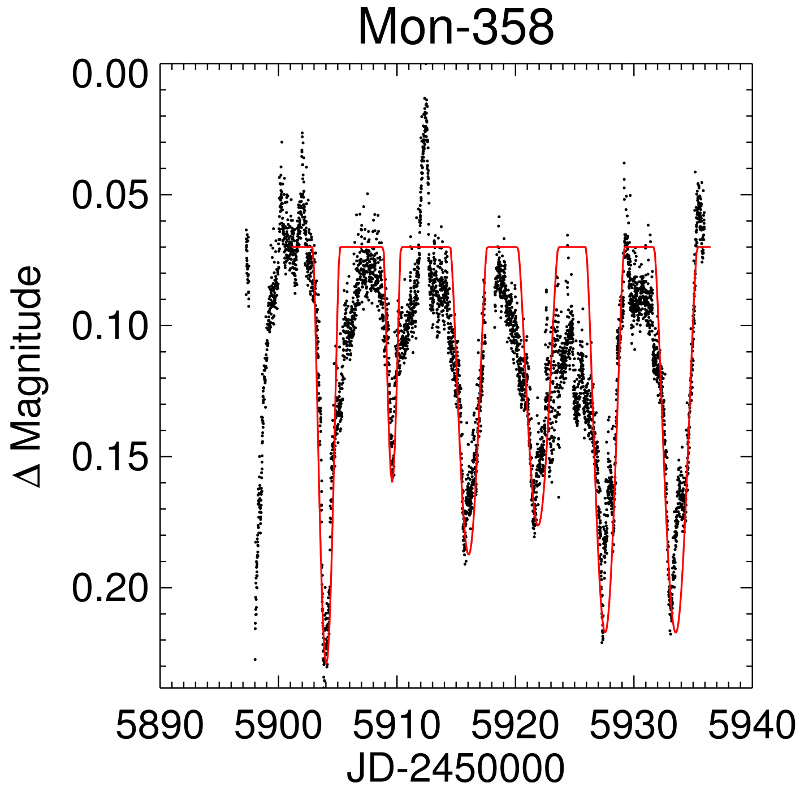}
\includegraphics[width=4.4cm]{Mon-379_SRa05_bouviermodeltest760_77.jpg}
\includegraphics[width=4.4cm]{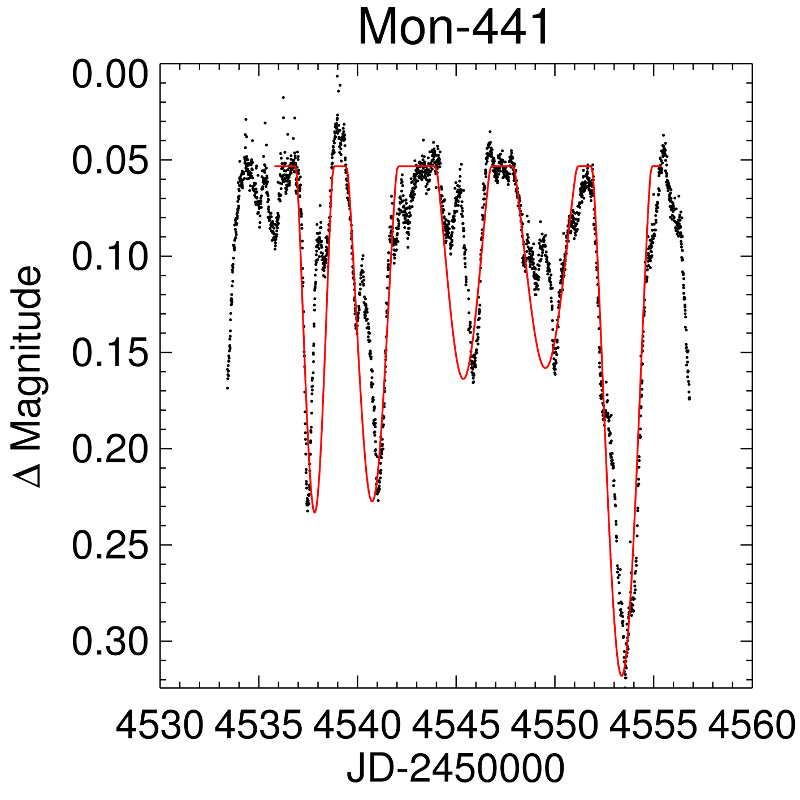}
\includegraphics[width=4.4cm]{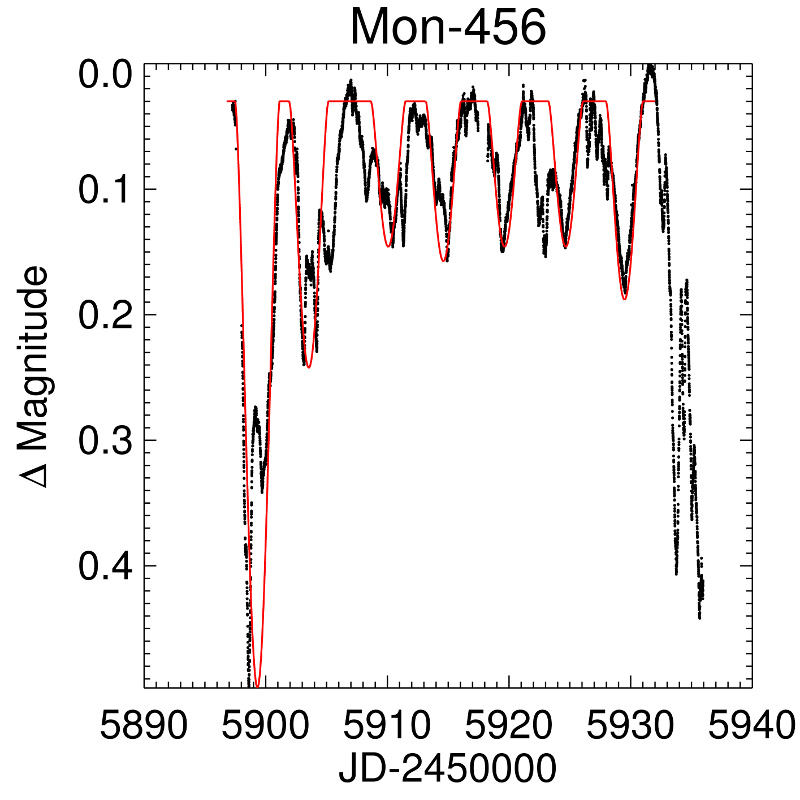}
\includegraphics[width=4.4cm]{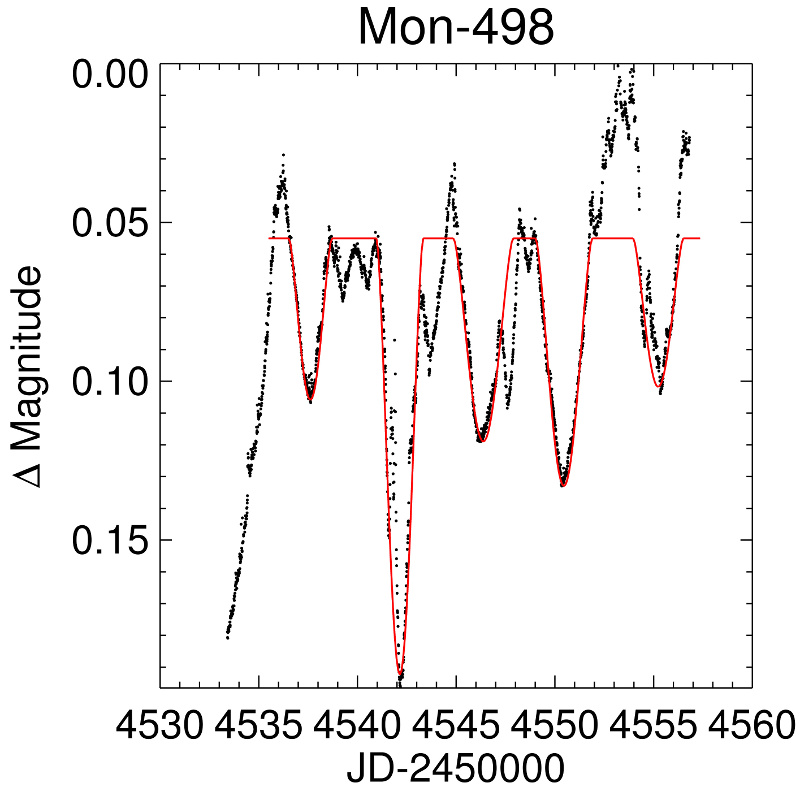}
\includegraphics[width=4.4cm]{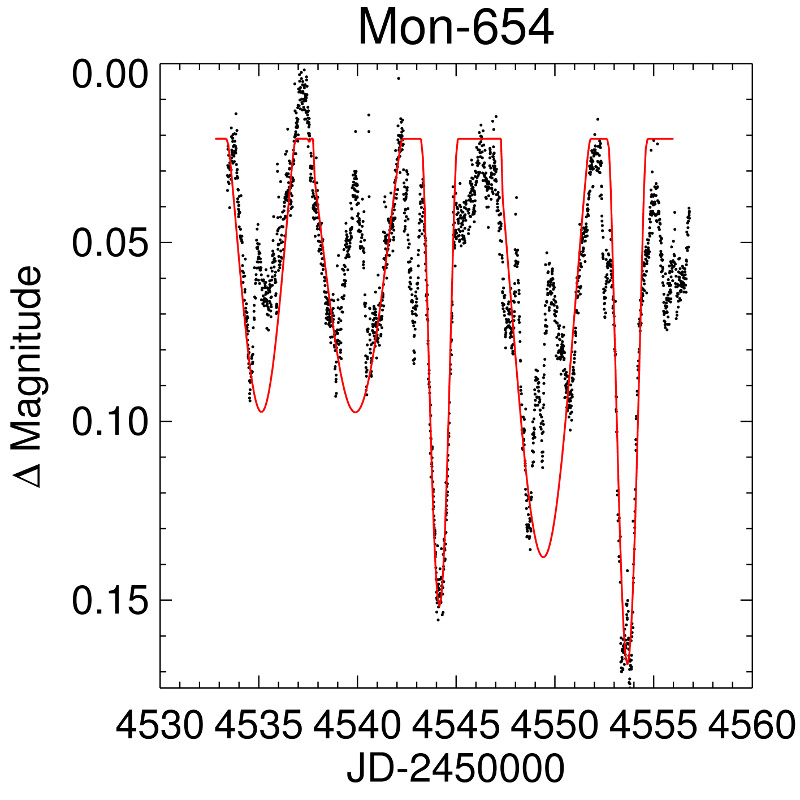}
\includegraphics[width=4.4cm]{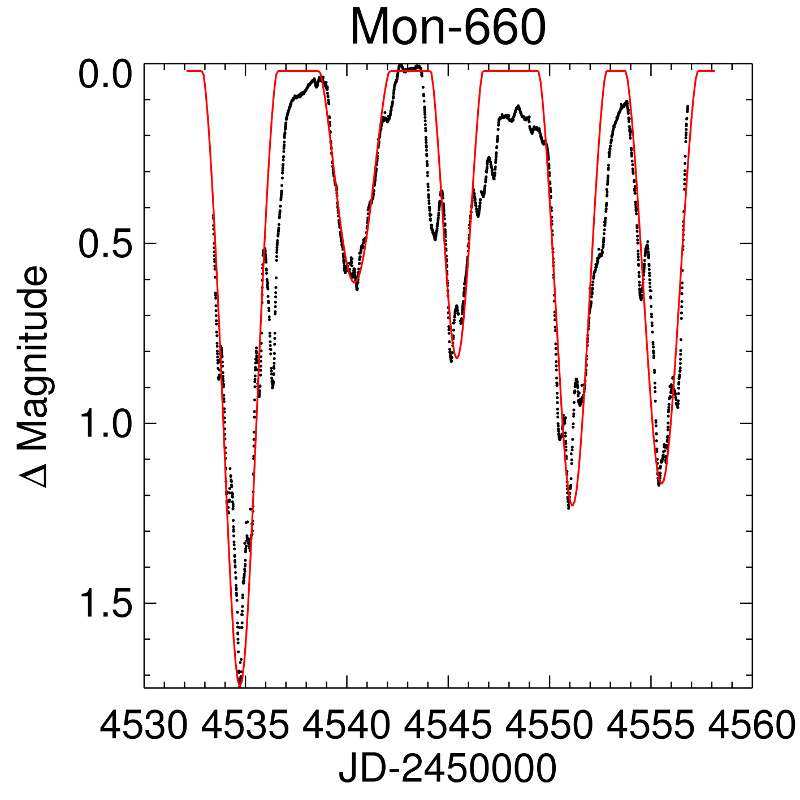}
\includegraphics[width=4.4cm]{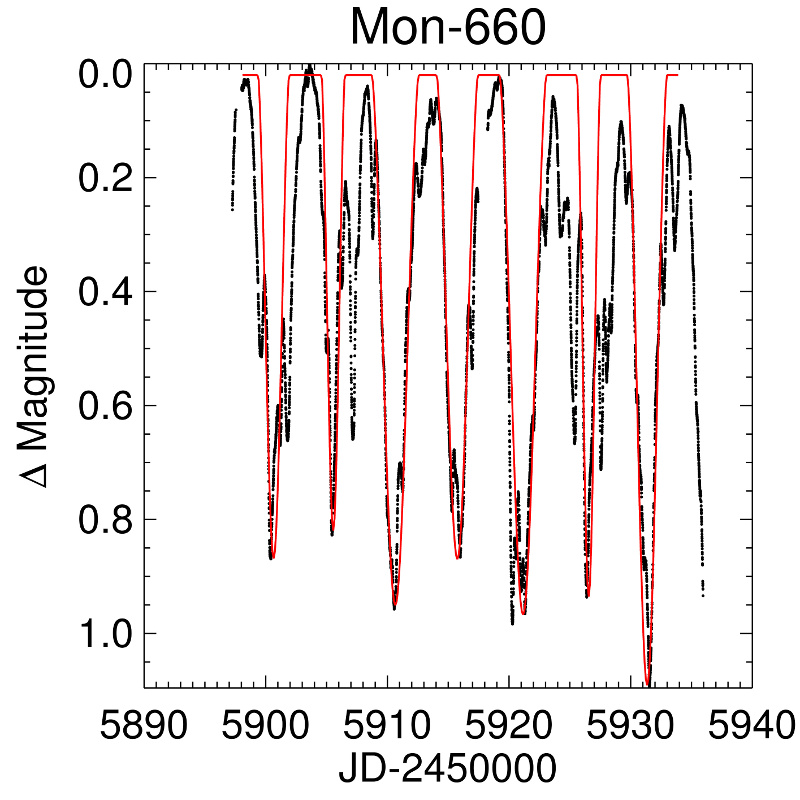}
\includegraphics[width=4.4cm]{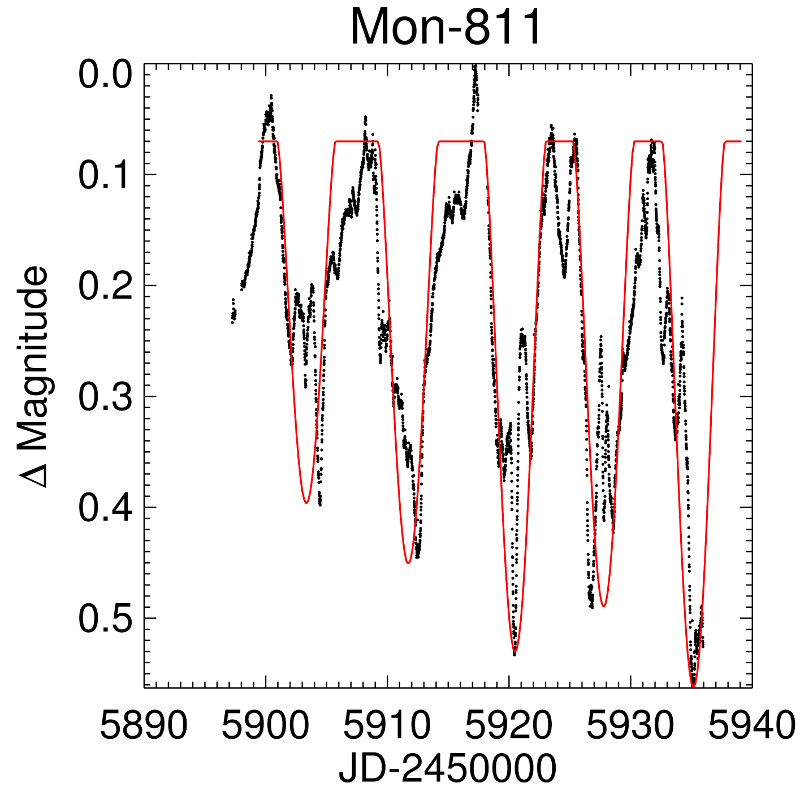}
\includegraphics[width=4.4cm]{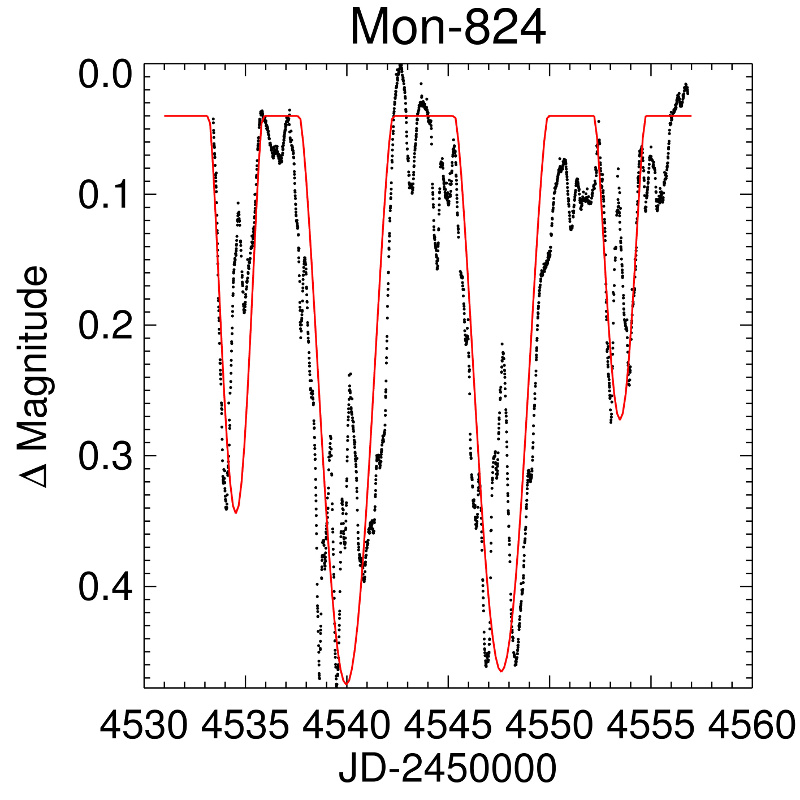}
\includegraphics[width=4.4cm]{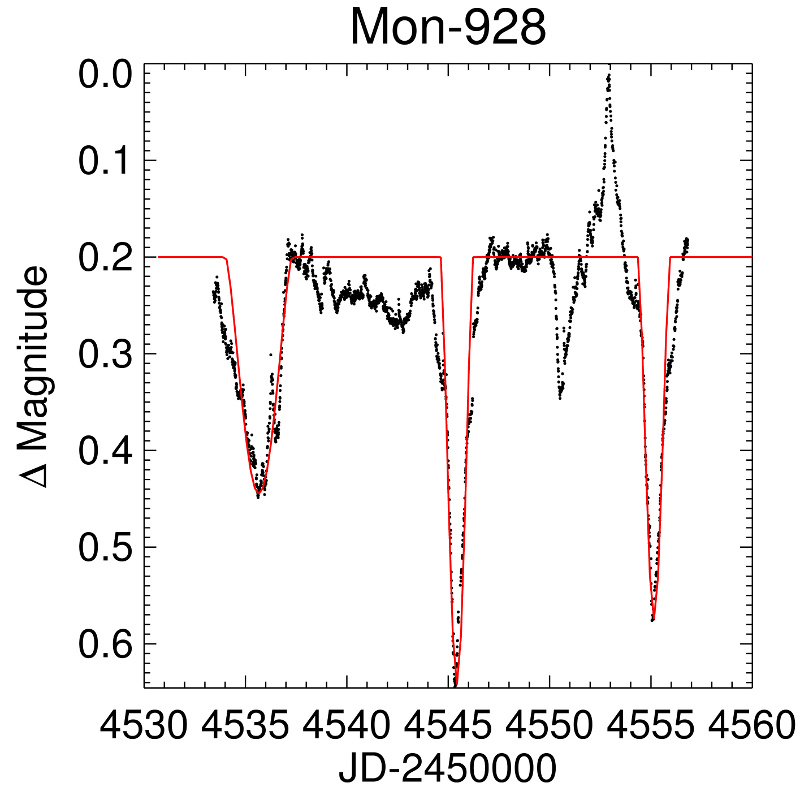}
\includegraphics[width=4.4cm]{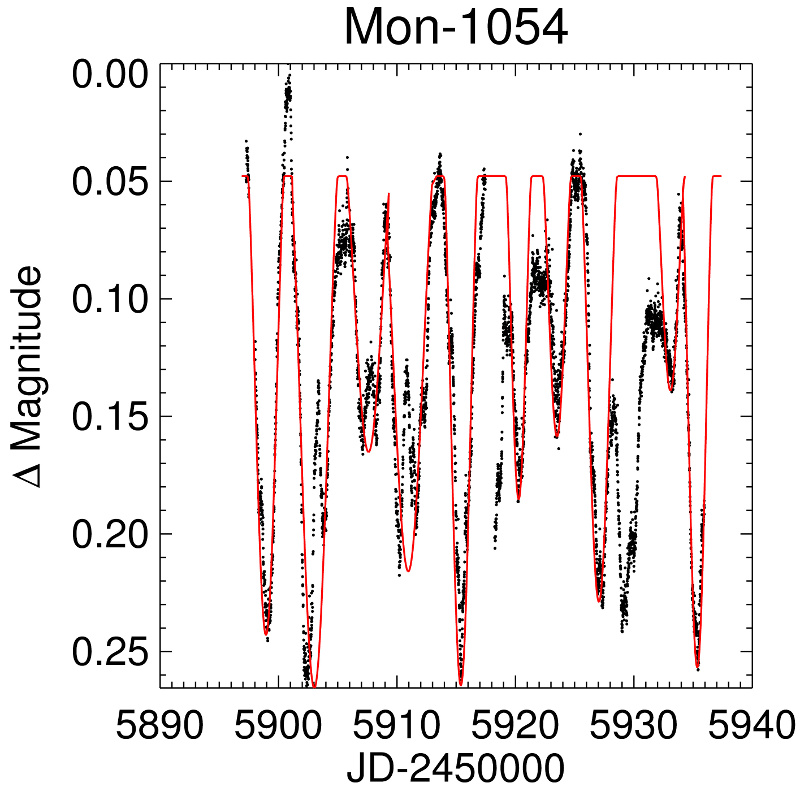}
\includegraphics[width=4.4cm]{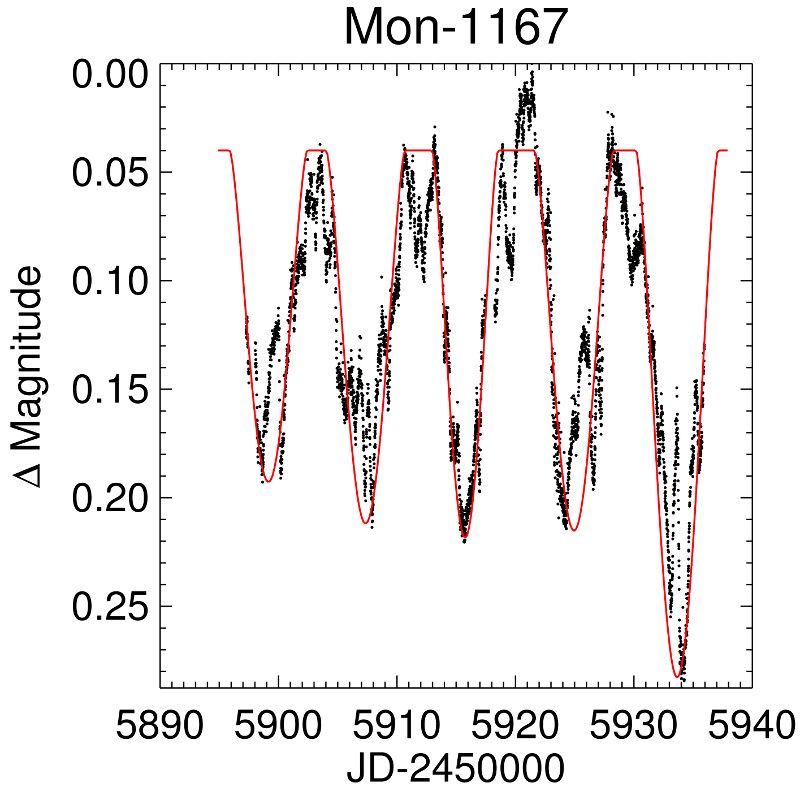}
\includegraphics[width=4.4cm]{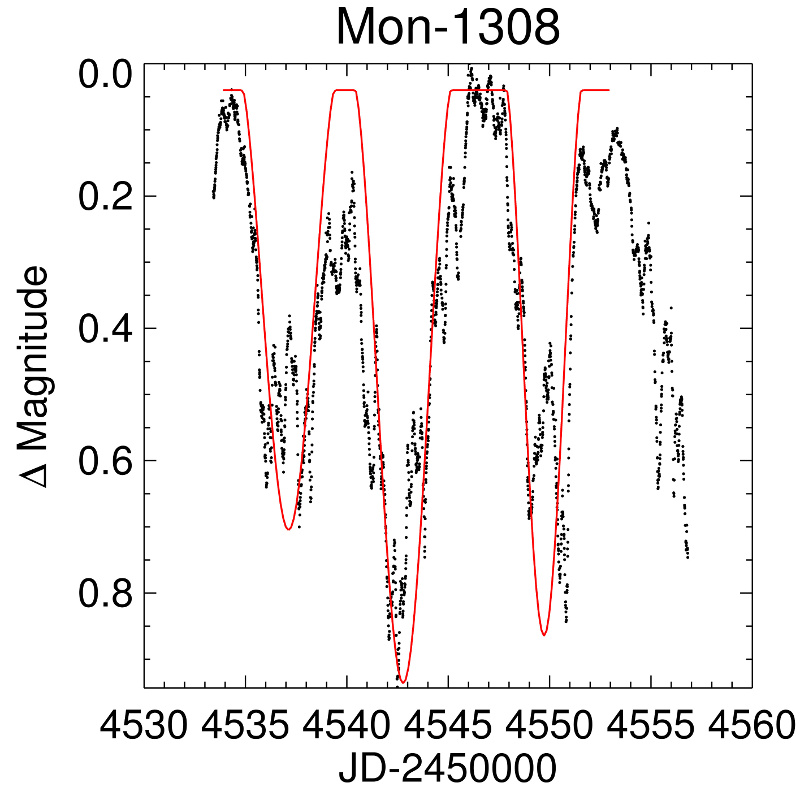}
\caption{Simulated light curves of AA Tau-like stars (red) plotted over CoRoT light curves (black).
}\label{fig:app3}
\end{figure*}
}

\onlfig{
\begin{figure*}[p]
\vspace*{1 cm}
\centering
\includegraphics[width=11.5cm]{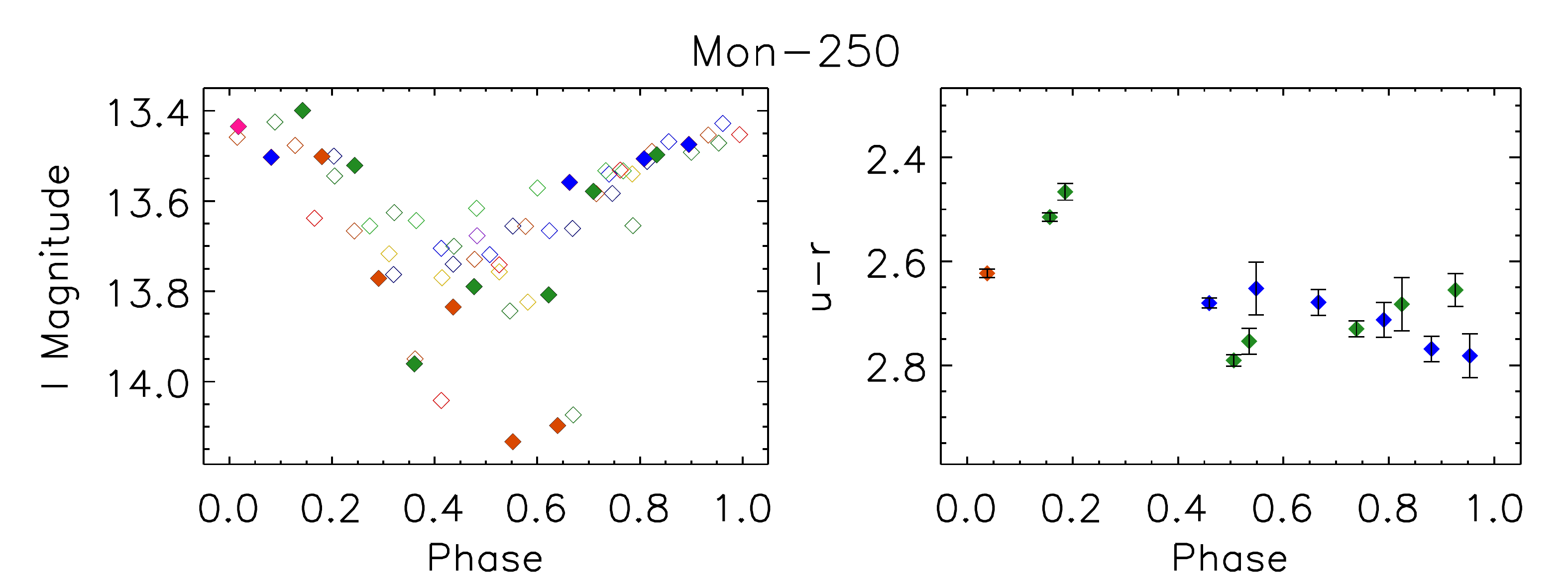}
\includegraphics[width=11.5cm]{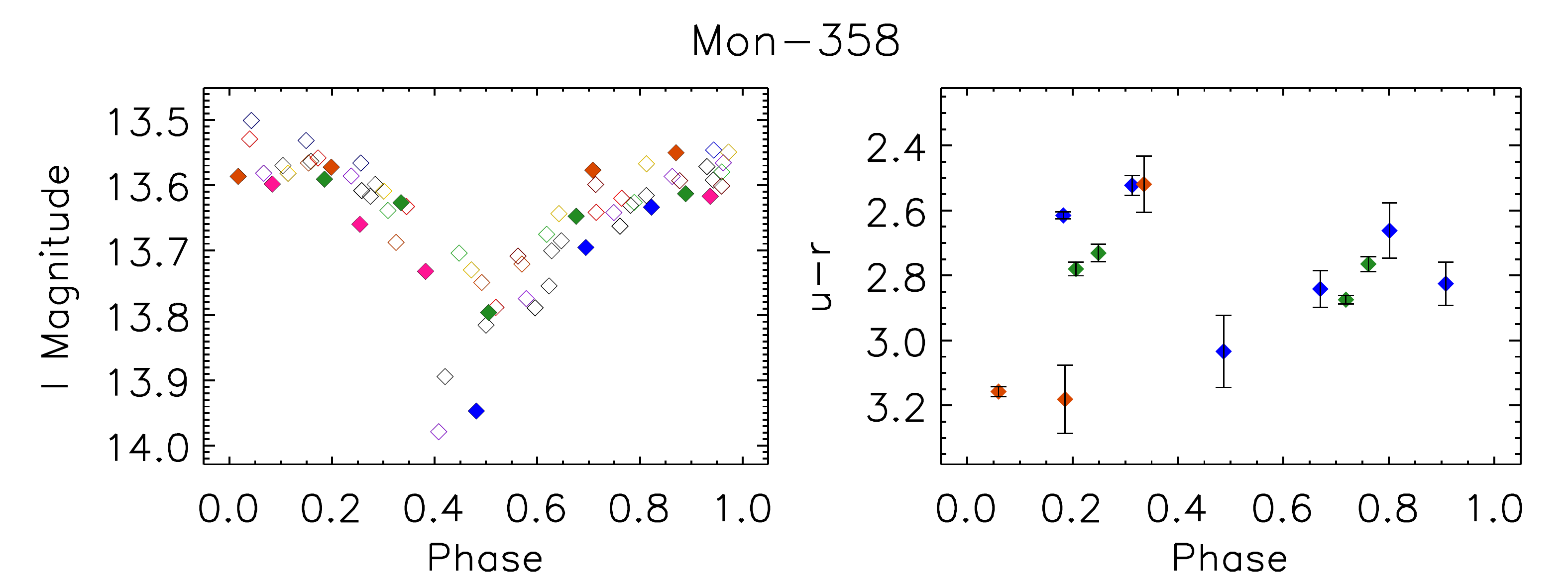}
\includegraphics[width=11.5cm]{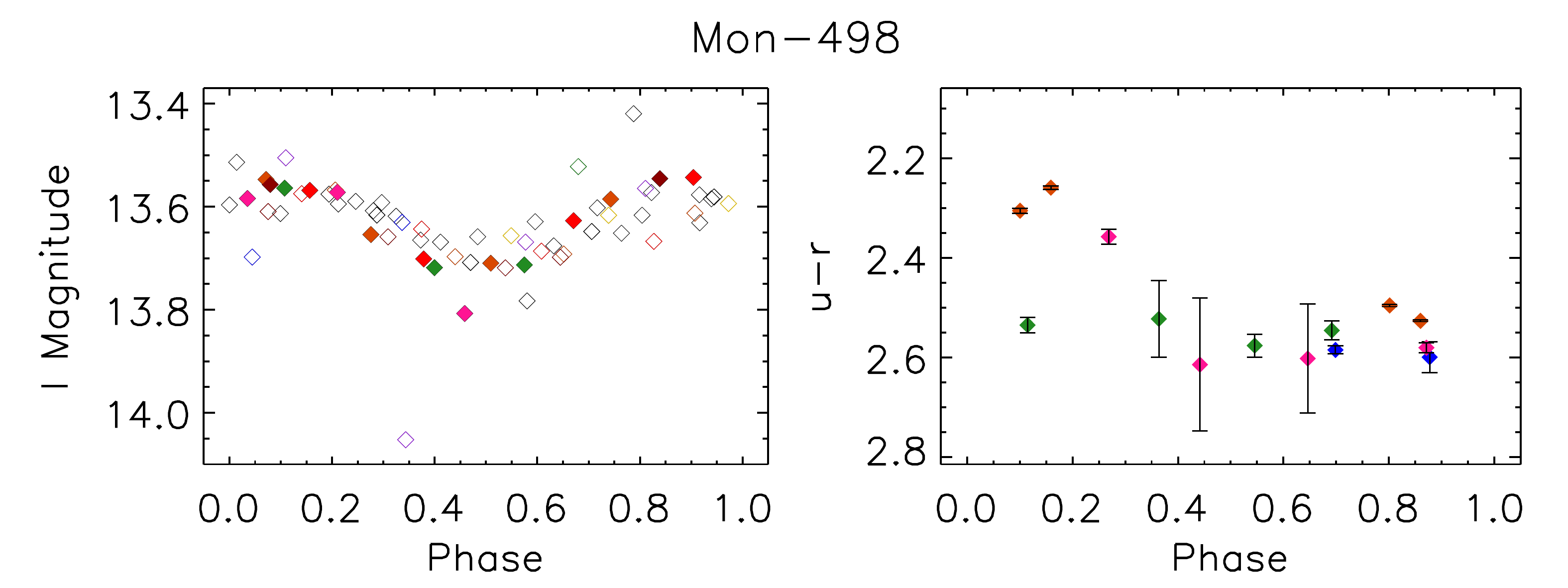}
\includegraphics[width=11.5cm]{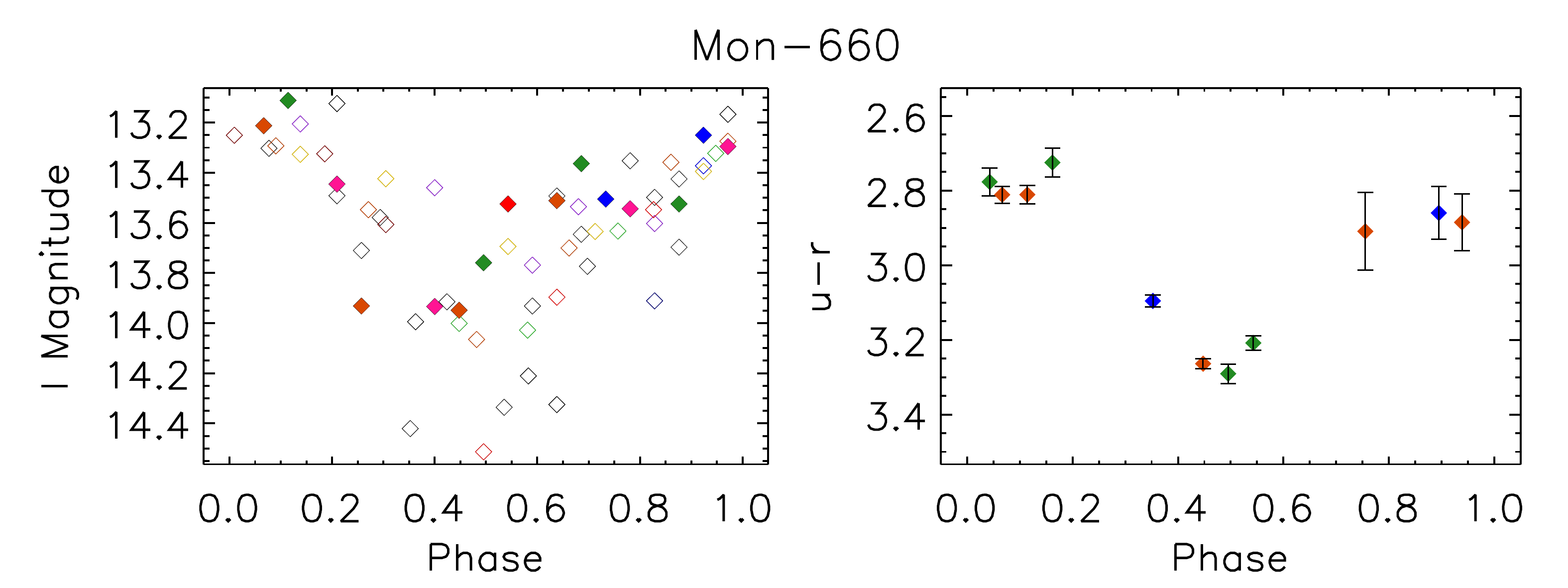}
\includegraphics[width=11.5cm]{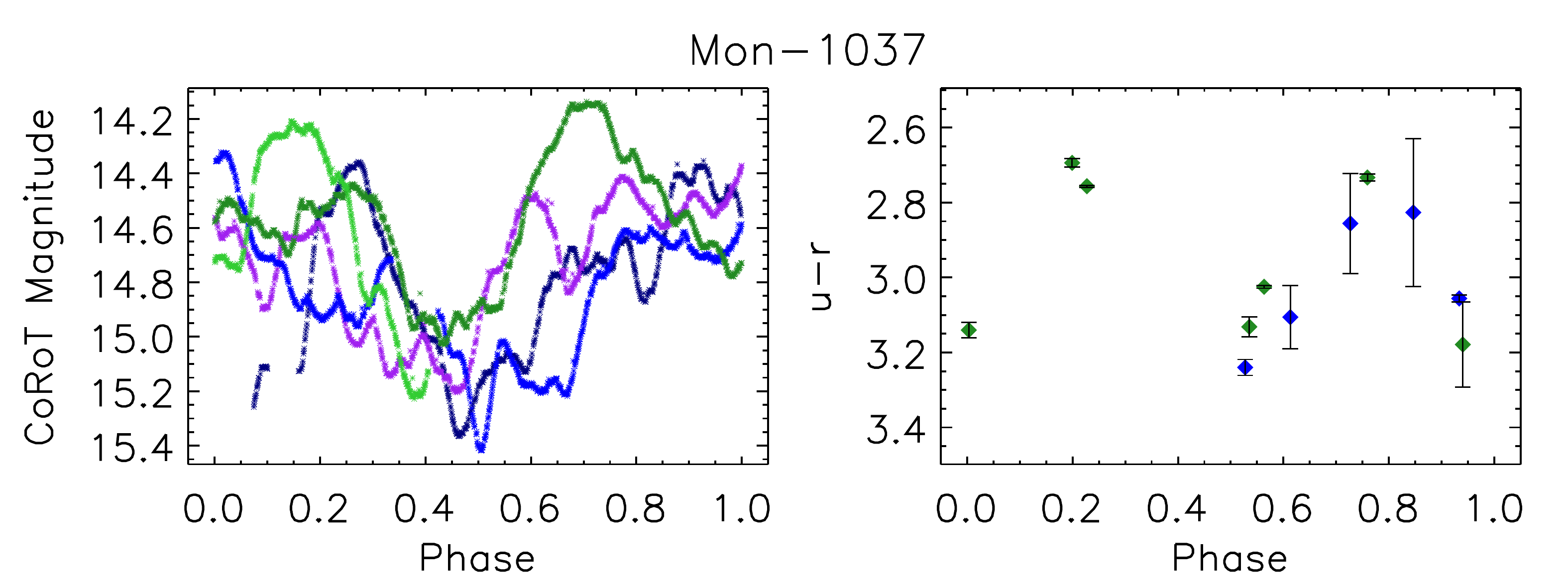}
\caption{I-band or CoRoT light curves and u-r color diagrams folded in phase. 
}
\label{fig:app4}
\end{figure*}

\begin{figure*}[p]
\vspace*{1 cm}
\centering
\ContinuedFloat
\includegraphics[width=11.5cm]{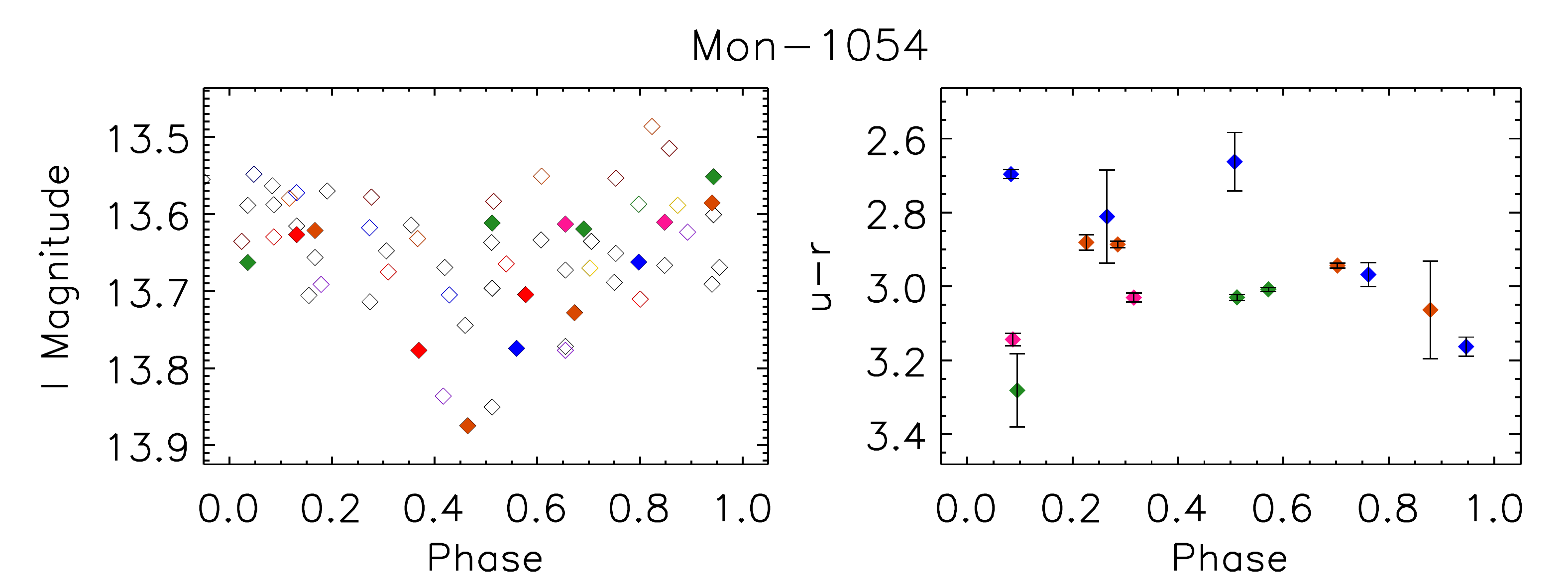}
\includegraphics[width=11.5cm]{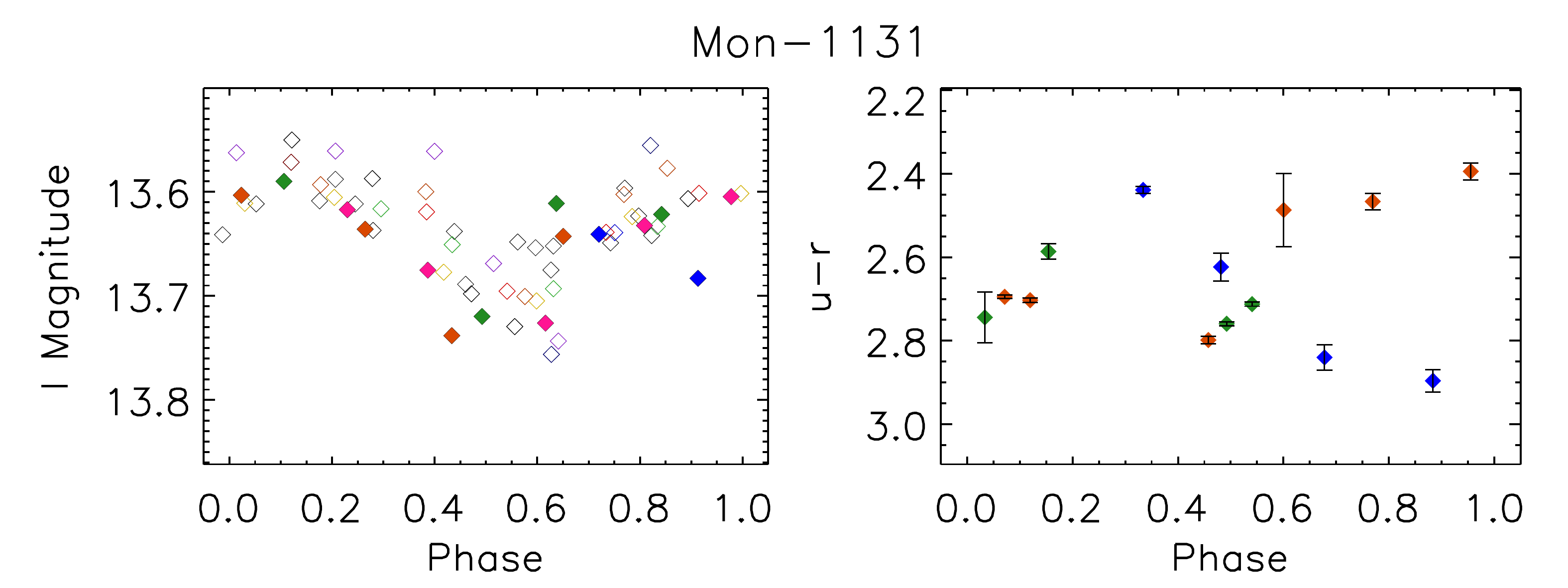}
\includegraphics[width=11.5cm]{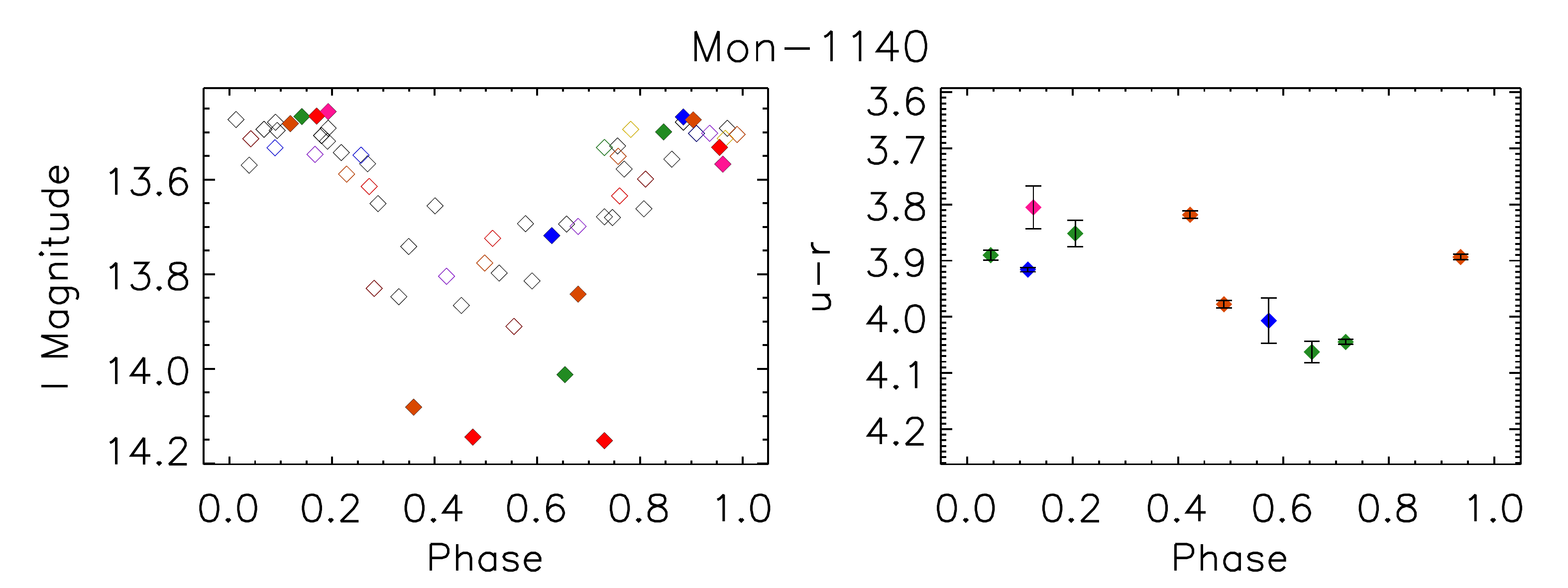}
\includegraphics[width=11.5cm]{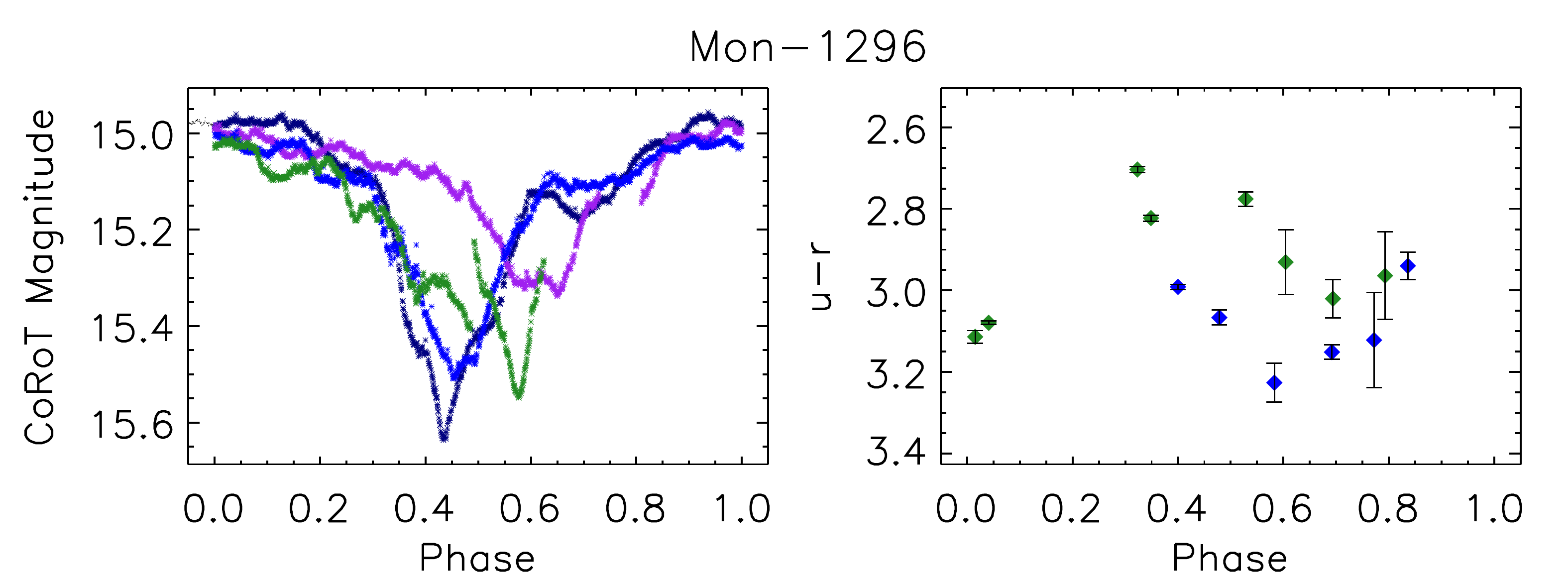}
\includegraphics[width=11.5cm]{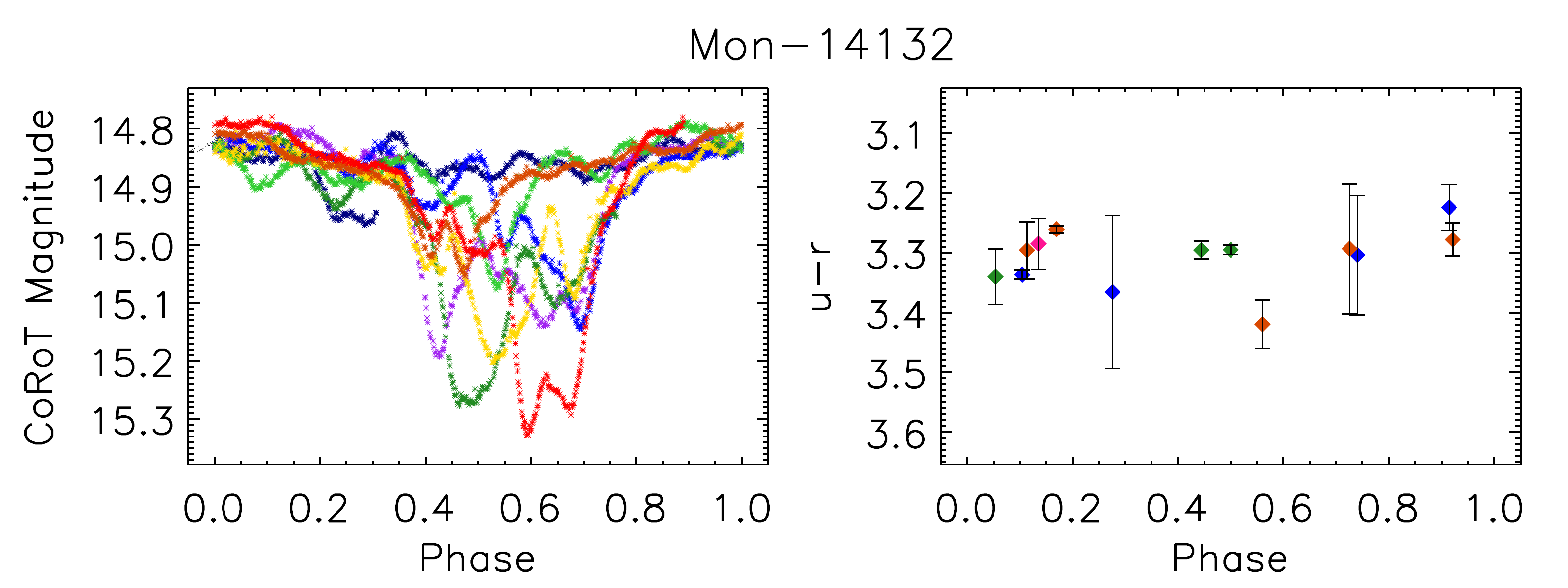}
\caption{Continued. 
}
\end{figure*}
}

\end{document}